\documentclass[a4paper,fleqn]{cas-dc}

\usepackage[authoryear]{natbib}
\usepackage{amsmath}
\usepackage{gensymb}
\usepackage{subfigure}
\usepackage[switch]{lineno}

\newcommand\aj{{AJ}}
\newcommand\apj{{ApJ}}
\newcommand\aap{{A\&A}}
\newcommand\icarus{{Icarus}}
\newcommand\mnras{{MNRAS}}
\newcommand\ssr{{SSRv}}
\newcommand\nat{{Nature}}
\newcommand\jgr{{J.~Geophys.~Res.}}
\newcommand\psj{{PSJ}}

\newcommand\maps{{M\&PS}}

\def\tsc#1{\csdef{#1}{\textsc{\lowercase{#1}}\xspace}}
\tsc{WGM}
\tsc{QE}
\tsc{EP}
\tsc{PMS}
\tsc{BEC}
\tsc{DE}

\begin{document}
\let\WriteBookmarks\relax
\def\floatpagepagefraction{1}
\def\textpagefraction{.001}
\shorttitle{The Dynamical Origin of Millimetre-Sized Sporadic Meteoroids}
\shortauthors{Do et~al.}

\title [mode = title]{The Dynamical Origin of Millimetre-Sized Sporadic Meteoroids}

\author[1]{Tam Do}[orcid=0009-0006-0795-970X]
\cormark[1]
\ead{tdo47@uwo.ca}
\ead[url]{https://github.com/tamidodo}

\author[1]{Peter Brown}

\author[2,3,4]{Petr Pokorn\'y}

\affiliation[1]{organization={Department of Physics and Astronomy, University of Western Ontario},
                addressline={1151 Richmond St.}, 
                city={London},
                state={Ontario},
                country={Canada},
                }

\affiliation[2]{organization={Astrophysics Science Division, NASA Goddard Spaceflight Center},
                addressline={8800 Greenbelt Rd}, 
                city={Greenbelt},
                state={MD},
                country={USA},
                }
        
\affiliation[3]{organization={Department of Physics, The Catholic University of America},
                addressline={ 620 Michigan Ave NE}, 
                city={Washington},
                state={DC},
                country={USA},
                }
        
\affiliation[4]{organization={Center for Research and Exploration in Space Science and Technology, NASA/GSFC},
                city={Greenbelt},
                state={MD},
                country={USA},
                }
                
\cortext[cor1]{Corresponding author}

\begin{abstract}
 Determining the relative contributions of cometary and asteroidal sources to the sporadic meteoroid population remains a longstanding challenge, particularly because commonly used orbit-based classification criteria have not been rigorously validated for meteoroids. We evaluate the efficacy of several established orbit-based criteria for meteoroid classification. These include the \citet{1954whipple} $K$-criterion, \citet{1967kresak} $Pe$-criterion, the \citet{1890Tisserand} invariant with respect to Jupiter (T$_J$), and a recent classification based on aphelion distance proposed by \citet{Borovicka2022physical}. Our validations suggest that $K$ and $Pe$ are most reliable at recovering whether a meteoroid was released from a cometary or asteroidal parent. We applied these criteria to a suite of 386 observed millimetre-sized meteoroids to try to constrain their original source populations. Our analysis used the observed orbit co-variances to backward integrate a suite of clones for each meteoroid to statistically evaluate their dynamical origin.

 We find that if meteoroids are released in the last ${\sim}150-200$ kyr, there is a dividing velocity of below 17 km/s where meteoroids in the millimetre to centimetre size range impacting Earth are predominantly asteroidal in origin, independent of the orbital criteria used. Above 17 km/s, the fraction of dynamically cometary meteoroids increases, although a definitively cometary dominated population does not arise until velocities of 27 km/s or higher. For ages older than 200 kyr, lower velocity meteoroids at Earth in the mm-sized range may be a mix of either cometary or asteroidal.

Meteoroids from different source populations experience distinct perihelion histories. Thermal processing provides an additional modifier of  physical properties making origin determinations from dynamical history or directly observed material properties challenging. We examined the role of past thermal processing in the apparent physical strength of meteoroids. The vast majority of meteoroids in our dataset which experience low perihelion are cometary in origin and exhibit higher $k_c$. This may reflect either that these meteoroids are young or that thermal cycling \citep{capek2012_streesesII} enhances fragmentation susceptibility. At the same time extreme thermal processing may preferentially remove the highest-$k_c$ (weakest) meteoroid population on cometary orbits, producing a survivor bias toward dynamically asteroidal meteoroids. If meteoroid ages are longer than 200 kyr, higher-$k_c$ populations may be dominated by cometary material that evolved onto asteroidal orbits, supporting the common assumption of distinct intrinsic compositional differences between (weak) cometary and (stronger) asteroidal meteoroids. 
  
\end{abstract}


\maketitle

\section{Introduction}
\label{intro}

The meteoritic complex has traditionally been divided into two main populations based on origin: particles that originate from asteroids and those that originate from comets \citep{1972millman}. Cometary sources can be further divided into Jupiter-family comets (JFCs), Halley-type comets (HTCs), and Oort-cloud comets (OCCs), where the HTC and OCC populations are sometimes grouped more generally as near-isotropic comets \citep{Nesvorny2011a,2017Nesvorny,2015Dones}. While it is generally assumed that most asteroidal particles are dense/strong and cometary particles are weaker \citep{1976Ceplecha,2016jenniskens}, there is recent evidence for weak asteroidal and strong cometary meteoroids \citep{Borovicka2022physical}. Thus material properties alone are insufficient to uniquely identify the parent body of a meteoroid. 

Identifying a common parentage for meteoroids is possible on short dynamical timescales. Once released from a small parent body in the inner solar system, meteoroids may be associated with a specific asteroid/comet, provided their orbit has not diffused significantly from the original parent. Such meteoroid streams are identified by similar encounter locations with the Earth over a short duration, as well as having common (similar) orbits \citep{jopek_meteoroid_2008, Jopek2017ProbabilityPairing}, usually measured with some form of orbital dissimilarity criterion \citep{Shober_Vaubaillon_2024}. Typical decoherence times for known meteoroid streams are of order 10-20 kyr \citep{Pauls2005, Shober_Courtot_Vaubaillon_2025}. Beyond this timescale (or even at similar timescales for weak showers), individual meteoroids are not easily associated either with each other or the original parent and become part of the diffuse, sporadic complex of dynamically unassociated meteoroids \citep{1998brown}. 

Quantifying the fraction of meteoroids impacting Earth originating from each parent body source is a long standing problem \citep{Whipple1967, Wiegert2009, 1985grun, Kortenkamp_2001, Sykes_Grün_Reach_Jenniskens_2004, 2024Shober}. Historical work has favoured either dominant cometary sources \citep{Whipple1967, 1970dohnanyi, Dohnanyi_1976}, asteroidal sources \citep{Dermott_Nicholson_1989,  Dermott_Durda_Grogan_Kehoe_2002} or potentially a more equal combination of both \citep{Durda_Dermott_1997}.

Dynamical and mass-balance models \citep{Wiegert2009, 2010nesvorny, Nesvorny2011a, 2020Carrillo-sanchez, 2024Pokorny} have generally found that most of the mass delivered to the Earth by meteoroids is from JFCs and that these encounter Earth at low speeds ($<35$ km/s). \citet{Wiegert2009} use radar meteor fluxes to constrain their model of the sporadic meteoroid complex, generating and integrating dust particles ranging from 10 $\mu$m to 10 cm in size, from both cometary and asteroidal parent bodies. They find that of the total meteoroid flux at Earth in their simulations, asteroidal dust comprises 1.3\% by number but that this is generally confined to low velocities. Of the low velocity ($v_\infty < 15$ km/s) sporadic meteoroid population, asteroidal material contributes around 4\%, however, they note that this is based on their assumption that asteroidal dust production in the main belt contributes a tenth of the dust that Comet 2P/Encke produces. If the main belt were to contribute one Encke-equivalent of dust per year on average, the contribution of asteroidal dust to the low velocity Earth impacting population becomes closer to 50\%. 

\citet{2010nesvorny, Nesvorny2011a} used a Wisdom-Holman integrator to track the orbits of particles in a model of the Zodiacal Cloud, with mid-infrared and radar flux measurements to constrain the contributions of each source population (HTCs, OCCs, JFCs, asteroids) to the overall cloud model. They apply the \"{O}pik collisional probability algorithm to their validated model to estimate the mass influx of particles impacting Earth and find that JFC particles have the highest rate of impact. If this dynamical model of the Zodiacal Cloud is correct, the material ablating in Earth's atmosphere should reflect the dominant contribution of cometary particles. \citet{2020Carrillo-sanchez} employ a Chemical Ablation MODel (CABMOD) that includes multiphase treatment of meteoroid ablation for both bulk silicate and Fe-Ni metal grains to show that for the Zodiacal Cloud Model proposed by \citep{Nesvorny2011a, 2010nesvorny}, the majority of the particles that ablate in Earth's atmosphere are consistent with the predicted influx and corresponding atomic ablation rate height profiles expected from JFCs with CI-like composition.

Across these studies, the inferred asteroidal contribution ranges from only a few percent to potentially comparable fractions with JFC input depending on model assumptions and particle size, highlighting substantial uncertainty in the relative contributions of cometary and asteroidal sources. It is likely that some of the discrepancy among these studies as to which minor body population dominates is rooted in variation in origin with size as suggested by some models \citep[e.g.][]{2016carrillo-sanchez}. Thus mass input may be JFC dominated (indicating that at the mass peak around 0.1mm diameter dust is JFC-derived) while at larger sizes asteroidal objects may predominate \citep{Hughes1994}. 

\citet{2002dermott} argue that the structure of the solar system dust bands discovered by IRAS show that asteroidal collisions are the dominant source of particles in the Zodiacal Cloud. It would then follow that the dominant fraction of the interplanetary dust particles in Earth's stratosphere also originated from asteroids and were formed from prolonged mechanical mixing in the deep regoliths of asteroidal rubble piles in the main belt. However, studies of micrometeorites that have been collected from the Earth's surface have identified a substantial population of primitive particles whose mineralogy and textures are more consistent with cometary sources, suggesting that cometary material may contribute significantly to the collected micrometeorite flux \citep[e.g.][]{Genge2008,Genge2017}.

Historically, one approach to classification has been to divide the meteoroid population based on orbital dynamics. At centimetre-to-metre sizes, non-gravitational forces are too weak to decouple meteoroids of this size from close encounters with Jupiter before such encounters recur so their orbital evolution over $\sim$10 kyr timescales should mirror that of their cometary parent bodies. Based on this argument that the chaotic orbit diagnostic of JFCs are inherited by genuinely cometary meteoroids and stable trajectories indicate asteroidal material diffused from the Main Belt, \citet{2021Shober} and \citet{2024Shober} backward-integrated sporadic Desert Fireball Network events on JFC-like orbits and found only $\sim$1-5\% of 646 fireballs to be dynamically consistent with JFCs, with the majority residing on stable orbits protected from close encounters by mean-motion and Kozai resonances.

Our analysis focuses on millimetre-sized meteoroids, corresponding to the size range most readily detected by ground-based optical systems at the velocities considered here \citep{2021Vida}. The goals of this paper are
\begin{enumerate}
    \item To evaluate the applicability of existing orbit-based criteria for classifying objects as asteroidal or cometary in the millimetre-size regime,
    \item To estimate the fractional contribution of asteroids versus comets to the sporadic meteoroid flux at the Earth at millimetre-sizes and low ($\leq$ 35km/s) speeds,
    \item To investigate the potential for a meteoroid's thermal processing history to constrain its dynamical origins via physical characteristics
\end{enumerate}

While some meteoroids arriving at Earth may be from reservoirs beyond Jupiter, we do not consider outer solar system sources such as Edgeworth-Kuiper belt objects (EKBOs), centaurs, or long-period comets in our analysis for several reasons. Among these are 
\begin{enumerate}

\item Dynamical evolution from outer solar system sources with starting perihelion distances beyond Jupiter (EKBOs, centaurs) takes over 10 Myr to evolve into meteor orbits that intersect Earth for the millimetre diameter range considered in this manuscript, much longer than typical collisional lifetimes at these sizes \citep{Poppe_2016}.

\item Meteoroids originating in Jupiter-crossing capable source populations (Halley-type and Oort Cloud comets) are vastly different from our observed data set which is limited to Earth impact speeds below 35 km/s (where HTC and OCC have $a > 10$ AU, $i > 45^\circ$) and dynamical models suggest that they are very unlikely to evolve into meteor orbits discussed in this work \citep{Poppe_2016}.

\item The Jupiter-barrier is more efficient in preventing larger outer solar system particles from entering the inner solar system, effectively diminishing the outer solar system contribution to the inner solar system budget in the millimetre size range to negligible amounts \citep{Poppe_2016}. 
\end{enumerate}

\section{Method}

Our approach extends these earlier works done to define orbit-based criteria for determination of meteoroid origins \citep{1890Tisserand, 1954whipple, 1967kresak, Borovicka2022physical, 2014Tancredi} by combining observational constraints with dynamical modelling to statistically infer meteoroid origins. To accomplish this, we start by computing orbital histories for 386 meteoroids observed as meteors in Earth's atmosphere with ground-based cameras. Of this total, we first validate the existing orbital criteria on 35 meteors associated with meteor showers as well as synthetic meteoroids starting on cometary orbits. Using the validated orbital criteria and additional ablative criteria, we then assign the remaining sporadic meteoroids either a cometary or asteroidal origin. This approach enables us to evaluate commonly used orbit-based classification criteria within a probabilistic framework and to assess how orbital evolution and epoch of ejection influences inferred parent body origin. By coupling these dynamical constraints with observable physical properties, we also investigate how processes such as thermal evolution may link meteoroid origin to material strength.

\subsection{Observational Datasets}
\label{obsdata}

\subsubsection{Cameras and Processing/Detection Software}

The first of the two instruments used to record our observational dataset is the Canadian Automated Meteor Observatory (CAMO) narrow-field fast-response mirror system, based on the AIM-IT system developed by \citet{2004gural}. CAMO is made up of two automated observatories with fixed-pointing camera systems, located at sites separated by a baseline of 50 km in Southwestern Ontario, Canada. The system is cued by a wide field camera which has a limiting detection magnitude of +5. We collected and measured a sample of 117 two-station meteors detected with CAMO's high precision mirror tracking system between 2016-2024, with uncertainties on the order of 10 m/s in speed and trajectory residuals under a metre. 

The second instrument, which contributed 269 additional meteors to our dataset collected between 2016-2024, is the Electron Multiplied CCD (EMCCD) camera system \citep{2022gural}. Although the EMCCD cameras have lower spatial and temporal precision than the CAMO cameras, they are sensitive to fainter magnitudes, having a peak magnitude detection limit of  +6.5 and a per-point detection limit of +8. Velocity uncertainties are on order of 10-100 m/s and trajectory residuals are under 10 m for our sample. A comparison of the CAMO and EMCCD system specifications can be found in \autoref{tab:cameraspecs}. We restricted our analysis only to sporadic meteors (except where mentioned in the validation section) using the shower radiant and speed association criteria summarized by \citet{Vida2021_GMN}.

\begin{table}[ht]
    \centering
    \begin{tabular}{m{15mm}|m{31mm}|m{22mm}}
        Instrument & CAMO (Narrow-Field) & EMCCD \\
        \hline
        Number of pixels & $1024\times1024$ & $512\times512$ \\
        \hline
        Frame rate & 100 fps & 32 fps \\
        \hline
        Bit-depth & 14-bit & 16-bit \\
        \hline
        Stellar limit & $+7^M$ & $+10.5^M$ \\
        \hline
        Meteor limit & $+5^M$ & $+6.5^M$ \\
        \hline
        Detection software & ASGARD & DetApp/ASGARD \\
        \hline
        Camera & Prosilica GX1050 & Hnu1024HS \\
        \hline
        Optics & 545 mm $f$/11 & 50 mm $f$/1.2 \\
        \hline
        Intensifier & 18 mm ITT FS9910 Gen-III &  \\
        \hline
        FOV size & $1.5\degree$ & $14.7\degree$ \\
        \hline
        Precision & 4 m at 100 km & 20 m at 100 km \\
        \hline
    \end{tabular}
    \caption{Summary of CAMO and EMCCD specifications, as described by \citet{2021Vida, 2013weryk, 2022gural}}
    \label{tab:cameraspecs}
\end{table}

To ensure the quality of the trajectory solutions for meteors used from the EMCCD dataset, we require events to be captured over a minimum of 6 frames for at least one camera, have a convergence angle between stations of greater than $5\degree$, include only meteors which begin and end entirely within the field of view of at least one camera, and exclude meteors captured at the edge of the frame (minimum and maximum $x,y$ position not within 50 pixel units of the edge of the frame). For further details on quality assurance measures, see \autoref{appendixb}.

While CAMO events are all manually reduced by members of the Western Meteor Physics Group, only 41 of the EMCCD events ($\sim 15\%$) used were manually reduced. All events have automatic trajectory solutions generated by the software \verb|pylig|, as described in \citet{pylig}\footnote{Publicly available at https://github.com/wmpg/WesternMeteorPyLib}. After some initial testing and comparison between manually and automatically reduced EMCCD events, we found that it was possible to use the automated detections with limited manual adjustment to achieve good initial velocity and radiant accuracy. 

\begin{figure}[h!]
    \centering
    \subfigure[]{\includegraphics[width=0.49\textwidth]{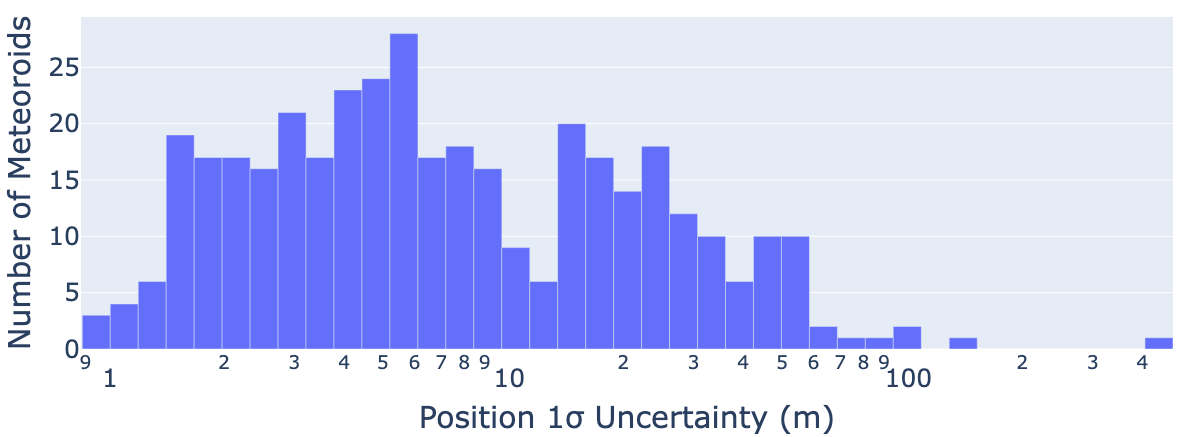}}
    \subfigure[]{\includegraphics[width=0.49\textwidth]{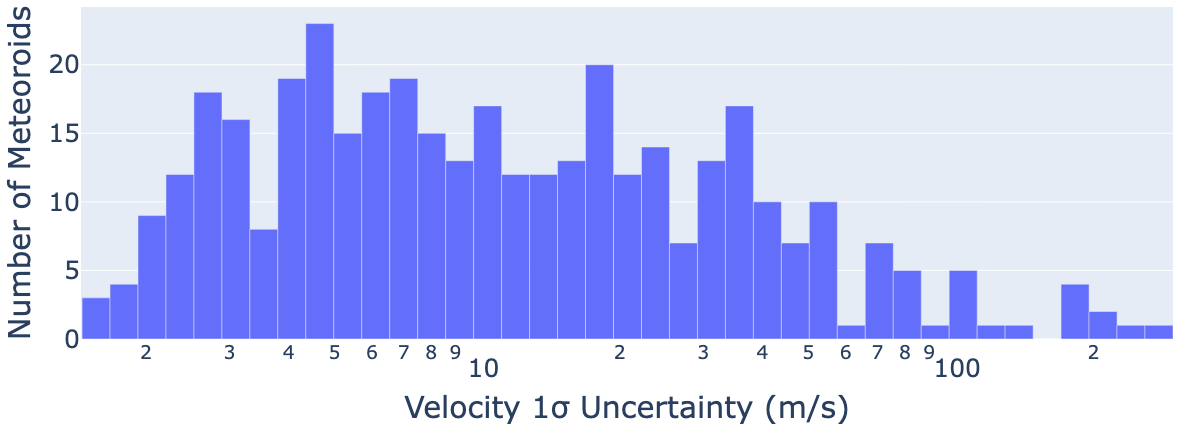}}
    \caption{The distribution of $1-\sigma$ uncertainties ($\sigma_r = \sqrt{\sigma_x^2+\sigma_y^2+\sigma_z^2}$, $\sigma_v = \sqrt{\sigma_{v_x}^2+\sigma_{v_y}^2+\sigma_{v_z}^2}$) for the 386 meteoroids in the combined CAMO/EMCCD dataset. Both panels have binning uniform in the log-space, with 14 (top panel) and 17 (bottom panel) bins per order of magnitude.}
    \label{fig:spread}
\end{figure}

We further improve the quality filtering for the EMCCD dataset by examining the lags and initial velocities produced by \verb|pylig| and readjusting the sliding window of points included in the calculation of the meteoroid's initial velocity at the top of the atmosphere. A full description of the secondary quality filtering can be found in \autoref{appendixb}, where we confirm that these adjusted automatic picks produce nearly identical initial velocity vectors to manual reductions.

\verb|pylig| produces a number of data outputs relevant to creating high fidelity simulations of meteoroid orbital evolution. Initial position and velocity state vectors, along with covariance matrices for their uncertainties derived directly from measurements, are provided in the Earth-Centered Inertial (ECI) frame, in the epoch of date. The \verb|REBOUND| integration code uses the Solar System Barycentric frame in J2000 when using the built-in code for pulling initial state vectors for other solar system bodies from the JPL Horizons API, so we convert the meteoroid's state vectors to the same frame and epoch. For each meteoroid, we generate 100 clones to sample the uncertainty space defined by the covariance matrices and integrate them backwards to show the spread in orbital uncertainty over the integration time. A sample visualization of the initial clone distribution in position and velocity space is shown in \autoref{fig:spread}.

The cameras also produce light intensity as a function of time, which is used to estimate the pre-atmospheric meteoroid mass. Together with estimates of bulk density, this helps define the ratio of radiation force to gravity (called $\beta$). This is used in the orbital evolution calculation and is different for each meteoroid. 

The luminosity equation, which describes the energy released by the meteoroid in the instrumental passband \citep{1998ceplecha}, from the combination of line emission of evaporated meteor atoms and atmospheric atoms is:

\begin{equation}
    I = \tau\frac{dE_k}{dt} = -\tau\left(\frac{v^2}{2}\frac{dm}{dt} + mv\frac{dv}{dt}\right),
\end{equation}

where $\tau$ is the fraction of kinetic energy lost by the meteoroid (luminous efficiency). In the case of small meteoroids where the deceleration is small and energy loss is dominated by fragmentation and subsequent ablation of released grains, the second term becomes negligible and the equation becomes simply: 
\begin{equation}
    I = \tau\left(-\frac{dm}{dt}\right)\frac{v^2}{2}.
\end{equation}

From there, assuming an average meteor speed and considering mass loss a positive quantity, we can integrate the equation to get the estimated total initial mass from 
\begin{equation}
    m = \int_0^m dm = \frac{2}{\tau v^2}\int_{0}^{t}Idt,
\end{equation}
where the luminous efficiency $\tau$ is estimated as a function of $v$.

\subsubsection{Population Demographics}\label{subsubsec:popdemo}

We compute orbital histories for a total of 386 meteoroids; 117 from CAMO observations and 269 from EMCCD records, of which 35 are associated with meteor showers. Shower associations are made using the meteor shower table from \citet{2018jenniskens}, with a fixed angular radius of association of $3\degree$ and a maximum geocentric velocity threshold of 10\%. The full raw dataset for both instruments can be found on Zenodo, as linked in \autoref{appendixa}. These meteoroids were selected for their observation quality and suitability for long-term backwards integrations (see \autoref{appendixb} for quality criteria). This large population of measured meteors helps us characterize the millimetre-sized meteoroid population that intersects the Earth at low speeds.

\begin{figure}[h!]
    \centering
    \includegraphics[width=0.48\textwidth]{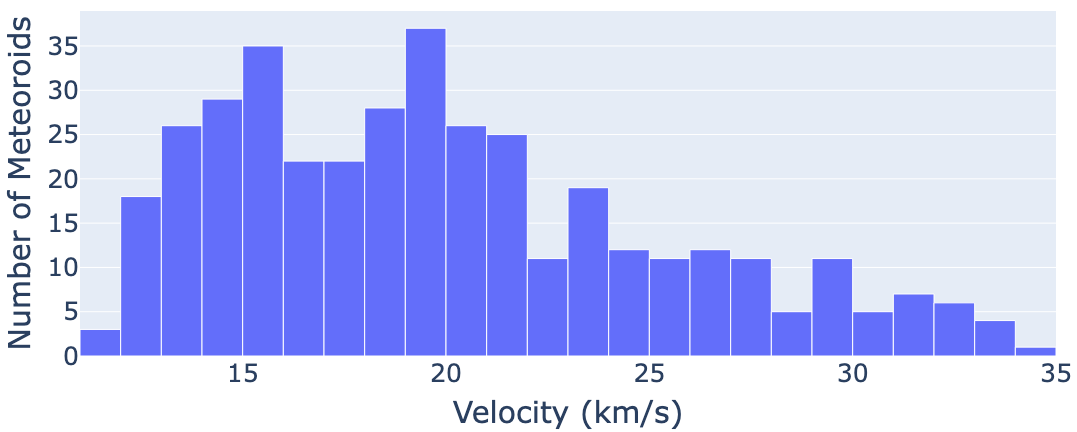}
    \caption{Histogram showing the initial speed for our dataset of 386 CAMO/EMCCD meteors. We explicitly do not sample above 35 km/s as this population is cometary-dominated on dynamical grounds.}
    \label{fig:vel-pop}
\end{figure}

Higher velocity meteors ($v\geq 35$ km/s) are almost certainly cometary in origin \citep{2019Vojacek} and although there may be a few outliers, it is highly improbable that the origin could be from debris released by Main Belt asteroids (MBAs). The main degeneracy between asteroidal and cometary meteoroids happens at lower velocities so we restrict our dataset to that range. A histogram showing the distribution of velocities in our dataset is shown in \autoref{fig:vel-pop}. Our dataset is heavily biased, as it only includes those with an orbit intersecting Earth's so our model will be incomplete. However, with the objective of determining origins for meteors (having sizes larger than 1 mm) observed by camera systems like CAMO and the EMCCDs colliding with Earth, our population of interest aligns with our dataset. 

To estimate the diameters of the meteoroids, we assume a spherical shape and compute their photometric mass. The photometric mass was found assuming a fixed luminous efficiency of $\tau=0.7$\% \citep{Subasinghe2017} for all automated solutions, while manually reduced events had masses computed using the erosion model of \citet{Borovička_2019_book} and the luminous efficiency formulation given in \citet{2021Vida}. 

\begin{figure}[h!]
    \centering
    \subfigure[]{\includegraphics[width=\linewidth]{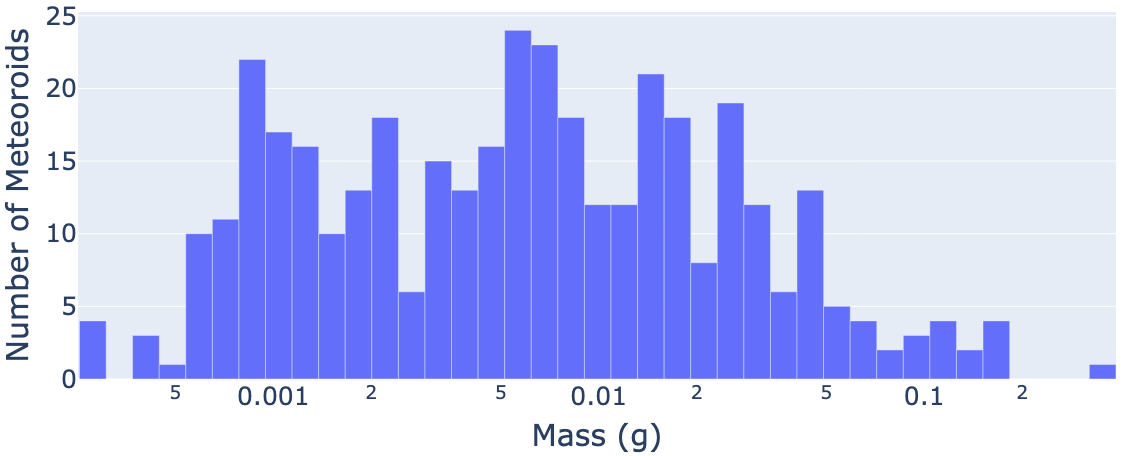}}
    \subfigure[]{\includegraphics[width=\linewidth]{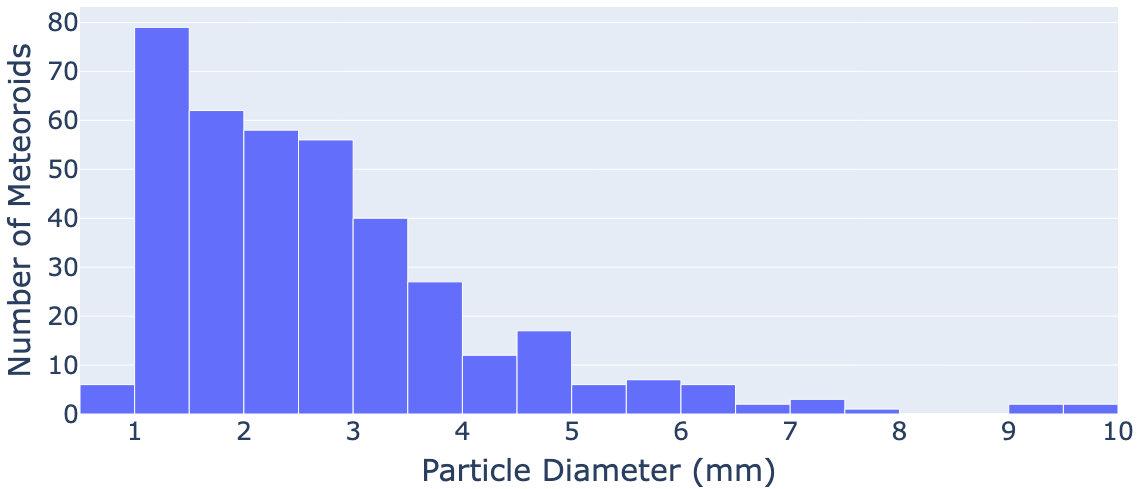}}
    \caption{Initial masses (a) and diameters (b) of the meteoroids observed by CAMO and EMCCD in our dataset. The majority of particles are 0.05g or less and around 1-5 mm in diameter.}
    \label{fig:mass-diam}
\end{figure}

We adopted a bulk density estimated from the meteor ablation behaviour. We assume the framework of density groupings using a one-dimensional parametrization, termed the $K_B$ parameter \citep{Ceplecha_1967}: 
\begin{equation}
    K_B = \text{log}\rho_B + 2.5\text{log}v_\infty - 0.5\text{log[cos(}z_R)]
\end{equation}
where the air density at the beginning of the luminous trajectory, $\rho_B$, can be calculated using the NRLMSISE-00 model of the atmosphere \citep{nrlmsise}, and $v_\infty$ (the initial velocity in cm/s) and $z_R$ (the zenith distance of the radiant) can be taken directly from the observation data. The estimated diameters and photometric masses for our population are shown in \autoref{fig:mass-diam}. We note that even a change in luminous efficiency by a factor of two will modify $\beta$ by $\approx$25\%.

\citet{1988ceplecha} proposed the following mapping based on assumed composition and parent body association, noting that the begin heights of meteors had distinct groupings when normalized by speed and entry angle:

\begin{itemize}
    \item Ordinary chondritic asteroidal meteors have $8.0 \leq K_B$ and a bulk density of $\rho = 3.7$ g/cm$^3$
    \item Group A (carbonaceous chondritic material from either asteroids or comets) have $7.3 \leq K_B < 8.0$ and a bulk density of $\rho = 2.0$ g/cm$^3$
    \item Group B (dense cometary material) have $7.1 \leq K_B < 7.3$; $q \leq 0.3$ AU and a bulk density of $\rho = 1.0$ g/cm$^3$
    \item Group C (regular cometary material) have $6.6 \leq K_B < 7.1$ and a bulk density of $\rho = 0.75$ g/cm$^3$
    \item Group D (soft cometary material) have $K_B < 6.6$ and a bulk density of $\rho = 0.27$ g/cm$^3$
\end{itemize}

\begin{figure*}[h!]
    \centering
    \includegraphics[width=\textwidth]{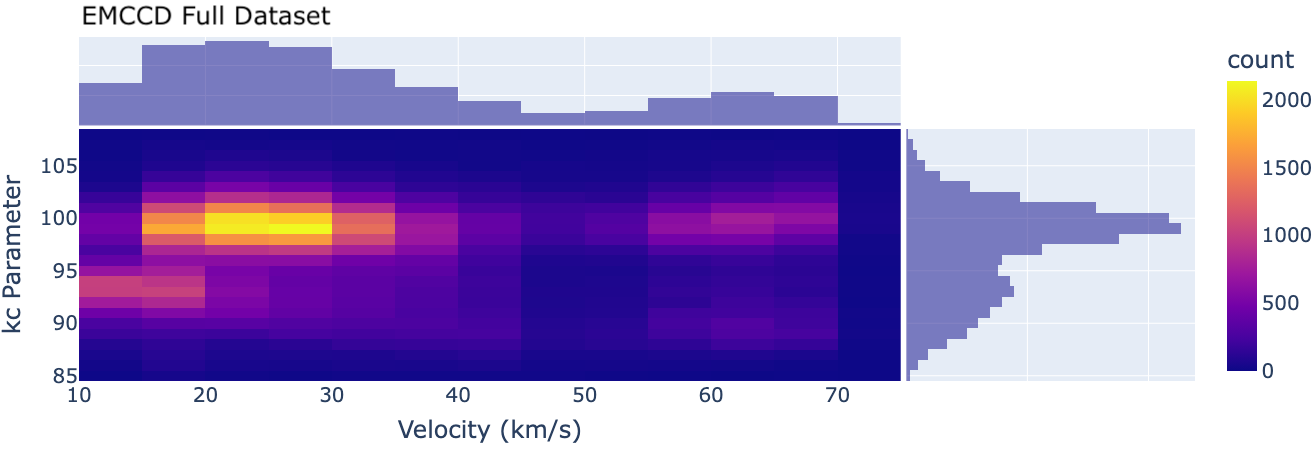}
    \includegraphics[width=\textwidth]{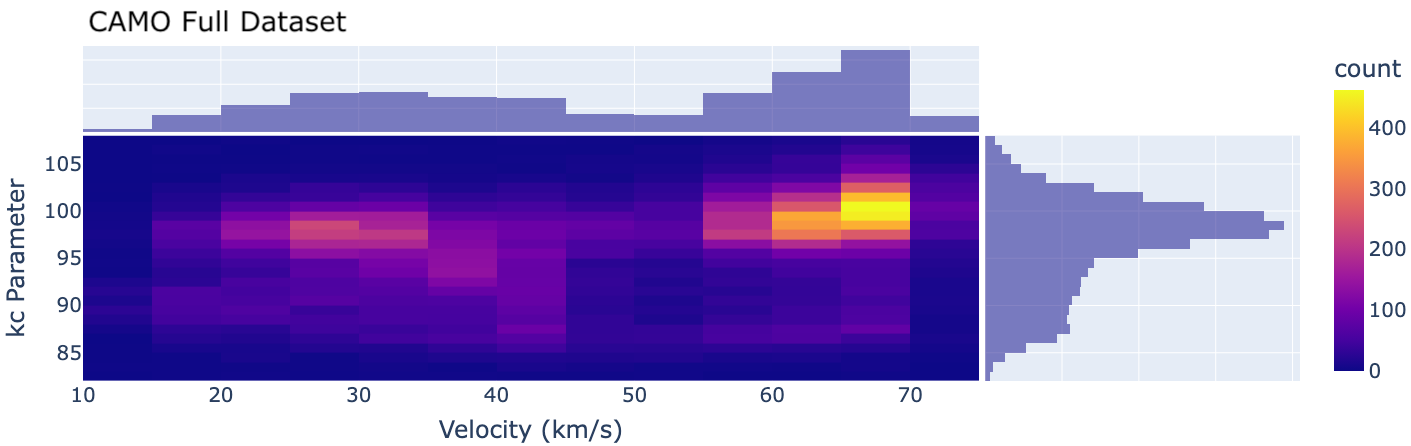}
    \caption{Density heat map of full EMCCD dataset of meteors ($N=113928$) (a) and full CAMO wide-field dataset of meteors ($N=18695$) (b) with histograms in both $k_c$ and velocity, restricted to the window $85 \leq k_c \leq 108$, and $12$ km/s $\leq V_\infty \leq 72$ km/s.}
    \label{fig:kc-group}
\end{figure*}

However, the Super Schmidt cameras that were used for the original classification \citep{Ceplecha_1967} have important spectral sensitivity differences from our cameras. To find analogous groups for the EMCCD and CAMO cameras, we plot the $K_B$ parameter against velocity for all observed EMCCD meteors (2016-2024, $N=113928$) and CAMO meteors (2010-2026, $N=18695$), not restricted to the low velocity meteoroids we study in this paper. We see similar clustering but the delineation between groups is not clear enough to create equivalent boundaries between groups. A later paper building on the characterization of these groups from \citet{2016jenniskens} uses a similar but improved parameter, $k_c$, defined as 
\begin{equation}
    k_c = H_b + \frac{2.86 - 2.00\log V_\infty}{0.0612}
\end{equation}
where $H_b$ is the beginning height (km) and $V_\infty$ is the initial velocity (km/s). The groups analogous to \citet{1988ceplecha}'s proposed by \citet{2016jenniskens} is as follows:
\begin{itemize}
     \item Ordinary chondritic asteroidal meteors have $k_c < 85$ and a bulk density of $\rho = 3.7$ g/cm$^3$
    \item Group A (carbonaceous chondritic material from either asteroids or comets) have $85\leq k_c < 91$ and a bulk density of $\rho = 2.0$ g/cm$^3$
    \item Group B (dense cometary material) have $91 \leq k_c < 95$ and a bulk density of $\rho = 1.0$ g/cm$^3$
    \item Group C (regular cometary material) have $95 \leq k_c < 100$ and a bulk density of $\rho = 0.75$ g/cm$^3$
    \item Group D (soft cometary material) have $100 \leq k_c$ and a bulk density of $\rho = 0.27$ g/cm$^3$
\end{itemize}

In \autoref{fig:kc-group}, we plot the $k_c$ parameter against velocity for the same full EMCCD and CAMO datasets. Comparing our plots to the criteria outlined by \citet{2016jenniskens}, we see a more direct correlation of our groupings to theirs, systematically shifted up in $k_c$. The high (A/B) and low (C/D) $k_c$ population is well separated.

Our final grouping criteria and the number of meteoroids from our sample in each category, matched to \citet{1988ceplecha}'s populations were chosen as follows:

\begin{itemize}
    \item Asteroidal group (0/386 meteoroids) has $k_c < 87$ for EMCCD meteors and $k_c < 85$ for CAMO meteors
    \item Group A (38/386 meteoroids) has $87 \leq k_c < 92$ for EMCCD meteors and $85 \leq k_c < 91$ for CAMO meteors
    \item Group B (78/386 meteoroids) has $92 \leq k_c < 96$ for EMCCD meteors and $91 \leq k_c < 95$ for CAMO meteors
    \item Group C (237/386 meteoroids) has $96 \leq k_c < 103$ for EMCCD meteors and $95 \leq k_c < 103$ for CAMO meteors
    \item Group D (33/386 meteoroids) has $103 \leq k_c$ for both EMCCD and CAMO meteors
\end{itemize}

From the $k_c$ values of our shower meteoroids (presented in \autoref{tab:stage1test} and \autoref{tab:stage1testSTA}), all observed shower meteoroids with known parent bodies (except one Encke meteoroid) fall into the cometary density groups B/C/D, making us confident in the accuracy of these ranges.

\subsection{Overview of Orbital Criteria}

\begin{figure*}[h!]
    \centering
    \subfigure[]{\includegraphics[width=0.49\linewidth]{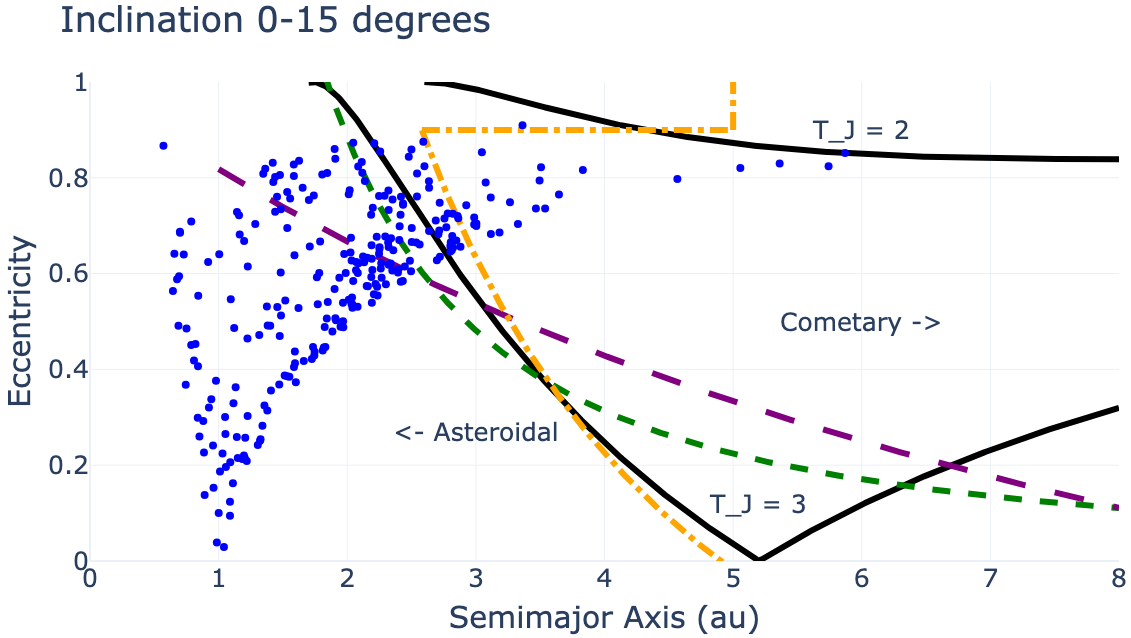}} \subfigure[]{\includegraphics[width=0.49\linewidth]{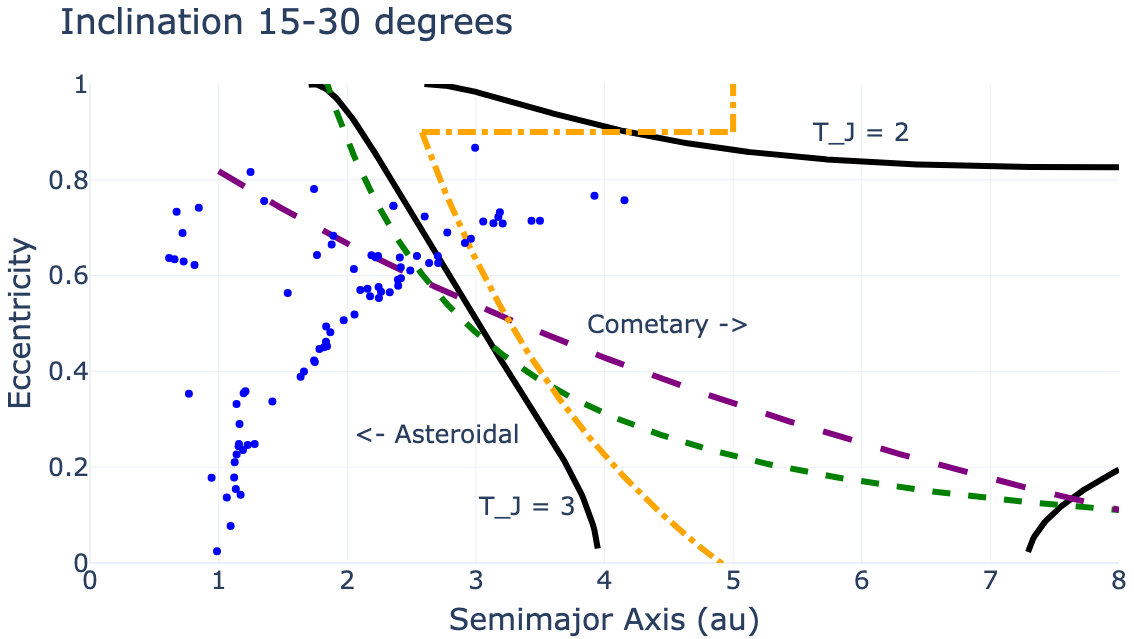}}
    \subfigure[]{\includegraphics[width=0.49\linewidth]{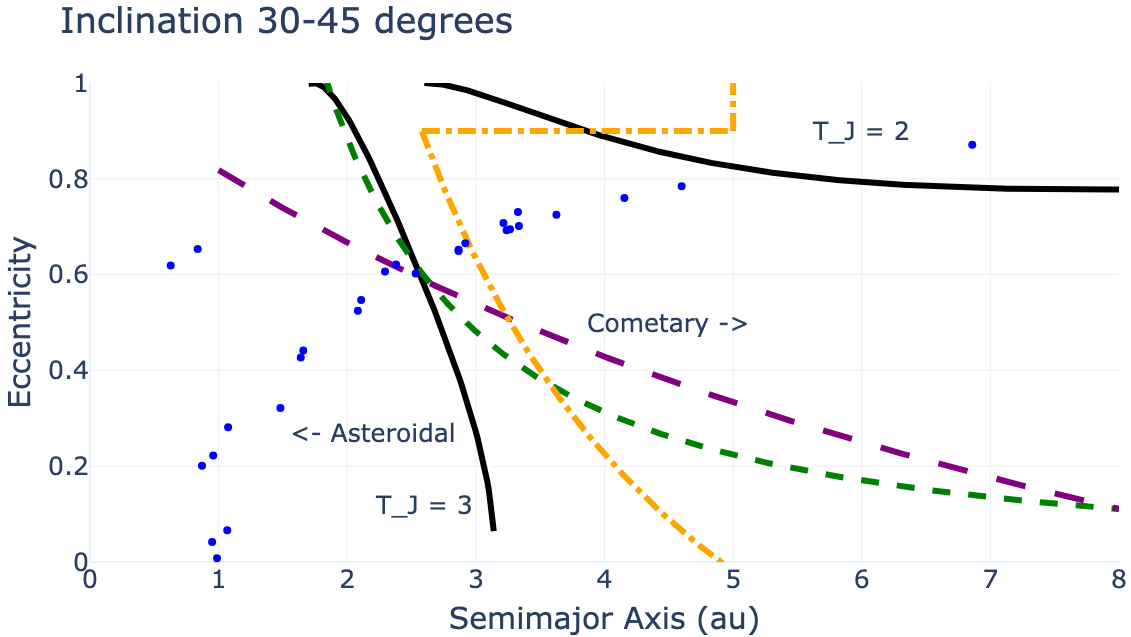}}
    \subfigure[]{\includegraphics[width=0.49\linewidth]{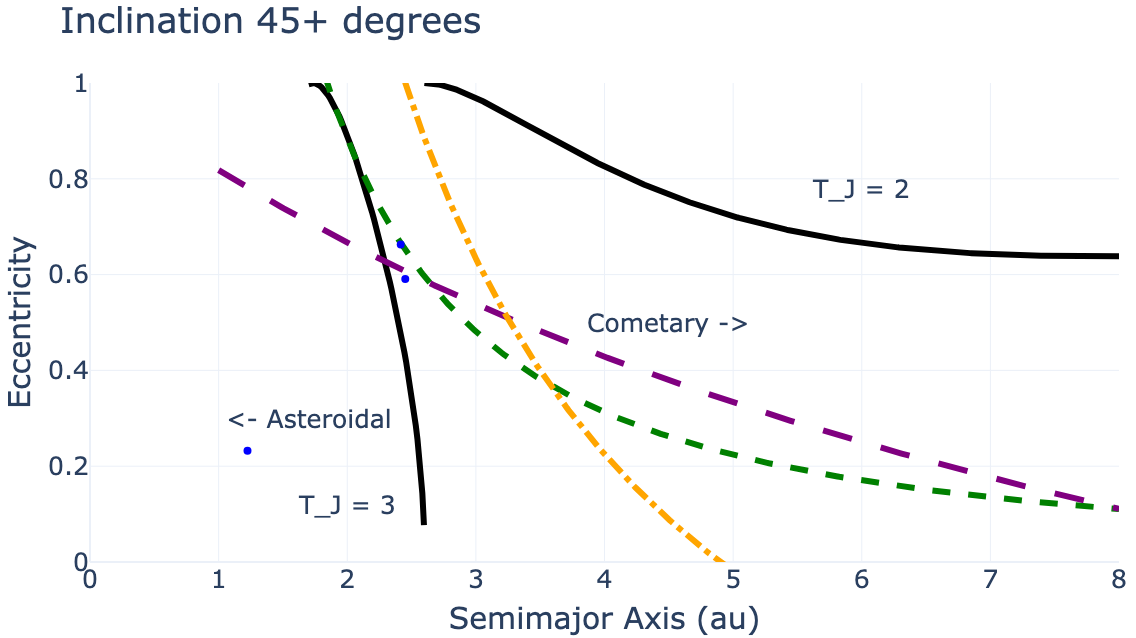}}
    \caption{A visual representation of the boundaries set by the \citet{Borovicka2022physical} Q cutoff criteria (orange, dash-dot), \citet{Whipple1967} $K$-criterion (purple, long dash), \citet{1967kresak} $Pe$-criterion (green, short dash), and Tisserand parameter with respect to Jupiter (black, solid) at 4 different inclination angles. Cometary objects are to the right of the coloured lines and asteroidal objects are to the left. The blue markers are the observed CAMO and EMCCD meteors in our dataset with the mean $e,a$ for each meteoroid's clones at the time of impact with Earth.}
    \label{fig:criteria-visual}
\end{figure*}

Although composition and material strength have been used to classify meteoroid origins, we have highlighted the potential for misclassification using this approach. Another option is to differentiate between asteroidal and cometary origins based on orbital dynamics. In this section, we describe the various criteria that have been suggested and evaluate their overall ability to classify small bodies. A visual representation summarizing the various criteria in $a-e$ space is shown in \autoref{fig:criteria-visual}. 

\subsubsection{Tisserand Parameter}

The oldest orbital discriminant for small body populations is the Tisserand invariant with respect to Jupiter, $T_J$ \citep{1890Tisserand}. This metric exploits the fact that Jupiter is normally the dominant gravitational body driving evolution of small bodies in the inner solar system. The Tisserand invariant generally describes the effects a planet has on a particle in a three-body problem with the Sun as the dominant body and the planet as a secondary influence \citep{1890Tisserand}. Because cometary bodies often evolve on orbits that have $a, e, i$ affected by Jupiter's gravitational influence, their Tisserand parameter with respect to Jupiter, $T_J$, is lower. Physically, this can be interpreted as cometary encounters with Jupiter tending to have greater changes in velocity compared to objects with higher $T_J$. The parameter is defined as 

\begin{equation}
    T_J=\frac{a_J}{a}+2\sqrt{\frac{a}{a_J}(1-e^2)}\cos{i}
\end{equation}
or equivalently:
\begin{equation}
    T_J=2\left(\frac{a_J}{q+Q}+\sqrt{\frac{2qQ}{(a_J)(q+Q)}}\cos{i}\right)
\end{equation} 
where $a_J$ is the semimajor axis of Jupiter and $a, e, q, Q, i$ are the semimajor axis, eccentricity, perihelion, aphelion and inclination of the meteoroid respectively. 

The main weakness of $T_J$ for our analysis is that it is invariant only under non-dissipative forces. For smaller meteoroids, where Poynting-Robertson (PR) drag plays a role, $T_J$ will vary to some degree over time. This is inherently a problem of all orbital criteria because they cannot include the effects of size and duration for dissipative forces unless the size and dynamical age of the particle are taken into account.

\subsubsection{Whipple's $K$-criterion}

\citet{1954whipple} used an early double station photographic camera setup to capture 144 meteors and derive their orbits. Among these, 95 were associated with meteor showers. They proposed a comet-asteroid orbital criterion to determine if any of their observed meteor streams could be asteroidal in origin. Based on the qualitative analysis of the known comet and asteroid orbits at the time, they noted that cometary orbits tend to have larger $e, a, i$ than asteroidal orbits but that many asteroids also have higher $i$ than that of many periodic comets. As a result, their empirical $K$-criterion does not depend on inclination and simply separates orbits based on $Q$ and $e$: 
\begin{equation}
    K = \text{log}_{10}\left[\frac{Q}{1-e}\right] - 1,
\end{equation}
with $K < 0$ for asteroids and $K > 0$ for comets. At the time the criterion was proposed, only 3 asteroids were known to violate the criterion which gave confidence in its empirical validity. 

However, there are now 875150 numbered asteroids and 607 numbered comets catalogued in the Jet Propulsion Laboratory (JPL) Horizons database \footnote{https://ssd.jpl.nasa.gov/horizons/, accessed January 6, 2026}. Applying the $K$-criterion to the full catalogue, we find the criterion misclassifies 1874 asteroids (0.21\%) and 198 comets (32.62\%). The majority of asteroids misclassified as comets were TransNeptunian Objects (TNOs) and the comets misclassified as asteroids were a mix of Encke-type comets (52/198) and JFCs (146/198). This suggests mixing of extinct comets among asteroidal populations, as well as the lack of consideration for the TNOs since at the time, Pluto was the only known TNO.

\subsubsection{Kresak's $Pe$-criterion}

Just over a decade after \citet{1954whipple} published their $K$-criterion, \citet{1967kresak} noted that there was no physical reasoning for their formula and other functions of $a$ and $e$ could be better used to differentiate orbits between asteroidal and cometary. In order to minimize the exceptions to the criteria, they proposed an updated comet-asteroid orbital classification in the form of the $Pe$-criterion:
\begin{equation}
    Pe = a^{3/2}e
\end{equation}
where orbits with $Pe < 2.5$ were asteroids and $Pe > 2.5$ were comets. Here, $P$ represents the orbital period (a proxy for $a$). 

Using the small body population known today, applying the $Pe$-criterion classifies 1970 asteroids and 91 comets incorrectly. Compared to the $K$-criterion, the $Pe$-criterion greatly improves the ability to separate JFCs from asteroids based on orbits alone. The $Pe$-criterion has a modern asteroid classification that is 99.77\% accurate.

Even with high classification accuracy, \citet{1967kresak} point out that both the $K$-criterion and $Pe$-criterion are not necessarily reliable indicators of meteor origin, as neither criterion takes into account the non-gravitational effects that affect the orbits of smaller particles significantly, such as PR drag and other radiation pressure effects. Particularly, cometary meteors from some source regions can pass Jupiter's perturbational barrier due to long dynamical lifetimes and significant PR drag \citep{2014Pokorny}, whereas comets are much less likely to be subject to forces that allow such a passage.

\subsubsection{Borovi\v{c}ka's $Q$-based Cutoff}

Accepting the notion that the influence of Jupiter is the driving factor in determining whether a small-body is dynamically asteroidal or cometary, \citet{1969kresakpt1} proposed a boundary based on the premise that objects captured by Jupiter into short-period orbits are prevented from having their aphelions reduce far inwards from Jupiter's orbit. They propose a boundary at $Q = 4.6$ with the understanding that orbits with $Q > 4.6$ could have started off on larger orbits and been captured by Jupiter. In contrast, orbits with $Q < 4.6$ would be dynamically isolated from close encounters with Jupiter and more stable and hence more similar to MBAs.

Building on this earlier work,  \citet{Borovicka2022physical} proposed a $Q$-cutoff criterion informed from an analysis of the physical properties of meteoroids. They found that objects with $Q > 4.9$ tend to be weaker than those on smaller orbits. They note that their limit of $Q = 4.9$ AU is very close to Jupiter's perihelion distance of 4.95 AU. They examined 824 fireballs in the centimetre and larger range based on the ``maximum dynamic pressure suffered by the meteoroid in the atmosphere" or $P_f$ which is a proxy for the global physical strength of the meteoroid. Correlating $P_f$ with orbital parameters, they proposed an empirical criteria that classified meteoroid orbits as follows \citep{2022Borovickacalculations}:

\begin{itemize}
    \item Asteroidal: $Q < 4.9$ AU
    \item Also asteroidal: $4.9$ AU $< Q < 7$ AU, $a < 5$ AU with EITHER $q < 0.25$ AU OR $i > 40\degree$
    \item Short-period cometary: $Q > 4.9$ AU, $e < 0.9$ OR $q>0.25$ AU, $i<40\degree$ AND $a < 5$ AU 
    \item Long-period cometary: $a > 5$ AU
    \item Unclassified: $Q > 4.9$ AU, $a < 5$ AU, $i < 40\degree$, $e > 0.9$ AND $q < 0.25$ AU 
\end{itemize}

It should be noted that unlike the $K$ and $Pe$ criteria, these dynamical boundaries were set based on observations of meteoroid susceptibility to ablation, with the important assumption that more resistant material is more likely to be of asteroidal origin (with the exception of irons that ablate easily and have low $P_f$). For example, short-period orbits with low perihelia are dynamically more likely to be cometary but only a subset of more resistant material can survive repeated low distance perihelion passages (see \autoref{sec:thermal}). 

Orbits with high inclinations and short periods also likely had low perihelia in the past due to the Kozai-Lidov effect \citep{1962kozai, 2021toliou, 2023toliou}. This was assumed to preferentially lead to the destruction of weaker material in those orbits and thus the surviving material is biased to be stronger, an effect also observed for larger meteoroids \citep{Shober2025perihelion,Borovicka2022physical}. They also note that there are a number of meteor streams of known cometary origin which do not follow the classification scheme and lie below the asteroidal $Q$ limit, including the $\alpha$-Capricornids (parent body 169P/NEAT \citep{2010jenniskens}) and Southern Taurids (parent body 2P/Encke \citep{jenniskensshowers}).

\subsubsection{Tancredi's Extension of Tisserand's Criterion}

To address the dynamical overlap between ``asteroidal orbits" and ``cometary orbits", \citet{2014Tancredi} presented a classification scheme, primarily based on $T_J, q, Q$ and the minimum orbital intersection distance (MOID). This criterion was formulated primarily to distinguish objects that were categorized as a comet or asteroid based on their observed activity or lack thereof, but had orbital dynamics resembling the opposite group, dubbed ``Asteroids in Cometary Orbit" or ``Comets in Asteroidal Orbit" (ACOs and CAOs).

However, this criterion is not suitable for this study for two main reasons. First, the \citet{2014Tancredi} classification scheme is designed for populations of known asteroids and comets, where the goal is to identify objects whose orbital parameters are inconsistent with their nominal classification. In this case, the objects being classified are meteoroids whose parent body class is not known a priori. For example, \citet{2014Tancredi} defines comets in asteroidal orbits as objects with $T_J > 3.05$ and $q < Q_J$. However, the majority of objects occupying this region of orbital element space are asteroidal, and without prior knowledge of the object's physical nature it is not possible to distinguish between an asteroid and a comet in this region using the \citet{2014Tancredi} criterion alone. 

Second, this classification was developed for large parent bodies, assuming that objects evolve primarily under gravitational perturbations and therefore occupy relatively well-defined regions of dynamical phase space. This study concerns millimetre-sized meteoroids which experience significant non-gravitational forces such as PR drag and radiation pressure, which cause secular changes in their semimajor axis and eccentricity. While this limitation also applies to the simpler dynamical criteria such as the $K$ and $Pe$ criteria used in this work, those criteria rely on fewer parameters and therefore introduce less sensitivity to small orbital perturbations.

\subsection{Critical Evaluation of Orbital Criteria}

All the criteria discussed above have limitations to various degrees. To try and determine which ones may best isolate meteoroids with cometary vs asteroidal origins, we quantitatively evaluate each criterion's ability to recover origins of meteoroids with known provenance. To serve as a dual validation of both the criteria and our \verb|REBOUND|-based simulation code, we explore each criterion's ability to properly classify origins for one real and two synthetic datasets including:

\begin{enumerate}
\item Observed meteors in our dataset from showers with known cometary parent bodies. We perform backwards integrations using our \verb|REBOUND|-based pipeline for these events and follow each criterion's classification over time,
\item Generating synthetic meteoroids by releasing particles from known near-Earth JFCs and forward integrating their orbits with a modified version of our \verb|REBOUND|-based pipeline to follow how their origins change for each criterion,
\item Backward integrations of virtual impactors selected from among those produced in item 2 but which have Earth-impacting orbits at the time of release. Again, this is done with our \verb|REBOUND|-based pipeline that we will use for the full observational dataset.
\end{enumerate}

These validation integrations include the same gravitational and non-gravitational forces (radiation pressure and PR drag) used for the full dataset, as described in \autoref{subsec:simsetup}.

\subsubsection{Observed Shower Meteors}
\label{sec:SM}
We first selected 35 meteoroids from our dataset that are associated with known showers using the shower association criteria described in \citet{Vida2021_GMN}. These consisted of 19 events from the CAMO dataset and 16 meteors from the EMCCD dataset. 

These observed shower meteoroids included 8 from the October Draconids (DRA), 9 from the $\alpha$-Capricornids (CAP), 2 from the Northern $\iota$-Aquarids (NIA), 2 from the $\tau$-Herculids (TAH), 7 from the Southern Taurids (STA), 2 from the November $\eta$-Taurids (NET), 1 from the $\beta$-Comae Berenicids (BCO), 2 from the Southern $\chi$-Orionids (ORS), 1 from the $\eta$-Virginids (EVI), and 1 from the February Leonids (FLO). 

Their measured velocities, $k_c$ parameter, mass and presumed parent body are given in \autoref{tab:stage1test} and \autoref{tab:stage1testSTA}. Note that we separate the Taurid complex of showers (STA, ORS, NIA, NET) into their own table as their parent body is generically believed to be 2P/Encke \citep{jenniskensshowers}. 2P/Encke has an orbit long recognized to be unusual for a short-period comet, such that Encke is often listed as being in an orbital class of its own \citep{Levison_1996}. Other meteors associated with showers having unknown parent bodies (FLO, BCO, EVI) are also included in \autoref{tab:stage1testSTA}. 

In our simulations, although we create an initial sample of 100 clones per meteor based on the measured covariance in each case (in addition to the nominal case), some clones end up on unphysical orbits (eg. those that end up colliding with the Sun or hyperbolic orbits with $e > 1$). We remove those clones at the time their orbits become unphysical. 

\begin{figure*}[h!]
    \centering
    \subfigure[]{\includegraphics[width=\linewidth]{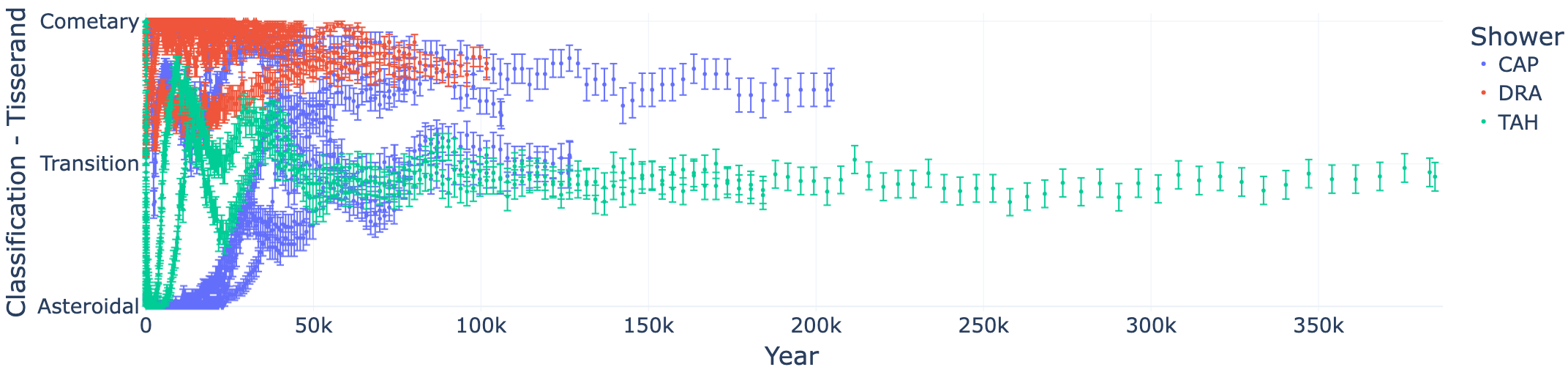}} 
    \subfigure[]{\includegraphics[width=\linewidth]{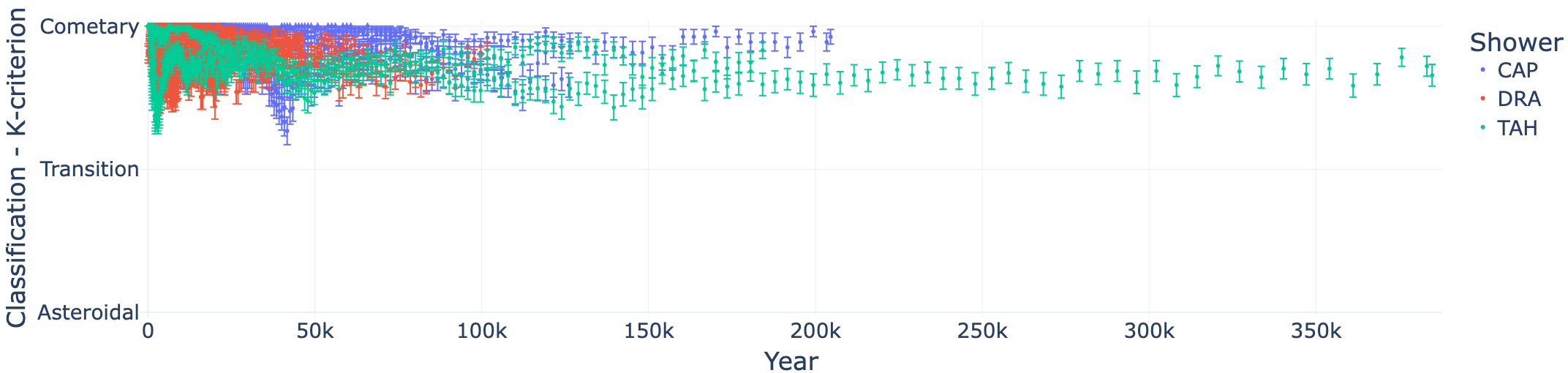}} 
    \subfigure[]{\includegraphics[width=\linewidth]{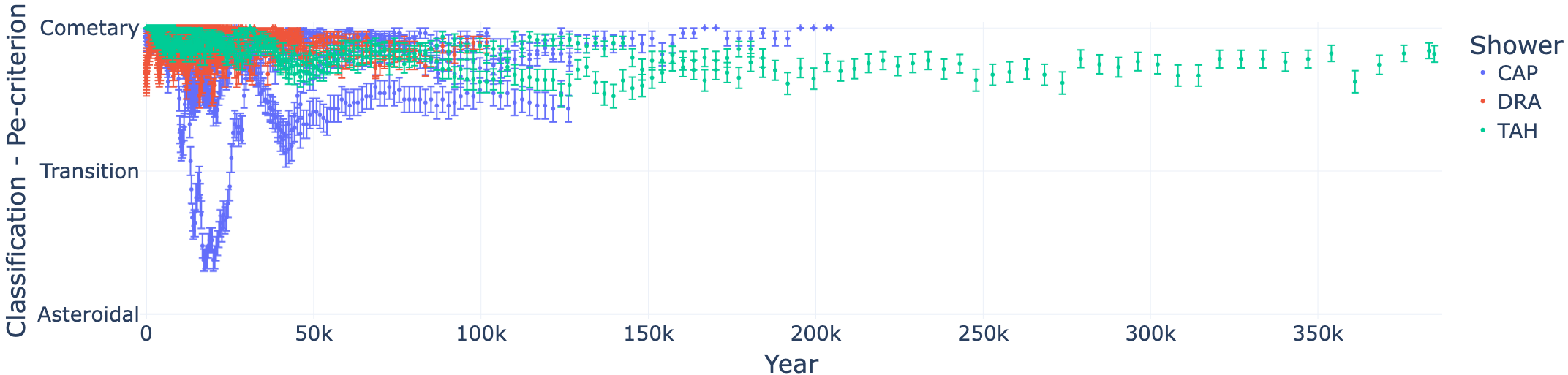}}
    \subfigure[]{\includegraphics[width=\linewidth]{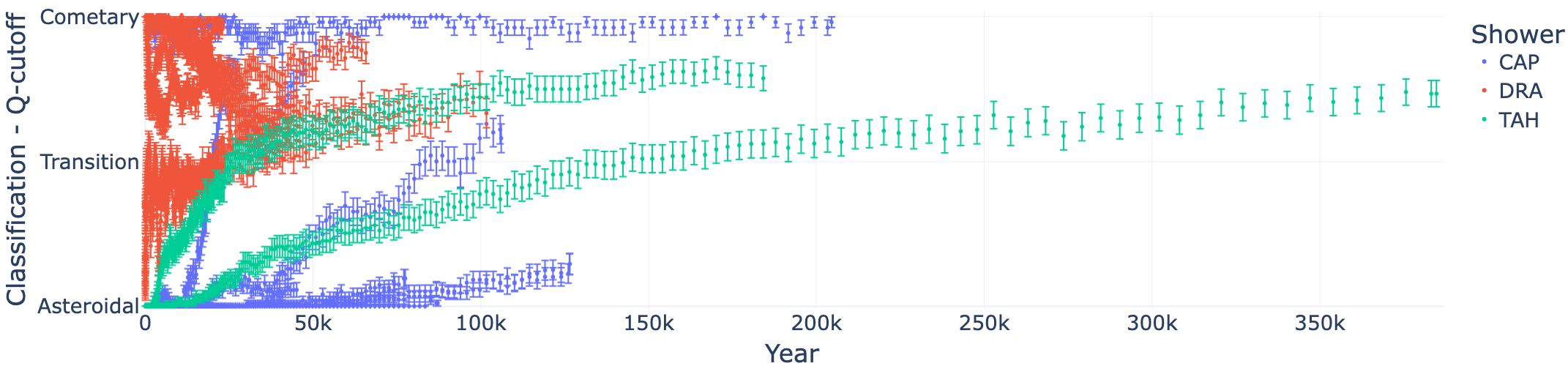}}
    \caption{The mean percentage of clones categorized as asteroidal vs cometary origin for the 19 shower meteoroids described in \autoref{tab:stage1test} that are known to be of JFC parentage as a function of backward integration time for each orbital classification criterion. The vertical axis indicates the relative number of clones categorized as either cometary or asteroidal. The $\alpha$-Capricornids are shown in purple, the October Draconids are shown in red and the $\tau$-Herculids in green. Error bars show the standard error in the mean among all clones at each time step.}
    \label{fig:crit-showers}
\end{figure*}

So as not to bias our analysis toward the more stable clone orbits, we restrict our analysis to the time window in which we have at least 95 surviving clones in the simulation for each meteoroid. We show each meteoroid's origins, coloured by shower stream in \autoref{fig:crit-showers} over the integration time. We see that both the $K$ and $Pe$ criteria consistently classify the meteoroids as cometary whereas the $Q$-cutoff criterion is more likely to classify $\alpha$-Capricornids and $\tau$-Herculids as asteroidal, and a significant fraction of October Draconids were also counted as asteroidal. 

\citet{1890Tisserand}'s criterion struggled to definitively classify the two $\tau$-Herculids as both meteoroids straddled the $T_J = 3.05$ boundary and although the October Draconids were correctly classified, the $\alpha$-Capricornids were generally in the asteroidal side of the $T_J$ boundary. It is worth noting that many JFC shower streams exhibit orbital histories that appear dynamically stable over $\gtrsim 10$ kyr which is more consistent with asteroidal dynamics than with the chaotic Jupiter-scattering histories typical of their parent comets. This stability is in part a selection bias: material released onto chaotic, Jupiter-encountering orbits disperses rapidly into the sporadic background, leaving the observable stream population preferentially composed of meteoroids that have, at least temporarily, decoupled from strong Jovian influence. As discussed at length in \citet{2024Shober} in the context of cm-scale shower fireballs, this means orbital criteria alone may misclassify genuinely cometary material that has undergone such dynamical decoupling. However, in the low-velocity regime studied here, where both cometary and asteroidal material occupy overlapping orbital phase space and no such velocity-based prior assumptions on composition are available, we validate the ability of dynamical criteria to recover the origins based only on orbits, thus the validation check using real meteors from 3 known JFCs suggests the $K$ and $Pe$ criteria are more reliable in recovering origins for cometary meteoroids than the $Q$-cutoff and $T_J$-based criteria.

\begin{table}[ht]
    \centering
    \begin{tabular}{c|c|c|c}
        Meteoroid & Velocity (km/s) & $k_c$ & Mass (kg) \\
        \hline
        TAH A & 15.99 & 97.65 & $1.7\times10^{-4}$ \\
        TAH B & 15.88 & 96.99 & $5.7\times10^{-5}$ \\
        CAP A & 23.8 & 96.27 & $5.7\times10^{-6}$ \\
        CAP B & 25.25 & 100.92 & $2.8\times10^{-5}$ \\
        CAP C & 24.09 & 99.48 & $2.5\times10^{-5}$ \\
        CAP D & 25.12 & 98.12 & $4.8\times10^{-5}$ \\
        CAP E & 28.51 & 99.13 & $8.31\times10^{-7}$ \\
        CAP F & 25.0 & 101.25 & $3.29\times10^{-6}$ \\
        CAP G & 24.04 & 99.36 & $1.5\times10^{-6}$ \\
        CAP H & 24.68 & 100.39 & $2.06\times10^{-6}$ \\
        CAP I & 25.51 & 99.50 & $5.85\times10^{-6}$ \\
        DRA A & 23.67 & 103.22 & $1.34\times10^{-6}$ \\
        DRA B & 23.29 & 107.98 & $1.73\times10^{-5}$ \\
        DRA C & 22.71 & 103.06 & $2.01\times10^{-6}$ \\
        DRA D & 21.6 & 109.03 & $1.54\times10^{-5}$ \\
        DRA E & 23.98 & 107.17 & $1.39\times 10^{-4}$ \\
        DRA F & 22.77 & 103.62 & $8.11\times10^{-7}$ \\
        DRA G & 24.51 & 101.81 & $1.45\times10^{-5}$ \\
        DRA H & 23.48 & 103.21 & $3.26\times10^{-5}$ \\
    \end{tabular}
    \caption{Characteristics of shower meteoroids used for criteria validation testing (all known to be cometary). Parent bodies for the showers are 73P/Schwassmann-Wachmann 3 (TAH) \citep{1963Southworth}, 169P/NEAT (CAP) \citep{2010jenniskens} and 21P/Giacobini-Zinner (DRA) \citep{1971Marsden}.}
    \label{tab:stage1test}
\end{table}

\begin{table}[ht]
    \centering
    \begin{tabular}{c|c|c|c}
        Meteoroid & Velocity (km/s) & $k_c$ & Mass (kg)\\
        \hline
        \textbf{STA A} & 27.42 & 95.49 & $4.8\times10^{-6}$ \\
        \textbf{STA B} & 26.16 & 99.96 & $1.6\times10^{-5}$ \\
        \textbf{STA C} & 27.91 & 94.97 & $3.5\times10^{-6}$ \\
        \textbf{STA D} & 29.61 & 100.21 & $1.6\times10^{-5}$ \\
        \textbf{STA E} & 28.92 & 99.08 & $1.5\times10^{-5}$ \\
        \textbf{STA F} & 29.51 & 97.66 & $6.5\times10^{-6}$ \\
        \textbf{STA G} & 26.32 & 97.27 & $4.5\times10^{-6}$ \\
        \textbf{ORS A} & 32.91 & 99.25 & $5.6\times10^{-6}$ \\
        \textbf{ORS B} & 29.66 & 105.38 & $5.3\times10^{-6}$ \\
        \textbf{NET A} & 29.80 & 98.24 & $5.2\times10^{-6}$ \\
        \textbf{NET B} & 31.50 & 100.57 & $2.5\times10^{-5}$ \\
        \textbf{NIA A} & 30.67 & 90.94 & $2.97\times10^{-7}$ \\
        \textbf{NIA B} & 29.72 & 98.09 & $3.03\times10^{-7}$ \\
        FLO A & 33.24 & 97.34 & $4.2\times10^{-6}$ \\
        BCO A & 29.26 & 100.58 & $2\times10^{-5}$ \\
        EVI A & 31.05 & 97.49 & $2.3\times10^{-5}$ \\
    \end{tabular}
    \caption{Characteristics of Taurid complex \citep{jenniskensshowers} meteoroids (shown in bold) and other shower meteoroids that have uncertain/asteroidal origins.}
    \label{tab:stage1testSTA}
\end{table}

\subsubsection{Virtual Particles Released From JFCs}
\label{sec:VP}
As another validation check, we generate 100 virtual particles that have the same characteristics as a typical cometary meteoroid in our observed dataset ($m=4.26\times10^{-6}$ kg, $d = 2.09$ mm, $\rho = 0.895$ g/cm$^3$, $\beta = 6.179\times10^{-4}$), then release them along the orbit of several known JFCs. We chose the orbits for four representative near-Earth JFCs from JPL Horizons \footnote{https://ssd.jpl.nasa.gov/horizons/, accessed May 21, 2025}: 182P/LONEOS, 197P/LINEAR, 209P/LINEAR and 21P/Giacobini-Zinner (the parent body for the October Draconids meteor shower). We then integrate the particles forward in time over 100 kyr into the future, then apply the orbital criteria to the virtual particles at regular time steps.   

\begin{figure*}[h!]
    \centering
    \subfigure[]{\includegraphics[width=0.98\linewidth]{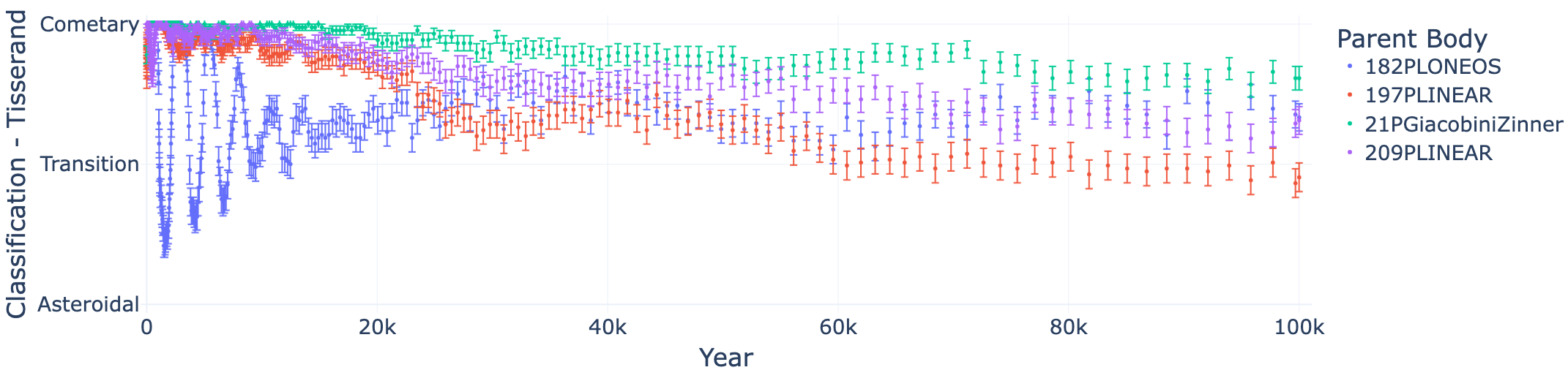}}
    \subfigure[]{\includegraphics[width=0.98\linewidth]{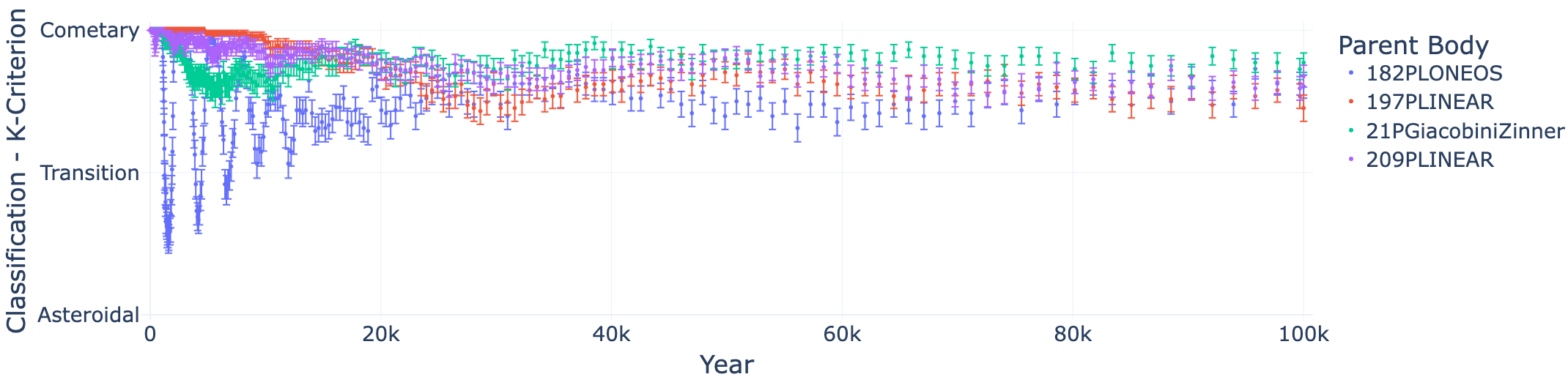}} 
    \subfigure[]{\includegraphics[width=0.98\linewidth]{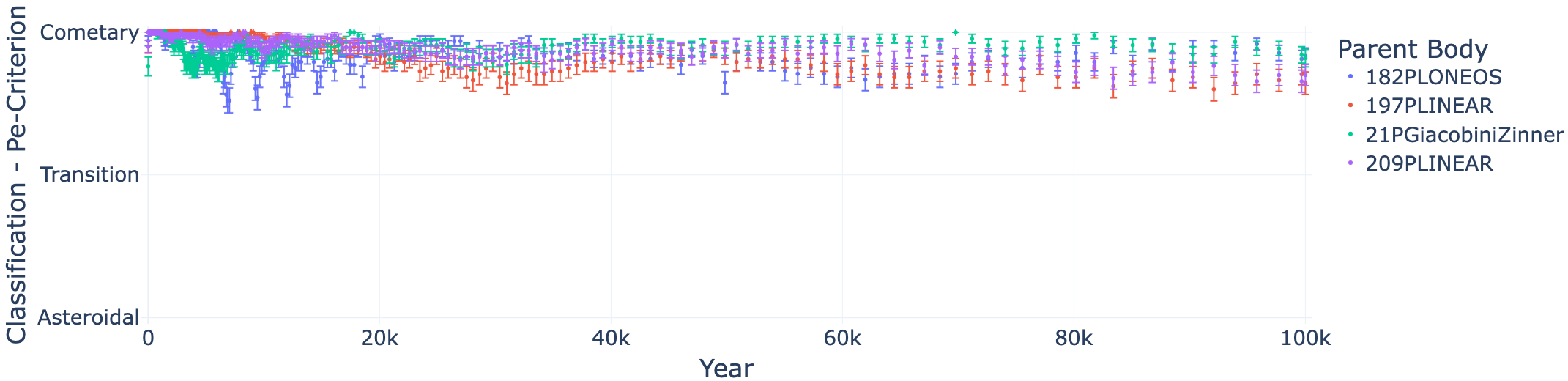}}
    \subfigure[]{\includegraphics[width=0.98\linewidth]{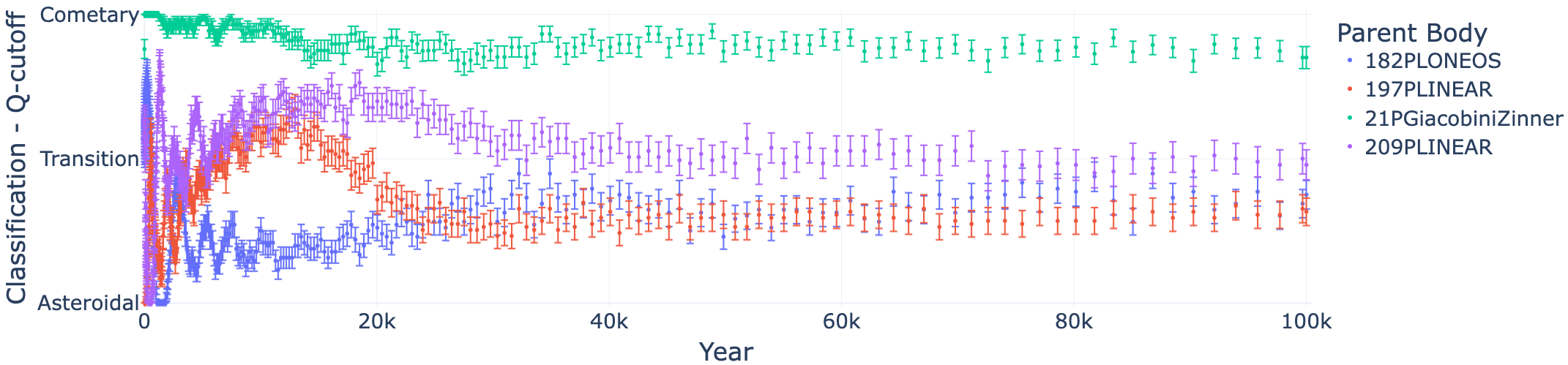}}
    \caption{Evolution of the classification as a function of time of 100 virtual particles released along the orbit for each of the JFCs 197P/LINEAR, 21P Giacobini-Zinner, 182P/LONEOS, and 209P/LINEAR. Shown is the percentage of asteroidal vs cometary dynamical origins over all clones, for \citet{1890Tisserand}'s criterion, \citet{1954whipple}'s $K$-criterion, \citet{1967kresak}'s $Pe$-criterion and \citet{Borovicka2022physical}'s $Q$-cutoff criterion. These are averaged at each time step with the standard error of the mean shown as error bars.}
    \label{fig:forwardsims}
\end{figure*}

The categorization according to each criterion for these meteoroids from each JFC are shown in \autoref{fig:forwardsims}. We see that for all 4 cases, the $Pe$-criterion very confidently classifies the particles as cometary, throughout the full integration history. The $K$ and $T_J$-based criteria also show the same basic trend with the exception of particles ejected from 182P/LONEOS which spread near the $K=0$ and $T_J=3.05$ boundary, resulting in an overall average that oscillates for about 20 kyr after ejection before settling into the expected cometary-origin orbit. It should also be noted that although the $T_J$-based criterion classifies over 50\% of the clones as cometary over the 100 kyr integration, it does so less confidently (50\%-70\%) for all parent bodies we analyze except 21P/Giacobini-Zinner, which is consistently characterized as a cometary source across all 4 criteria. The $Q$-cutoff criterion only correctly recovers 1 out of the 4 comets' particles as being mainly of cometary origin over the full integration history. This appears to further highlight the limitation of the $Q$-cutoff criterion in orbital classification for JFC related meteoroids, based purely on orbital behaviour.

\subsubsection{Virtual Cometary Earth Impactors}\label{subsubsec:virtual-impactors}
 Using the virtual particles generated in Section \ref{sec:VP} we identify any particles that come close to Earth during any period of the forward integration and treat them as virtual Earth impactors. We adopt \citet{Wiegert2009}'s threshold (minimum distance less than 0.1 AU) for defining a population of Earth impacting meteoroids; this represents a compromise between accuracy and the need for sufficient numbers.

For each virtual impactor, we adopt a covariance matrix typical for those observed in our dataset of shower meteoroids and generate 100 clones per impactor. We then use the velocity and position at the time of minimum distance as the ``observed" state vector. All clones for each impactor are backward integrated for 100 kyr following the same procedure we do for each real observed meteor. Among these virtual impactors, we then check if all criteria correctly classify the virtual impactors as being of cometary origin at the known time of ejection. We also check the stability of the classification further in the past to estimate the robustness of the association with differing potential ejection epochs. 

\begin{table}[h!]
    \centering
    \begin{tabular}{c|c|c}
        Meteoroid & Velocity (km/s) & Age (years)\\
        \hline
        197P/LINEAR Clone 74 & 16.43 & 602.22 \\
        197P/LINEAR Clone 93 & 15.19 & 513.99 \\
        209P/LINEAR Clone 43 & 15.21 & 13.99 \\
        209P/LINEAR Clone 47 & 15.49 & 89.98 \\
        209P/LINEAR Clone 63 & 17.60 & 11.03 \\
        209P/LINEAR Clone 89 & 15.46 & 19.98 \\
        209P/LINEAR Clone 90 & 16.50 & 9.99 \\
        209P/LINEAR Clone 93 & 15.62 & 4.99 \\
        209P/LINEAR Clone 97 & 17.93 & 251.97 \\
    \end{tabular}
    \caption{Characteristics of virtual impactors released from comets 197P/LINEAR and 209P/LINEAR. The impact velocity is the theoretical speed the virtual impactor would impact the top of the Earth's atmosphere (including the effects of gravity). Here, age refers to the time from release until the clone met our selection criteria as a virtual impactor (see text for details).}
    \label{tab:virtual-impacts}
\end{table}

\begin{figure*}[h!]
    \centering
    \subfigure[]{\includegraphics[width=0.98\linewidth]{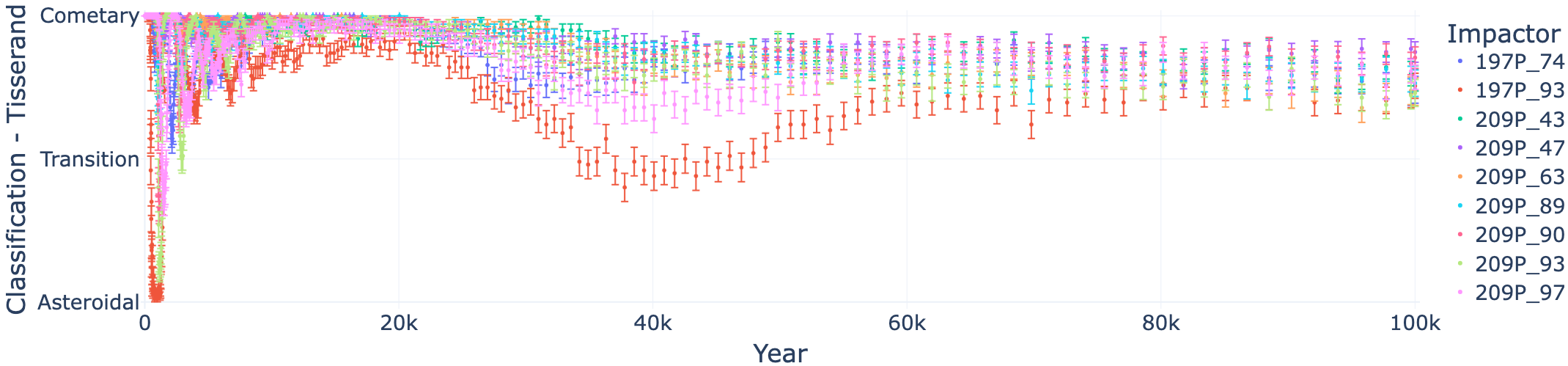}} 
    \subfigure[]{\includegraphics[width=0.98\linewidth]{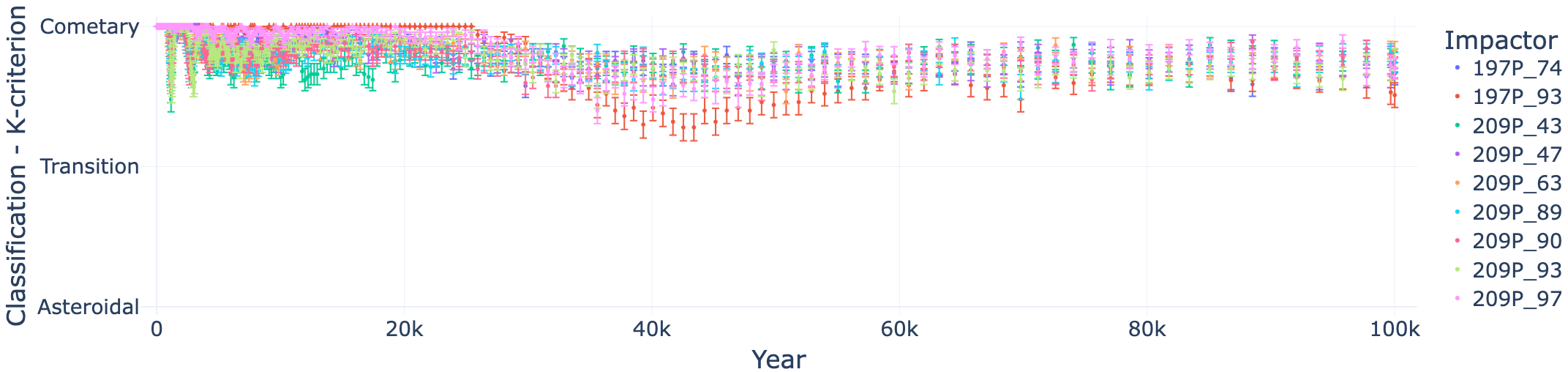}} 
    \subfigure[]{\includegraphics[width=0.98\linewidth]{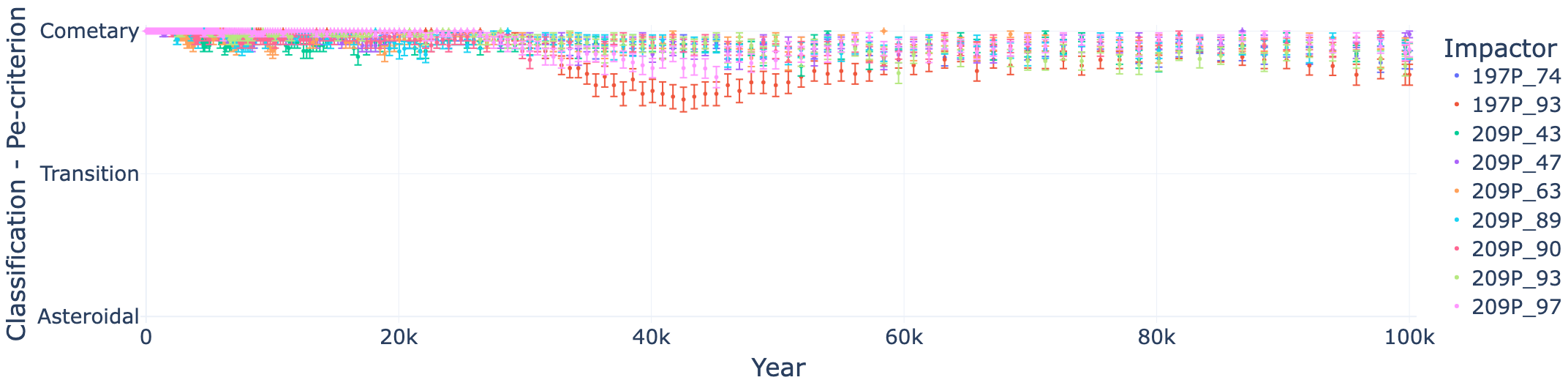}}
    \subfigure[]{\includegraphics[width=0.98\linewidth]{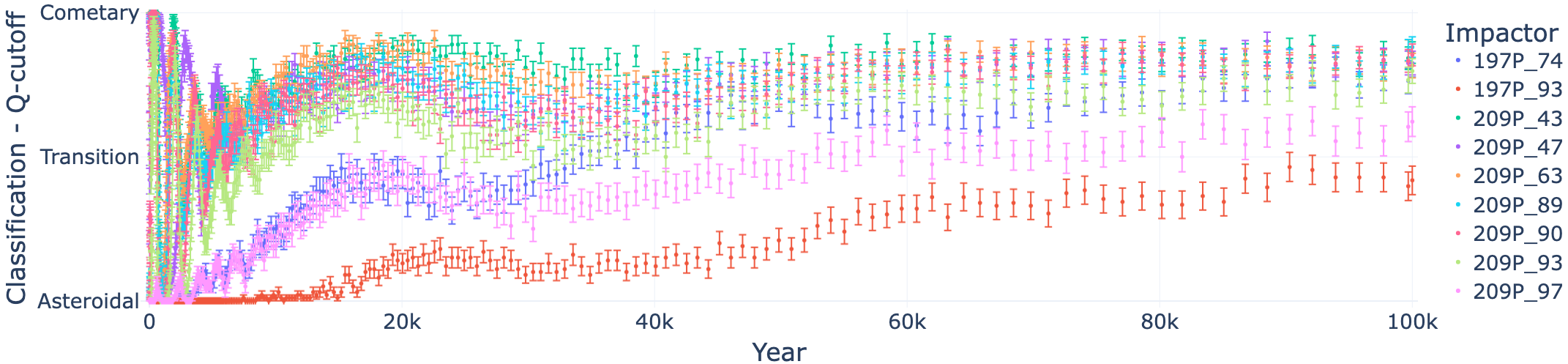}}
    \caption{The percentage of the 100 clones which show cometary or asteroidal origin for each of 9 virtual impactors that were released from the 4 test comets. The origins of 100 clones for each impactor are analyzed for \citet{1890Tisserand}'s criterion, \citet{1954whipple}'s $K$-criterion, \citet{1967kresak}'s $Pe$-criterion and \citet{Borovicka2022physical}'s $Q$-cutoff criterion, and averaged at each time step over 100 kyr (standard error shown in error bars).}
    \label{fig:virtualimpactorsims}
\end{figure*}

The theoretical velocity of impact with Earth and ages of the virtual particles we selected as impactors are given in \autoref{tab:virtual-impacts}. Consistent with the general assumption that cometary particles tend to be young, all impactors had an age of less than 1000 years, with the youngest particle having been ejected just 5 years before it reached our distance threshold to be considered a virtual impactor with Earth. This is as expected since the comets we chose were near-Earth JFCs. The results of the asteroidal vs. cometary classification according to each criteria for the virtual impactors and their clones are shown in \autoref{fig:virtualimpactorsims}. Here, we use the mean and standard error of the 100 clones per impactor for their origin determination.

As was the case for our validation checks in sections \ref{sec:SM} and \ref{sec:VP} we find that the $K$ and $Pe$ criteria were better at recovering the origin of the virtual impactors as cometary compared to the $Q$-cutoff and $T_J$ criteria. Although \citet{1890Tisserand}'s criterion generally correctly classifies cometary particles, it does so less confidently. We therefore only use $K$ and $Pe$ for orbital classification of meteoroids as being cometary or asteroidal in origin for the remainder of this work. 

The results across \autoref{sec:SM}-\ref{subsubsec:virtual-impactors} are based on our \verb|REBOUND/X| pipeline, which was independently validated and importantly, the orbital criteria themselves are algebraic functions of the osculating orbital elements; given correctly integrated orbits, their evaluation is deterministic. 

\subsubsection{Long-term Evolution of Criteria}
\label{sec:petrsims}

\begin{figure*}[h!]
    \centering
    \subfigure[]{\includegraphics[width=\textwidth]{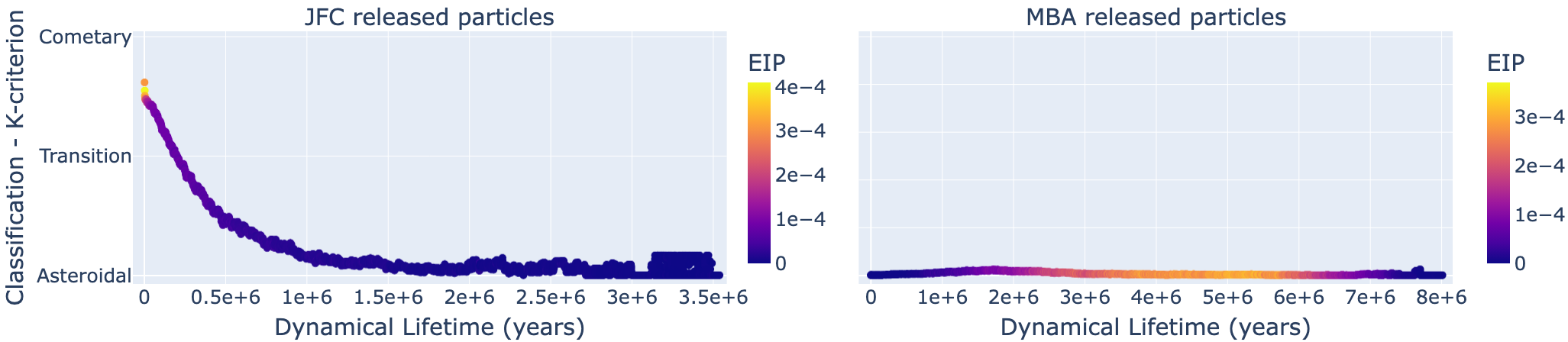}}
    \subfigure[]{\includegraphics[width=\textwidth]{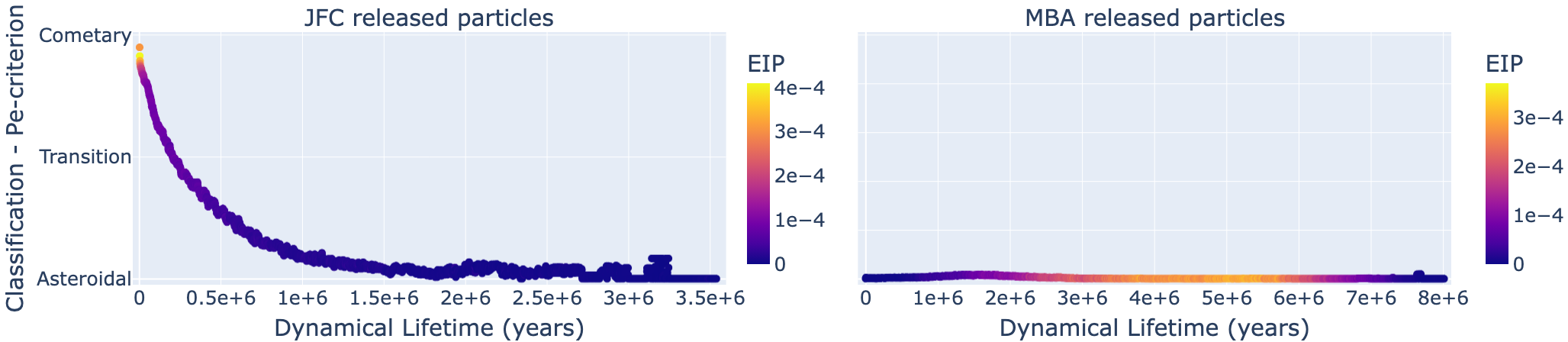}}
    \caption{Classification distribution of particles released from JFCs (left) and MBAs (right) as a function of time. Shown is the $K$-criterion classification \citep{1954whipple} (top plots) and $Pe$-criterion classification \citep{1967kresak} averaged for all particles. Here, the average Earth impact probability (EIP) for all surviving particles is shown from the time of particle release.}
    \label{fig:petrsims}
\end{figure*}

As an independent validation approach, we also used a different integration code already validated in previous studies \citep{2024Pokorny} to examine the evolution of two meteoroid parent populations: JFCs and MBAs. The goal was to analyze the global behaviour and evolution in time of the criteria for meteoroids in our size range of interest released from these two distinct populations. Both population models are adopted from \citet{Pokorny2019} where they used N-body simulations of $D=1000~\mu$m particles originating from JFCs (a purely cometary source) and MBAs (a purely asteroidal source). This size was chosen as it is the mode for our observed meteor population (see plot b in \autoref{fig:mass-diam}.)

In this simulation, meteoroids are released from large source populations with $N>5000$, different combinations of $a,e,i$ and randomized values of $\Omega, \omega, M$. These particles are then subjected to gravity of all eight planets, radiation pressure, PR drag, solar wind drag (30\% of the PR-drag component), and are also collisionally evolved in time. All meteoroids are numerically traced until they are removed from the simulation via three main routes: a) being too close to the Sun ($r<0.005$ AU), b) impacting one of the planets, c) being too far from the sun ($r>10^4$ AU). All numerical simulations in these global models were performed using \verb|SWIFT-RMVS3| \citep{2013Levison}. The Earth Impact Probability (EIP) is calculated for each meteoroid/clone separately using the \citet{1981kessler} method assuming that Earth is on a coplanar circular orbit. We also compared our EIP to values estimated using the \citet{2013pokorny} method and generally obtained similar results. As can be seen in \autoref{fig:petrsims}, the EIP is high for JFCs near the time of release as these are on orbits much closer to the Earth. The MBA particles, in contrast, must spiral inward to the Earth from the main belt through PR drag. This takes 3 - 5 Myr, explaining why the EIP peaks for this population in this interval.

The dynamical evolution of the JFC population in \autoref{fig:petrsims} show that 1 millimetre particles stay in cometary orbits, as classified by both the $K$ and $Pe$ criteria, for around 200 kyr and after that, the mean classification switches to asteroidal. This is an expected result due to the effect of PR drag that decreases the particle semimajor axis and circularizes their orbits. This result gives an approximate age limit for meteoroids that are classified as cometary because for material older than 200 kyr, both $K$ and $Pe$ criteria would classify everything as asteroidal. 

The MBA meteoroid population, in contrast, stay correctly classified for their entire dynamical lifetime. From these results, if we see a meteoroid's recent history as being classified as asteroidal but transitions to a cometary-like orbit further back in the integration, it is likely to be an older cometary meteoroid (of at least 200 kyr), unless a shorter lifetime limitation applies (see \autoref{subsec:timelimits}).

The results of our validation analysis are that overall $K$ and $Pe$ are better at classifying cometary meteoroids based on orbital properties and that we expect an effective average dynamical backward ``memory" for millimetre-sized cometary particles to be 200 kyr. The qualitative agreement between both code frameworks supports the interpretation that the observed criterion performance reflects the mathematical structure of the criteria rather than any artifact specific to the \verb|REBOUND/X| implementation.

\subsection{Orbital Integration Setup}
\label{subsec:simsetup}

Having measured our meteor trajectories and established their covariances, we apply the $K$ and $Pe$ orbital criteria backward in time to classify each meteoroid. To accomplish the backward integrations, we use the open source package \verb|REBOUNDX| \citep{tamayo} that interfaces with the \verb|REBOUND| N-body integration package \citep{rein} to accurately incorporate radiation pressure and PR drag to trace our meteoroids orbital evolution backward in time. 

We ingest our measured trajectory covariances directly into \verb|REBOUND| and produce clones for each event. We validated that the resulting orbits produced results consistent with the analytic heliocentric orbits using the WMPG software suite \citep{2020Vida} and against the independent numerical orbital integration package described in \citet{2011dave}.

We use the Integrator with Adaptive Step-size control, 15th order (IAS15) included in \verb|REBOUND|. It is a very high order, non-symplectic integrator which can handle arbitrary forces and is accurate down to machine precision (16 significant decimal digits) in most cases. The implementation of radiation forces on a particle included in the \verb|REBOUNDX| package \citep{tamayo} is based on the description given by \citet{1979burns}. The parameter $\beta$ used to calculate the radiation forces is given by 
\begin{equation}
    \beta = \frac{F_R}{F_G} = \frac{Q_{pr}3L_\odot}{16\pi GM_\odot c\rho r},
\end{equation}
where $L_\odot = 3.85\times10^{26}$ W, $M_\odot = 1.989\times10^{30}$ kg, $G = 6.6754\times10^{-11}$ m$^3$kg$^{-1}$s$^{-2}$, $c = 3.0\times10^{8}$ m/s, and $r$ is the particle radius (in m).
For the purpose of our simulations, we take the radiation pressure efficiency factor, $Q_{pr}$, to be 1, which holds true for $r \gg \lambda_{rad}$, and we use the bulk density $\rho$ per measured meteoroid as estimated by \citet{2016jenniskens}'s $k_c$ mappings.

We apply the $K$ and $Pe$ criteria to determine the original parent body at a set of time steps over the total integration time. The maximum time we follow the particle backward in time is set physically by its collisional lifetime (discussed in Section \ref{subsec:colltime}) and/or its thermal processing/disruption lifetime (Section \ref{sec:thermal}). Independent of these timescales we also explore how orbital diffusion and chaos affect the timescale over which we can meaningfully expect to track orbits.

To ensure all meteoroids have their orbital elements output with the same time steps for analysis, we generate a list of times with growing intervals to save disk space over the collisional lifetime (see Section \ref{subsec:colltime}) using a precomputed geometric series: 
\begin{equation}
    \begin{split}
        t_k = t_iR^k,\\ 0<k<n_\mathrm{max},\\
        n_{max} = \log_R(t_\mathrm{coll} / t_i),
    \end{split}
\end{equation} 
where $t_i$ is the smallest increment of time which we set to 1 day and $R$ is the ratio which we set to 1.02. We also check the criteria at the maximum collisional lifetime, $t_{coll}$. Note that these are not the time steps used for the integration as the IAS15 integrator uses adaptive time-stepping automatically, these are just the times we log information from the simulations.  

\begin{figure*}[h!]
    \centering
    \includegraphics[width=0.37\linewidth]{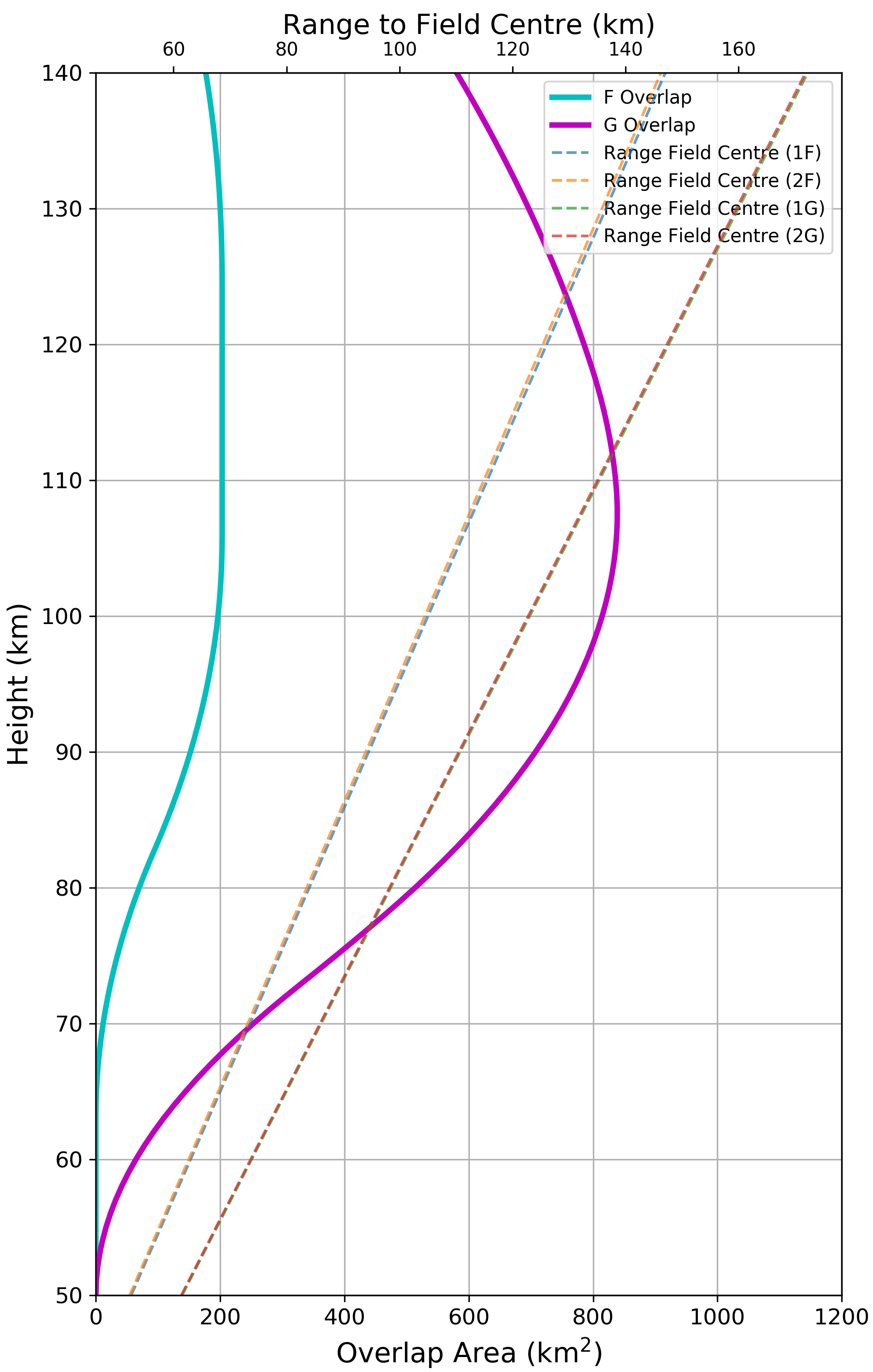}
    \includegraphics[width=0.55\linewidth]{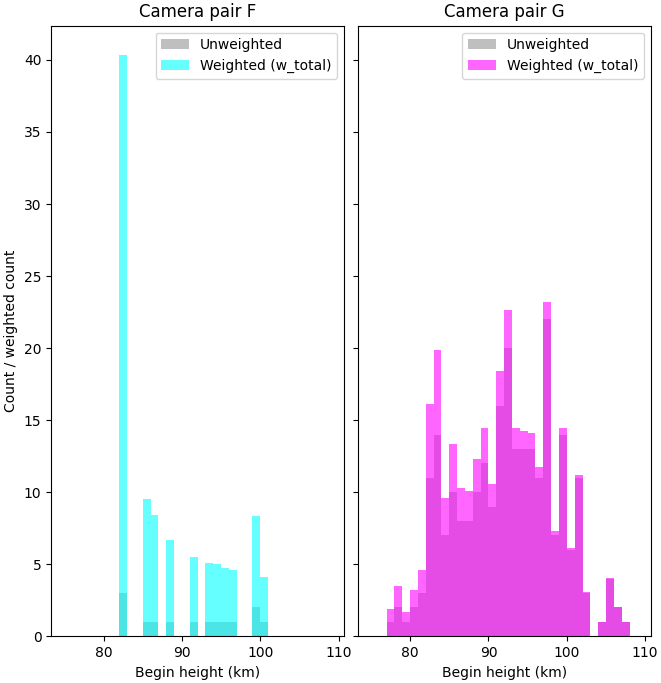}
    \caption{Collecting areas as a function of height for each EMCCD camera pair (F, G) (bottom axis, solid lines) and the range to the field center (top axis, nearly overlapping dashed lines) are shown on the left plot. The corresponding re-weighted observation data is shown on the right plot. Note that because the G cameras have a larger collecting area at all heights, we normalize the weighting such that the peak G collecting area has a total weighting of 1 and the F cameras with a smaller collecting area are scaled up by the ratio of overlap area at each height.}
    \label{fig:height-bias}
\end{figure*}

Finally, our observed dataset is restricted to meteors observed by at least two-stations which requires an overlapping collecting area. Due to the geometry of the EMCCD camera station positions and pointings, this collecting area changes with height which can bias our dataset. To complicate things further, the two pairs of EMCCD cameras (F pair and G pair) have different collecting areas as a function of height, as can be seen in \autoref{fig:height-bias}. To debias our observations, we re-weight the data in two steps: first, we debias the height effect by the peak collecting area for the individual camera pairs (F,G), and then we do a cross-pair normalization using the height-dependent area ratio between the two pairs. The meteors detected by each of the pairs are shown with and without the weighting applied in \autoref{fig:height-bias}. There are a handful of events from the F camera pair which are weighted by a factor of $\gtrsim10$; each event is in a velocity and $k_c$ bin that is dominated by dynamically asteroidal origins and the heavier weighting of these individual events does not significantly change the analysis of their respective bins.

In contrast to the EMCCD systems, the narrow-field CAMO cameras that capture meteor events in higher precision are cued by a wide-field system. This has a much wider field of view such that over the meteor height range of 80 - 120 km there is negligible height bias as the collecting area variation is much smaller. With this in mind, as well as the relative contributions to the full combined EMCCD/CAMO dataset, we do not re-weight the CAMO meteor observations.

\subsection{Integration Time Limitations}
\label{subsec:timelimits}

\subsubsection{Collisional Lifetimes}
\label{subsec:colltime}

\begin{figure*}[h!]
    \centering
    \includegraphics[width=0.97\textwidth]{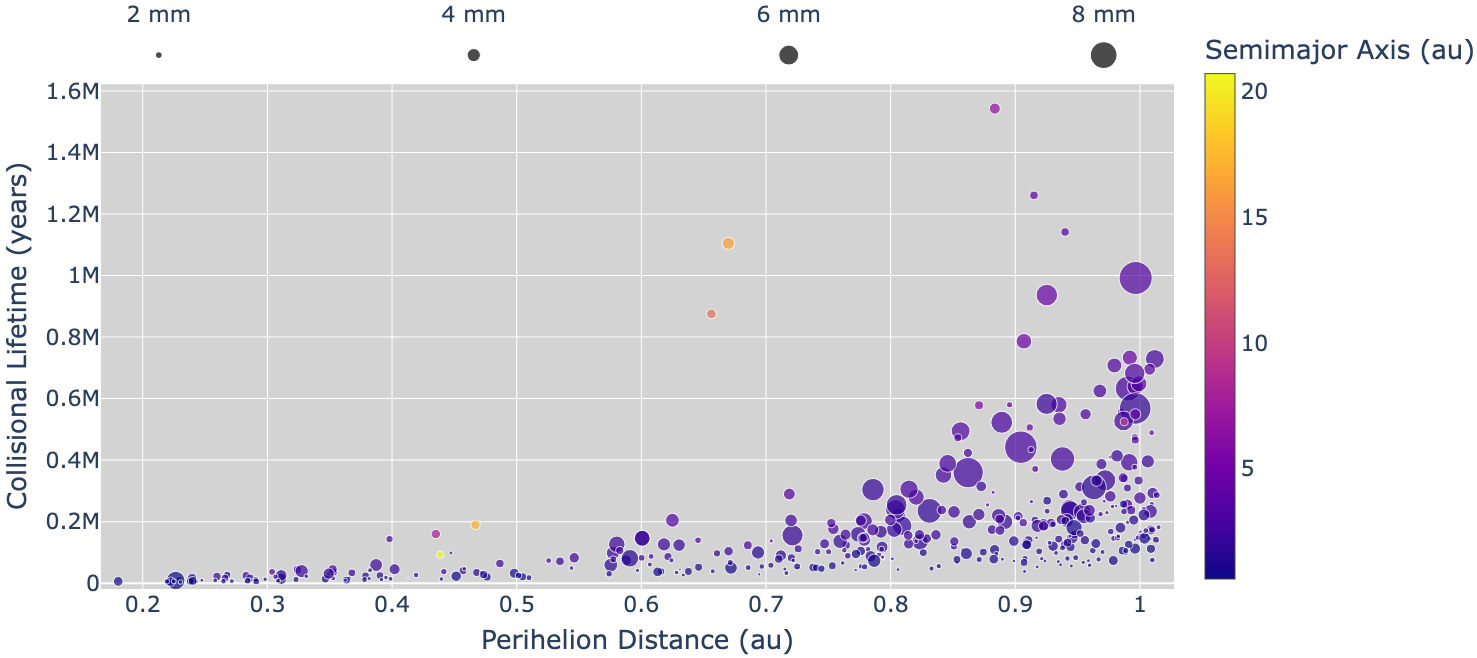}
    \caption{All 386 CAMO/EMCCD meteor events in our dataset, showing collisional lifetime as a function of perihelion distance in AU  coloured by semimajor axis. The marker size reflects particle diameter (see top of plot for symbol scale sizes).}
    \label{fig:coll-life}
\end{figure*}

It is difficult to define a meteoroid's time of ejection from its parent body and therefore impractical to interpret a particle's origins at one specific time in its orbital history. As such, we use the collisional lifetime for the orbit intersecting Earth's as the upper limit to our integrations. From the time of observed Earth impact backward to the estimated collisional lifetime we save orbital information to apply the $K$ and $Pe$ orbit-based criteria. In this way, we constantly determine the particle's origin, presuming it was released at any one of the regular intervals integrations are saved between its observed time of Earth impact and estimated collisional lifetime. 

The meteoroid collisional lifetime is defined to be the maximum time a particle can go without having a catastrophic collision \citep{1970dohnanyi}. This can be thought of as the inverse of the collision rate as a function of meteoroid mass $M$: 
\begin{equation}
    \tau_{cc}(M) = \frac{1}{\int_{m/\Gamma}^{M_\infty}v\sigma_cS(M)dM}
\end{equation}
where $v$ is the velocity of the particle, $S$ is the spatial density of particles for mass $M$, $M_\infty$ is the largest mass in the system being considered, $\sigma_c$ is the collision cross-section and $\Gamma$ is the factor defining the minimum size of an incident projectile that would cause catastrophic fragmentation.

Since \citet{1970dohnanyi}'s original description, two analytic methods incorporating an assumed shape, density, and velocity distribution for particles in the Zodiacal Cloud have been widely adopted: \citet{1985grun} and \citet{1986steel}. Building on these earlier analytic methods, \citet{2024Pokorny} recently developed publicly available code that implements a numerical ``collisional grooming" algorithm originally created by \citet{2009Stark}. This method ingests a ``collisionless seed model" which is an historical record of an $N$-body simulation that has been evolved dynamically through time. Particles can be removed via particle sinks other than collisions between particles, such as being flung out of the system, falling into the central star or hitting a planet or moon. The particles are then replaced with a probability pathline that has an initial number density, which decreases along the path due to collisions until a steady-state solution is reached.

Using this code, we calculated the collisional lifetimes for our meteoroids, given their $a, e, i, \Omega, \omega,$ and diameter at the time of their impact with Earth. We note that this may in general be a lower limit as asteroidal particles in the millimetre size range (appropriate to our sample) evolving under PR drag from the main belt may have much longer collisional lifetimes. However, as JFCs have typical dynamical lifetimes of order 10$^5$ years \citep{Levison_1997, fernandez2002}, backward integrations of this order would be sufficient to distinguish JFC origins from MBA origins for meteoroids, the primary goal of this work. 

We show the collisional lifetimes for our dataset of meteor observations from both CAMO and EMCCD instruments as a function of their perihelion distance, coloured by their semimajor axis, and marker size scaled by their diameter in \autoref{fig:coll-life}. We see a general trend of increasing collisional lifetime with perihelion distance, with a few outliers that have large semimajor axes. The mean collisional lifetime in our dataset is approximately $1.8\times10^5$ years with a standard deviation of $2\times10^5$ years. The median is slightly lower at $1.2\times10^5$ years. In summary, the majority of our observed meteoroids have collisional lifetimes on the order of $10^5$ years with a few outliers having much longer lifetimes approaching or exceeding $10^6$ years.

\subsubsection{Physical Lifetimes - Thermal Processing}
\label{sec:thermal}

Close approaches to the Sun expose meteoroids to intense thermal environments that can drive progressive physical and chemical alteration over their dynamical lifetimes, measurable in the spectra of millimetre sized meteoroids \citep{2005Borovicka}. Quantitative modelling by \citet{2009Capek} demonstrates that thermal processing becomes particularly efficient for meteoroids with perihelion distances $q\leq0.2$ AU, where peak equilibrium temperatures exceed ${\sim}700$ K for millimetre to centimetre sized bodies. At these distances, volatile elements such as sodium undergo rapid depletion through a sequence of thermally activated processes, including solid-state diffusion within mineral grains, thermal desorption from grain surfaces, and subsequent escape through interconnected pore networks. 

Their model predicts that sodium loss is strongly limited to the weeks spent near perihelion, due to the steep temperature dependence of diffusion coefficients. Over timescales of several thousand years, meteoroids composed of fine-grained aggregates can experience near-complete depletion of sodium when repeatedly exposed to low-perihelion conditions \citep{2009Capek}. This establishes a clear threshold in perihelion distance below which thermal processing rates abruptly increase. A similar threshold has recently been identified experimentally by \citet{2026Tsirvouils}, who found that carbonaceous chondritic material can be eroded by solar irradiance alone at heliocentric distances of roughly $r \leq 0.2$ AU, implying that thermally driven mass loss may affect the survival of dark, carbonaceous parent bodies on low-perihelion orbits.

Beyond volatile loss, thermal cycling at small perihelion distances also induces significant internal stresses that may modify rock structure and bulk physical properties \citep{2014Delbo, Libourel2021}. As shown in \citet{2010capek}, thermal stresses in small meteoroids may lead to fragmentation or to fracturing and formation of a regolith-like surface layer on meteoroids with low perihelion distances. In extreme cases, complete disruption may occur. Such a process has been proposed to explain the lack of low-q Near Earth Objects \citep{Granvik2016} found by telescopic surveys compared to model predictions. The recent irradiance experiments from \citet{2026Tsirvouils} also support the interpretation that purely thermal mechanisms can disrupt primitive bodies at small perihelion distances, as it was demonstrated that their meteorite simulants were destroyed within minutes under irradiation equivalent to heliocentric distances of up to 0.2 AU.

\citet{2010capek} analytically evaluated thermally induced stress fields in small (millimetre to centimetre sized) meteoroids and found that thermal stresses increase rapidly with decreasing heliocentric distance, scaling approximately with the inverse square root of solar distance. For perihelia $q < 0.15-0.2$ AU, the resulting stresses can locally exceed the tensile strength of ordinary and carbonaceous chondrite–like materials. While these stresses may lead to fragmentation in some cases, they can also promote crack closure, sintering, or reorganization of pore space during repeated heating cycles, particularly in the outer layers. When considered alongside volatile depletion, these processes suggest a pathway by which meteoroids evolving onto small $q$ orbits may undergo progressive reduction in porosity and corresponding increases in bulk density over time before potentially disintegrating. However, should the meteoroid survive, the model suggests fracture and a weaker surface layer may develop.

For our population of observed meteoroids, we calculate a ``Thermal Processing Coefficient" (TPC) to quantify the amount of thermal processing a meteoroid has undergone over its entire orbital history, as the perihelion of the measured meteoroid at the time of Earth impact may not reflect its cumulative past thermal exposure. For each meteoroid which is backward integrated, the osculating orbital elements are recorded at each time step and the orbital dynamical origin criteria are applied (see \autoref{subsec:simsetup} for the geometric series). Following \citet{2010capek}'s results for meteoroids of millimetre sizes wherein thermal processing is expected to be significant at a threshold of 0.2 AU, we calculate a proxy for how many perihelia passages at a threshold distance below 0.2 AU ($q_{thresh}$) or less each meteoroid experiences by considering each time interval between output times, $\Delta t_i = |t_{i+1} - t_i|$, and the perihelion at the start and end of each interval, $q_i, q_{i+1}$, for each of the surviving clones per meteoroid.

The contribution of a given interval to the total number of years a clone spent with $q < q_{thresh}$ was defined as the fraction 
of $\Delta t_i$ during which the perihelion lay below the threshold, multiplied by $\Delta t_i$:

\begin{itemize}
    \item If both $q_i$ and $q_{i+1}$ were less than $q_{thresh}$, the meteoroid was assumed to remain inside the threshold for the entire interval, and the full $\Delta t_i$ was added to the total.
    
    \item If both $q_i$ and $q_{i+1}$ were greater than $q_{thresh}$, the meteoroid was assumed to remain outside $q_{thresh}$, and the interval did not contribute to the total.
    
    \item If the perihelion crossed the threshold within the interval (i.e., one of $q_i$ or $q_{i+1}$ is below $q_{thresh}$ and the other is above), we used a linear interpolation and  the time at which $q = q_{thresh}$ within the interval was computed. Only the portion of $\Delta t_i$ on the $q < q_{thresh}$ side of the crossing contributed to the total. 
\end{itemize}

The TPC is evaluated at each output time as the cumulative sum of all interval contributions up to that time, averaged over all surviving clones for that meteoroid, and then divided by the mean orbital period at that time step. As both the total time spent with $q<q_{thresh}$ and the orbital period are measured in years, this gives us an estimate of the number of perihelion passages the meteoroid has made where the perihelion distance was below the threshold where we assume thermal processing occurs. It should be noted that using this method, the mean perihelion for a meteoroid's set of clones may never be measured to be below $q_{thresh}$, however the contributions of some clones having $q < q_{thresh}$ gives the meteoroid a non-zero TPC.

To balance the need for manageable file sizes with the higher time resolution applied closer to the time of Earth impact, our $\Delta t_i$ increases with $i$. This is done such that our approach of using linear interpolation becomes less reliable with larger $\Delta t$. Our sampling rate becomes on the order of 100 years per clone around $\sim$ -5 kyr rising to 1000 years at $\sim$ -50 kyr. 

Furthermore, some meteoroid orbits can experience Lidov-Kozai oscillations from Jupiter's perturbations, causing $q$ to vary significantly on a short cycle. As the sampling rate approaches the period of the Lidov-Kozai cycle, we become less confident in our TPC calculation because perihelion oscillations are no longer sufficiently resolved to robustly count threshold crossings. In order to restrict our analysis to windows of time where we have confidence in our TPC values, we calculate the component of each clone's orbital angular momentum parallel to the angular momentum of the Sun/Jupiter orbit \citep{2007Kinoshita}, 
\begin{equation}
    L_z = \sqrt{1-e^2} \text{cos}(i),
\end{equation}
at each time output.  If $L_z < \sqrt{0.6}$, we expect the oscillations to occur. For circular orbits ($e = 0)$, the critical inclination angle is 
\begin{equation}
    i_{crit} \ge \text{arccos}\left(\sqrt{\frac{3}{5}}\right) \approx 39.2^\circ.
\end{equation}
We calculate the approximate timescale associated with these oscillations using the analytic solution described in \citet{2007Kinoshita} and restrict our TPC calculations to the time steps in which our sampling rate is below $T_{Kozai}$. This is done on the nominal solution for each meteoroid, rather than all its clones, as it is expected that the variation between clones is unlikely to affect the Lidov-Kozai oscillations significantly.

\section{Results and Analysis}

\subsection{Meteoroid Orbital Classification by Ejection Epoch}

\begin{figure*}[h!]
    \centering
    \subfigure[]{\includegraphics[width=0.9\linewidth]{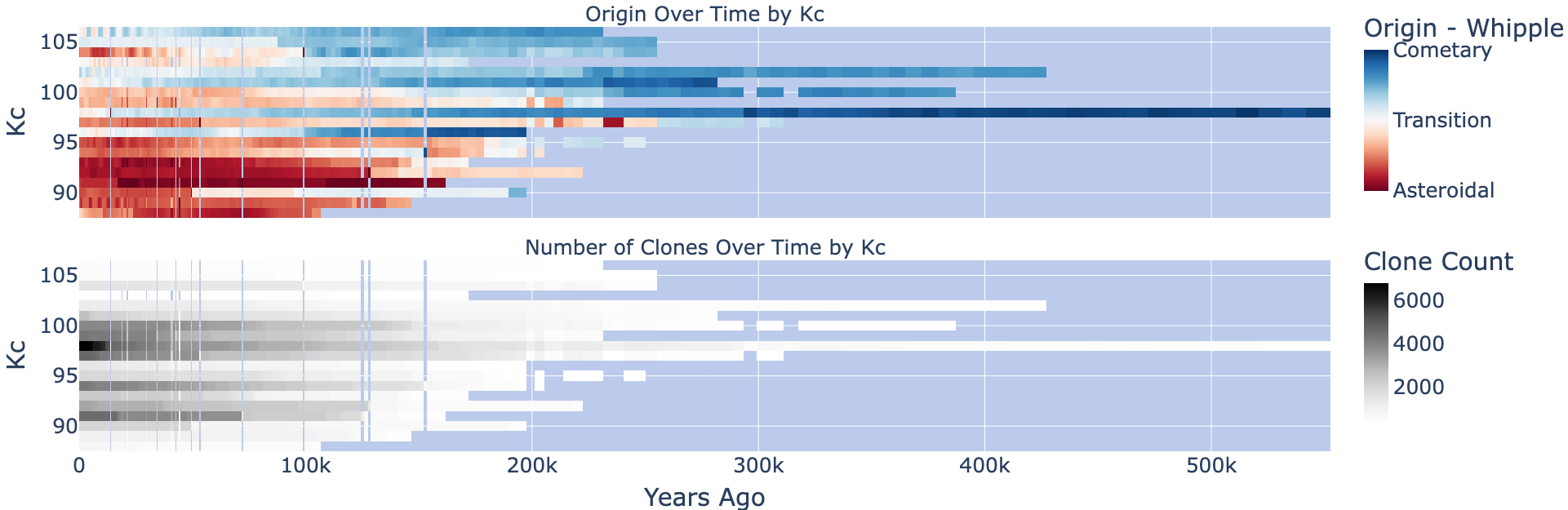}}
    \subfigure[]{\includegraphics[width=0.9\linewidth]{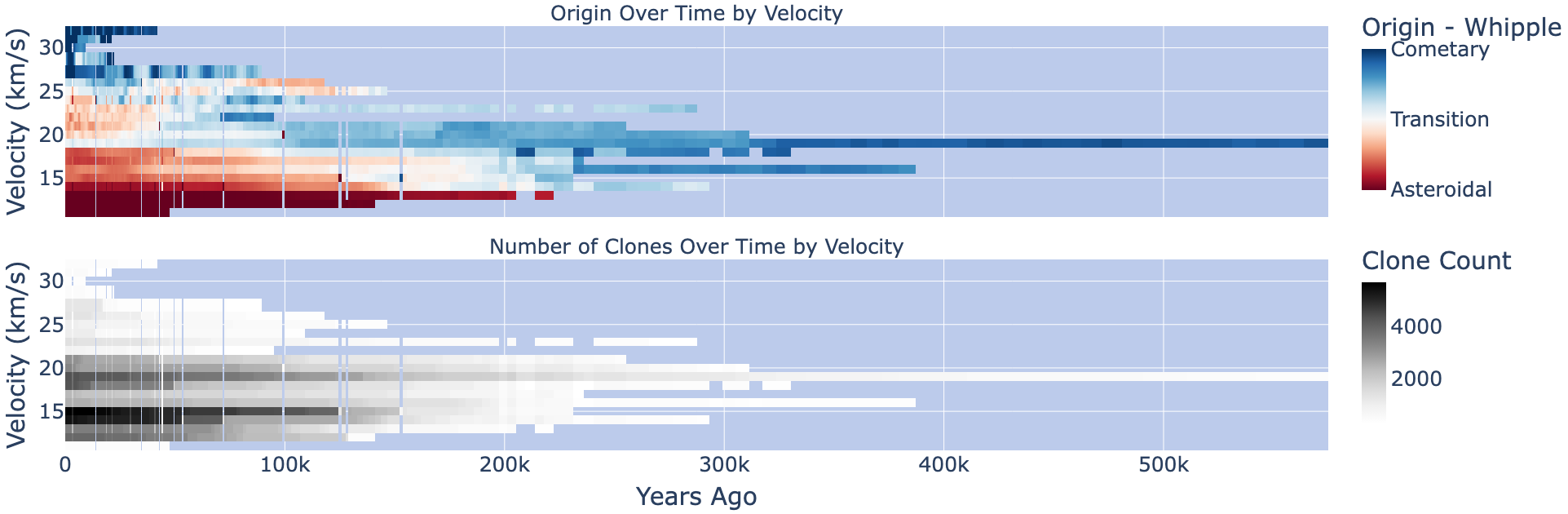}}
    \subfigure[]{\includegraphics[width=0.9\linewidth]{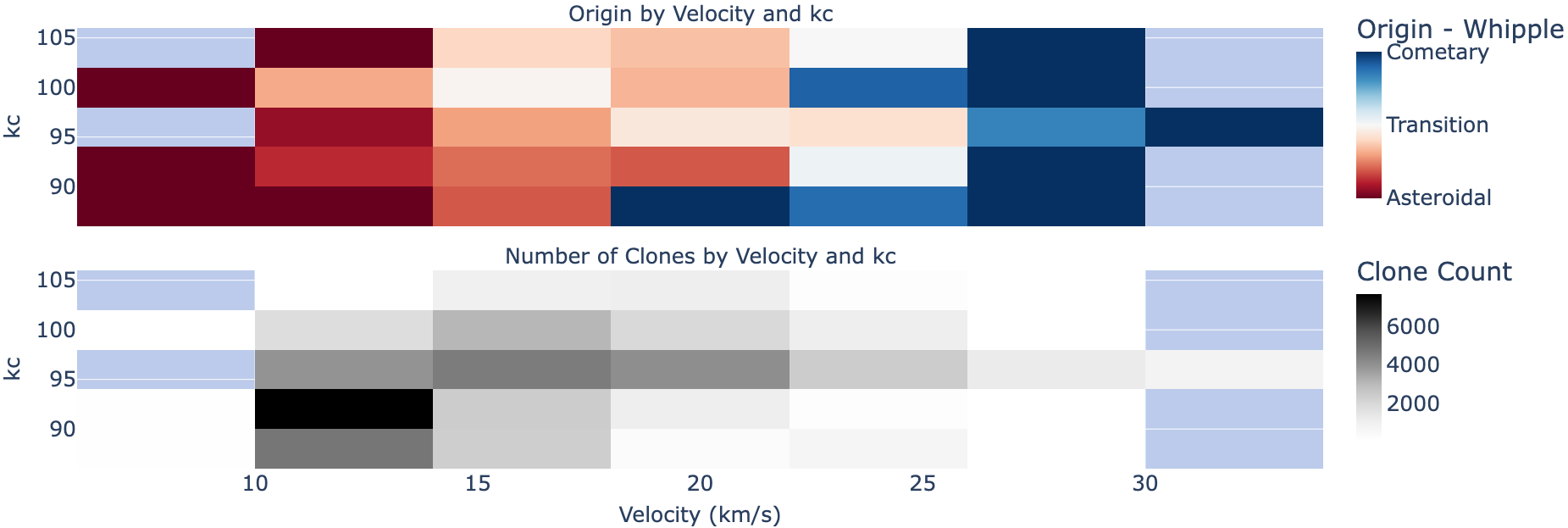}}
    \caption{Meteoroid origin, as determined by \citet{1954whipple}'s $K$-criterion, binned by $k_c$ parameter in increments of 1 (a) and velocity binned in increments of 1 km/s (b) over simulation time for the full dataset (CAMO and EMCCD). For a time close to the start of the simulation ($\sim -1$ years), we plot origin in a heat map with $k_c$ on one axis and velocity on the other (c). For all three plots, we show in a black-white scale how many clones are in each bin.}
    \label{fig:total-results-whipple}
\end{figure*}

\begin{figure*}[h!]
    \centering
    \subfigure[]{\includegraphics[width=0.93\linewidth]{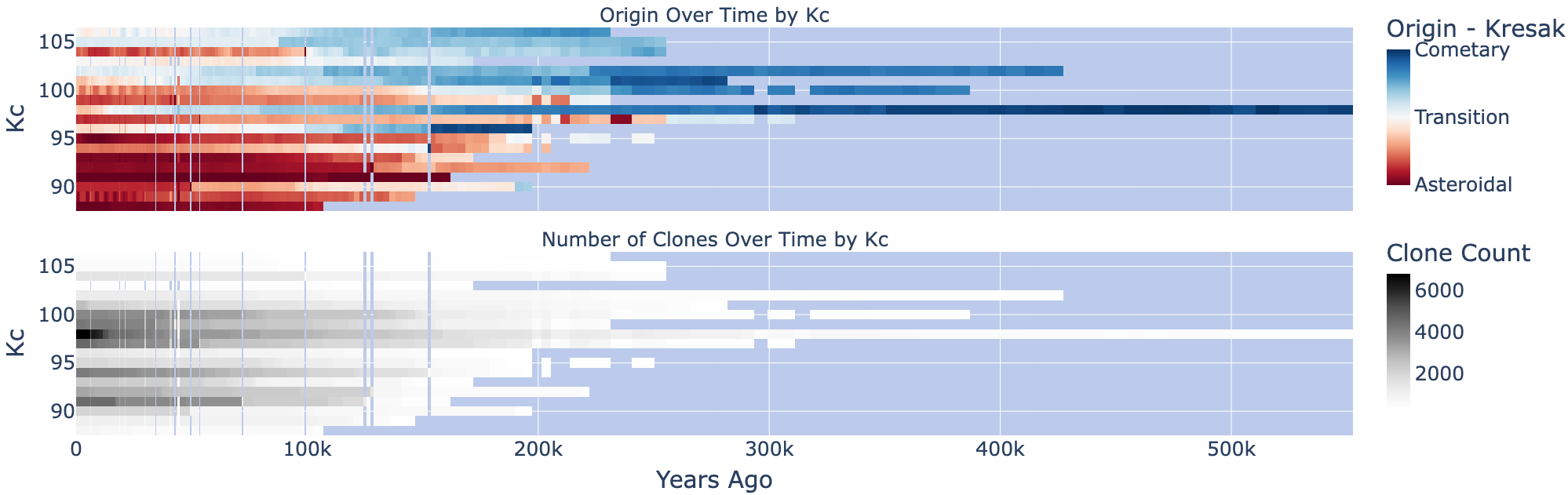}}
    \subfigure[]{\includegraphics[width=0.93\linewidth]{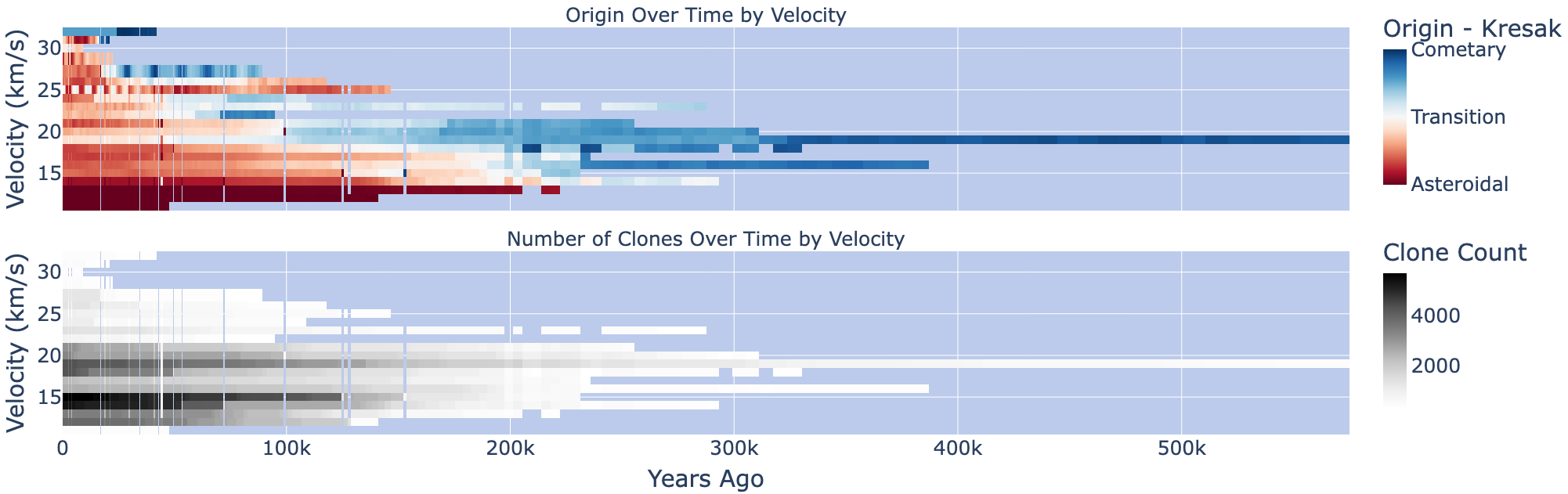}}
    \subfigure[]{\includegraphics[width=0.93\linewidth]{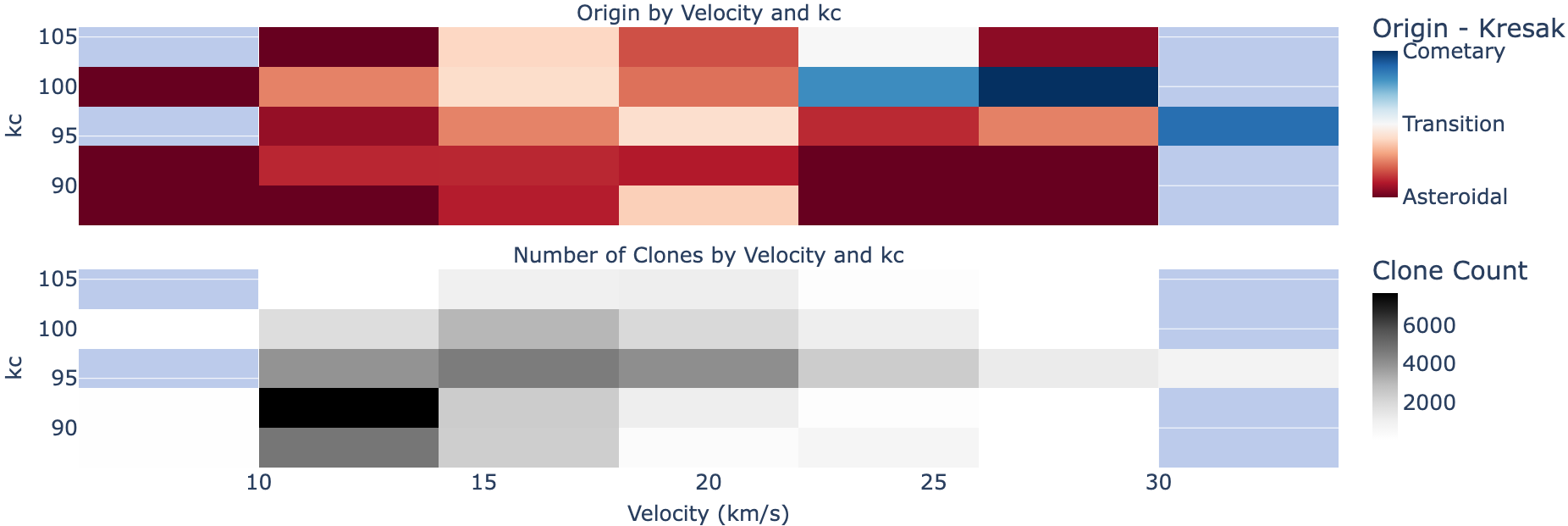}}
    \caption{Meteoroid origin, as determined by \citet{1967kresak}'s $Pe$-criterion, binned by $k_c$ parameter in increments of 1 (a) and velocity binned in increments of 1 km/s (b) over simulation time for the full dataset (CAMO and EMCCD). For a time close to the start of the simulation ($\approx-1$ years), we plot origin in a heat map with $k_c$ on one axis and velocity on the other (c). For all three plots, we show in a black-white scale how many clones are in each bin.}
    \label{fig:total-results-kresak}
\end{figure*}

\begin{figure*}[h!]
    \centering
    \subfigure[]{\includegraphics[width=\linewidth]{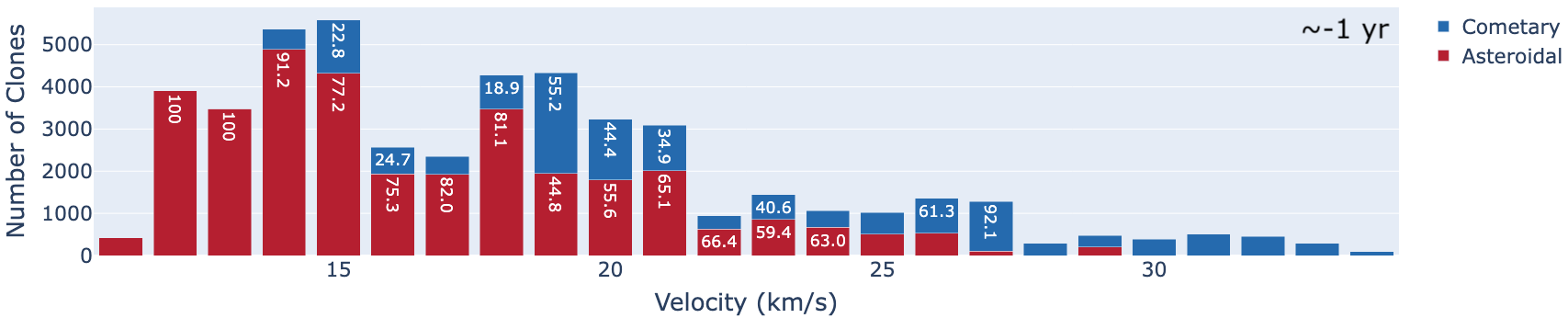}}
    \subfigure[]{\includegraphics[width=\linewidth]{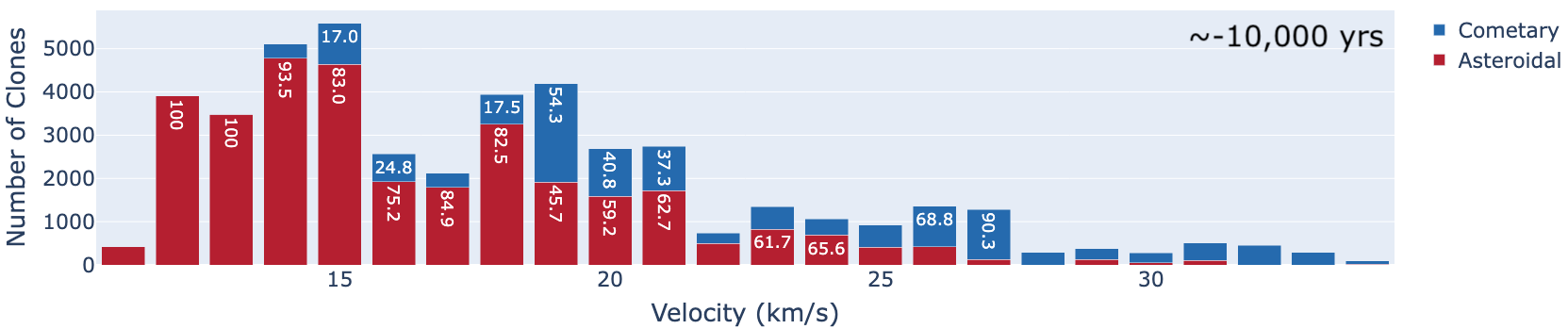}}
    \subfigure[]{\includegraphics[width=\linewidth]{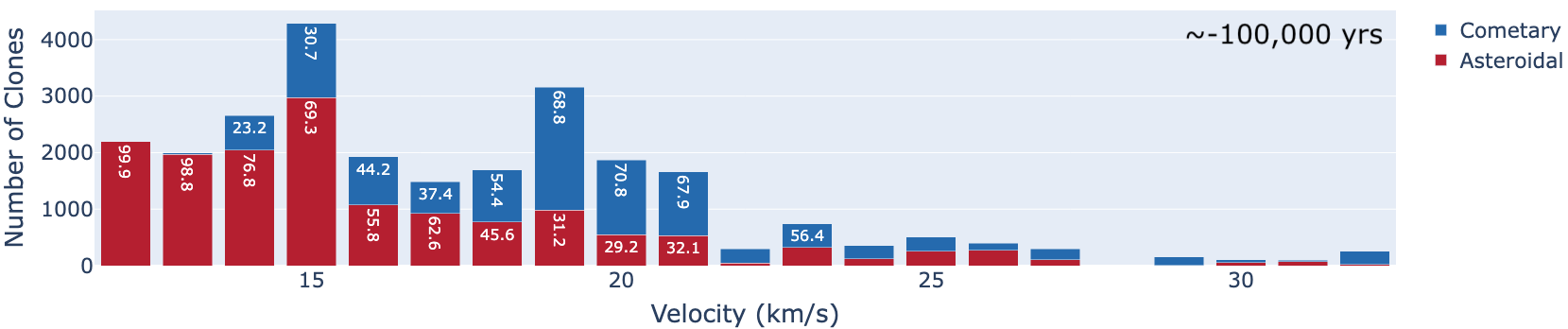}}
    \subfigure[]{\includegraphics[width=\linewidth]{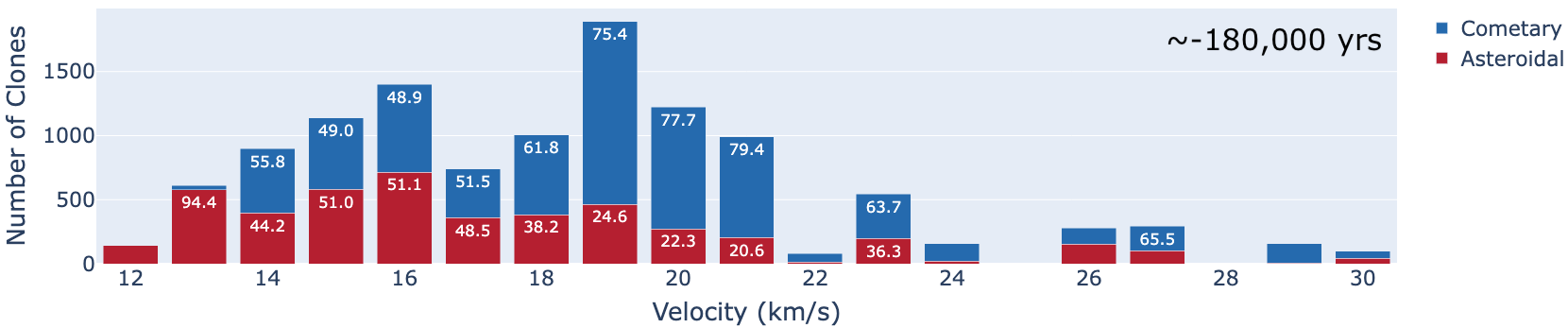}}
    \caption{Meteoroid origin, as determined by \citet{1954whipple}'s $K$-criterion, binned by velocity in increments of 1 km/s for discrete simulation times for the full dataset (CAMO and EMCCD). We show snapshots at a) $\sim -1$ year, b) $\sim -10$ kyr, c) $\sim -100$ kyr, and d) $\sim -180$ kyr. Shown in white on each bar (of reasonable size) are the percentages of cometary/asteroidal meteoroids in that bin.}
    \label{fig:whipple-time-slices}
\end{figure*}

\begin{figure*}[h!]
    \centering
    \subfigure[]{\includegraphics[width=\linewidth]{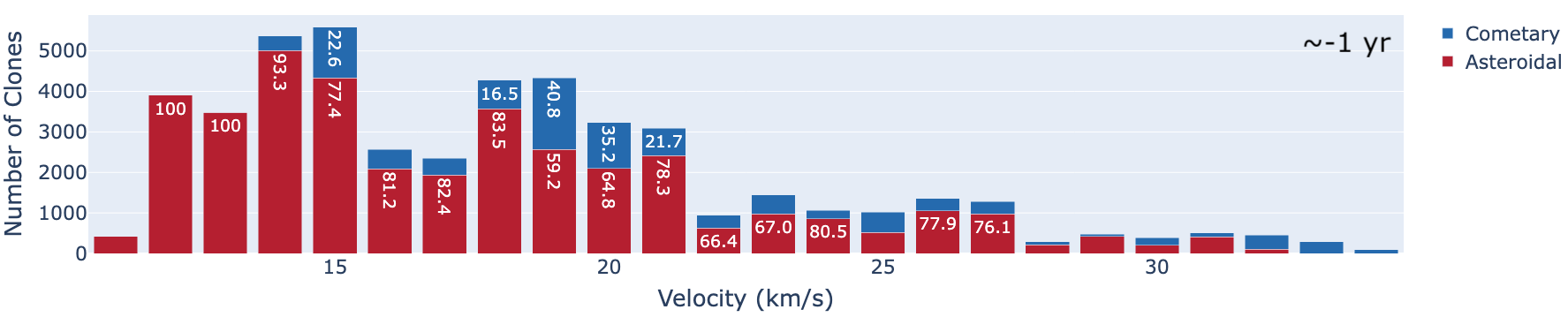}}
    \subfigure[]{\includegraphics[width=\linewidth]{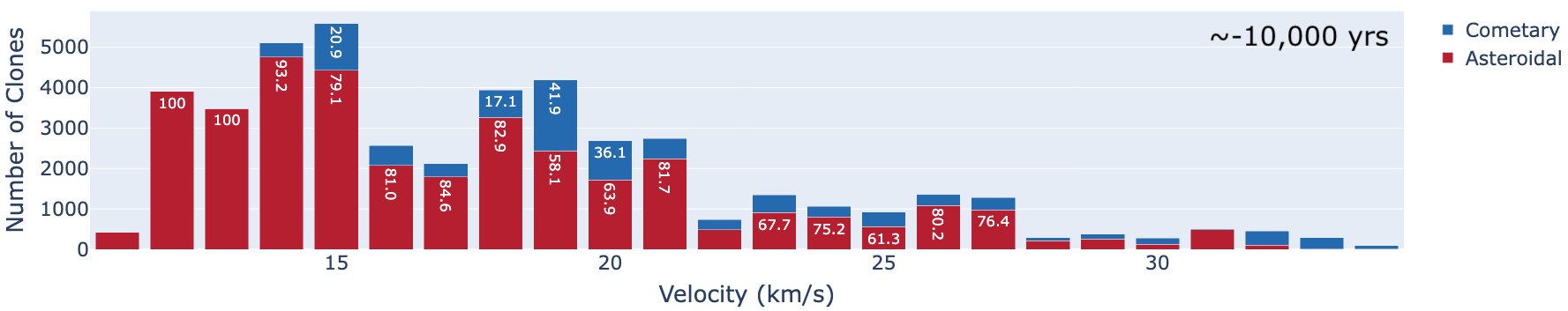}}
    \subfigure[]{\includegraphics[width=\linewidth]{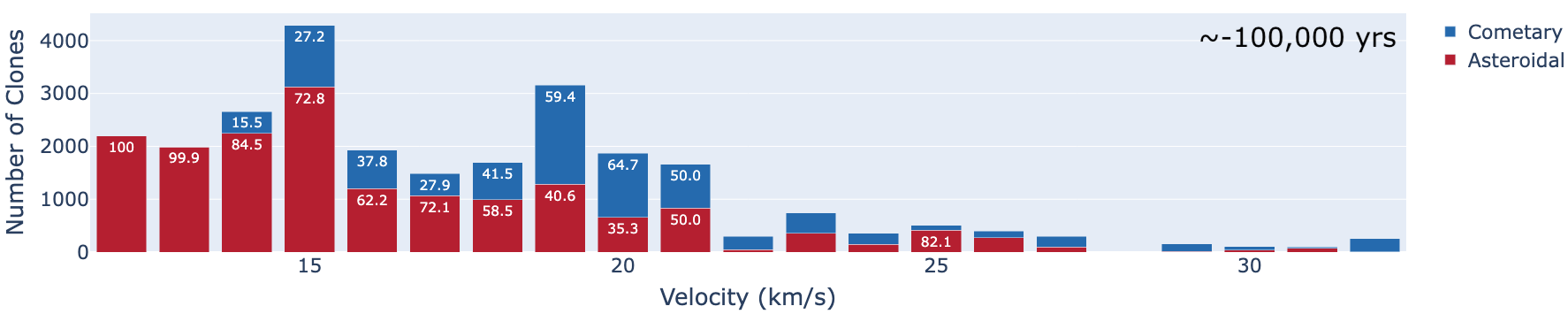}}
    \subfigure[]{\includegraphics[width=\linewidth]{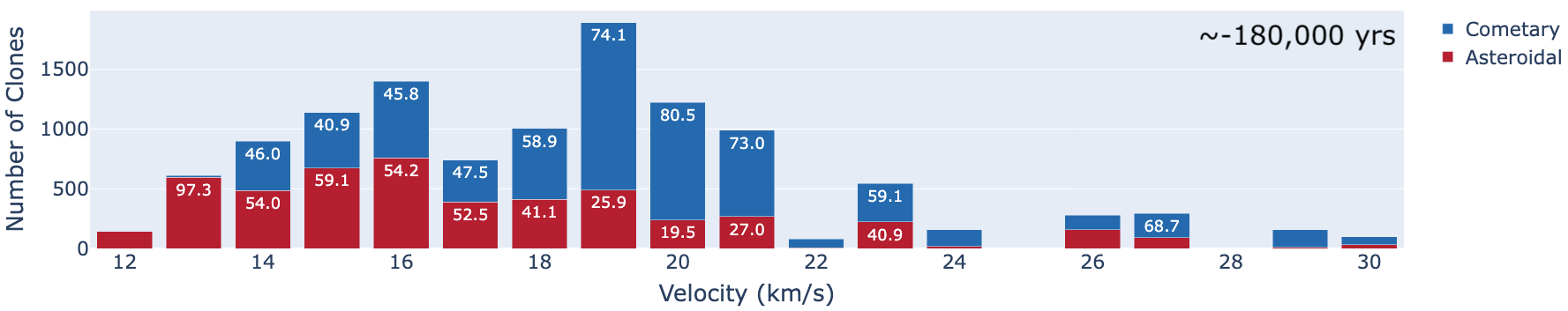}}
    \caption{Meteoroid origin, as determined by \citet{1967kresak}'s $Pe$-criterion, binned by velocity in increments of 1 km/s for discrete simulation times for the full dataset (CAMO and EMCCD). We show snapshots at a) $\sim -1$ year, b) $\sim -10$ kyr, c) $\sim -100$ kyr, and d) $\sim-180$ kyr.}
    \label{fig:kresak-time-slices}
\end{figure*}

One goal of our work is to characterize the distribution of meteoroids with asteroidal or cometary origin as a function of their impact velocity with Earth and time of ejection as most observational biases in meteor measurements are velocity-dependent and meteoroid orbits evolve over their lifetimes. The estimated parent body origin using $K$ and $Pe$ over a full range of possible ejection epochs is shown in \autoref{fig:total-results-whipple} and \autoref{fig:total-results-kresak}. The same results at four fixed ejection epochs are presented in \autoref{fig:whipple-time-slices} and \autoref{fig:kresak-time-slices}. These figures show that low-velocity meteoroids are predominantly asteroidal in origin (provided meteoroid ages are less than 200 kyr), while the fraction of cometary material increases with velocity, consistent with expectations from dynamical models of the sporadic meteoroid complex (e.g., \citealt{Wiegert2009, Nesvorny2011a}).

As prior work has suggested that asteroidal $k_c$ should generally be lower than cometary $k_c$ \citep{Borovička_2019_book, 2016jenniskens} based on parent body properties, we also examine correlation of our orbital history results with $k_c$. The percentage of meteoroids classified as asteroidal or cometary over time as a function of the $k_c$ parameter is shown in \autoref{fig:total-results-whipple} for the $K$-criterion and in \autoref{fig:total-results-kresak} for the $Pe$-criterion. Also shown as the bottom sub-plot in each figure is a snapshot in time about 1 year before Earth impact (ie. assuming the meteoroids are very young), which shows a heatmap of the orbital origins as a function of $k_c$ and velocity.

Examining subplot b in \autoref{fig:total-results-whipple} we find that \citet{1954whipple}'s $K$-criterion produces a dividing line between asteroidal and cometary orbits at 17 km/s for origin times less than $\sim -150$ kyr. If the age of the meteoroid is older, we do see a crossover where meteoroids appear to have been cometary upon ejection and evolved into asteroidal-like orbits over time. This can be seen clearly in subplot d of \autoref{fig:whipple-time-slices}, in which we see more ambiguous and slightly cometary-dominated velocity bins down to 14 km/s at a time slice taken at $\sim -180$ kyr.

There is also a clear divide, independent of age, where meteoroids at velocities of higher than 26 km/s are definitively cometary, and between 18 km/s and 26 km/s a potential mix of origins, leaning more towards cometary. However, the meteoroids we included at these higher velocities had shorter collisional lifetimes of $\sim10^4$ years so our cometary result here is limited in time. Based on orbital considerations alone, such high speeds are generally only accessible from cometary-type orbits so this is likely a secure result. 

Sub-plot a) in the same figure shows that the $k_c$ parameter tends to correlate strongly with asteroidal origins for $k_c<93$ while at higher $k_c$, we see a mix of both cometary and asteroidal origins. Superficially this suggests that the strongest (lowest $k_c$) are likely asteroidal in origin, but note that we have no significant number of surviving clones beyond a few hundred kyr, limiting this conclusion to short lifetimes. The mixing that is present at higher $k_c$ suggests that ablative strength is not necessarily indicative of parent body; there can be weakened asteroidal material. 

From the $k_c$ vs velocity heat map in the bottom panel of \autoref{fig:total-results-whipple}, we see that velocity seems to be a stronger indicator of parent body. The origin is consistently divided by velocity but there is overlap of both asteroidal and cometary origins across the $k_c$ axis, though this conclusion only applies assuming very recent ejection epoch, as the heat map is a snapshot of the origins a single year before the meteor observation at Earth. It is unlikely most of the sporadic background would be very young (less than the decoherence time of tens to hundreds of thousands of years for meteoroid streams), so the trend towards higher $k_c$ bins being dominated by cometary particles at older ejection epochs is more likely to be representative of this population (see the top panel in \autoref{fig:total-results-whipple}). 

In \autoref{fig:total-results-kresak}, we perform the same analysis with \citet{1967kresak}'s $Pe$-criterion. We see agreement with \citet{1954whipple}'s $K$-criterion about the dividing line between asteroidal and cometary orbits being at the 17 km/s bin, with older meteoroids having cometary origin and evolving into asteroidal orbits over time, although the transition age is closer to 200 kyr ages for the $Pe$-criterion. 

Unlike the results using \citet{1954whipple}'s $K$-criterion, we do not see a lower velocity limit above which all meteoroids are strongly cometary in origin. Generally, there is a mix of origins above 17 km/s, consistent with our findings in the validation section that the $Pe$-criterion tends to classify more towards asteroidal in the shower meteoroid cases than the $K$-criterion. 

Exploring the $k_c$ parameter, we see a stronger asteroidal origin signature compared to the $K$-criterion for $k_c < 95$ while at higher $k_c$, we see a mix of both cometary and asteroidal origins (sub-plot a). We still see from the $k_c$ vs velocity heat map that velocity is the stronger dividing factor for origin than $k_c$. As with the $K$-criterion, these patterns are limited by our lack of clones backward in time, particularly for high speeds and min/max $k_c$. This reflects the generally short collisional lifetimes of millimetre-sized particles in these orbits.

Overall, we find that the lowest speeds and $k_c$ values are likely populated by asteroidal origin meteoroids providing they are relatively young ($\leq \sim 150-200$ kyr). As noted in section \ref{sec:petrsims}, if meteoroids are older than 200 kyr, whether released from the main belt or JFCs, they would appear in asteroidal-like orbits at the current epoch. 

\subsection{Meteoroid Thermal Processing History and $k_c$}

\begin{figure*}[h!]
    \centering
    \setcounter{subfigure}{3}
    \subfigure[]{\includegraphics[width=\textwidth]{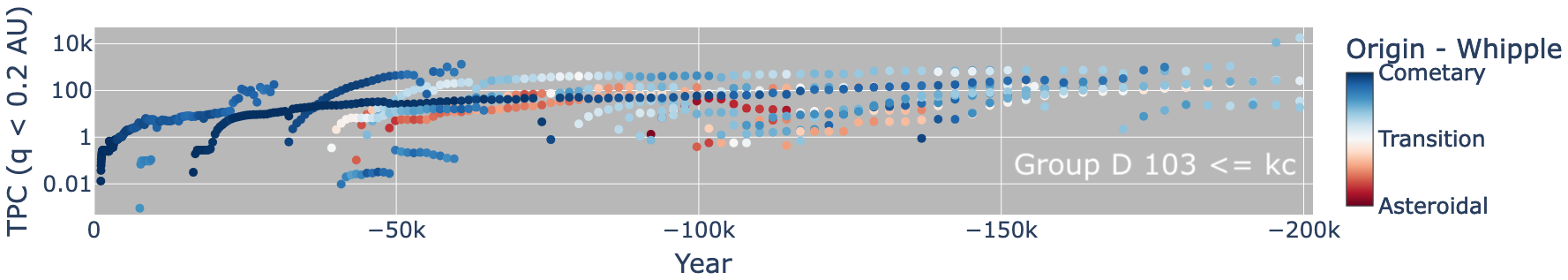}}
    \addtocounter{subfigure}{-2}
    \subfigure[]{\includegraphics[width=\textwidth]{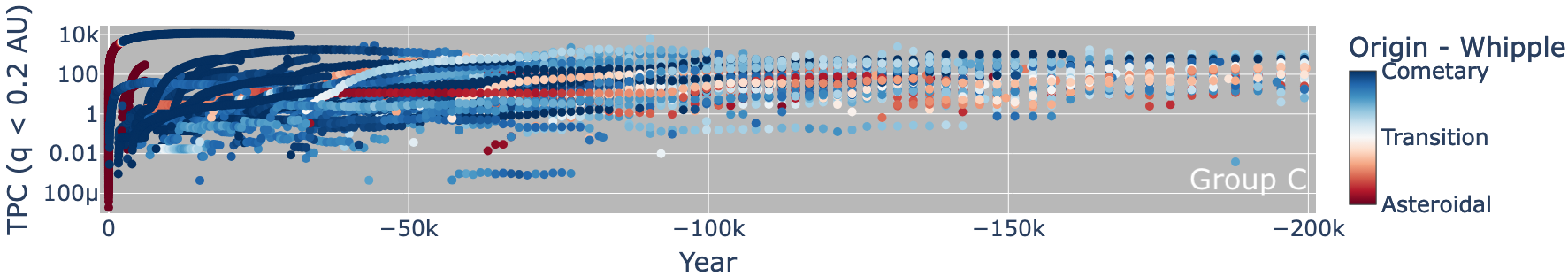}}
    \addtocounter{subfigure}{-2}
    \subfigure[]{\includegraphics[width=\textwidth]{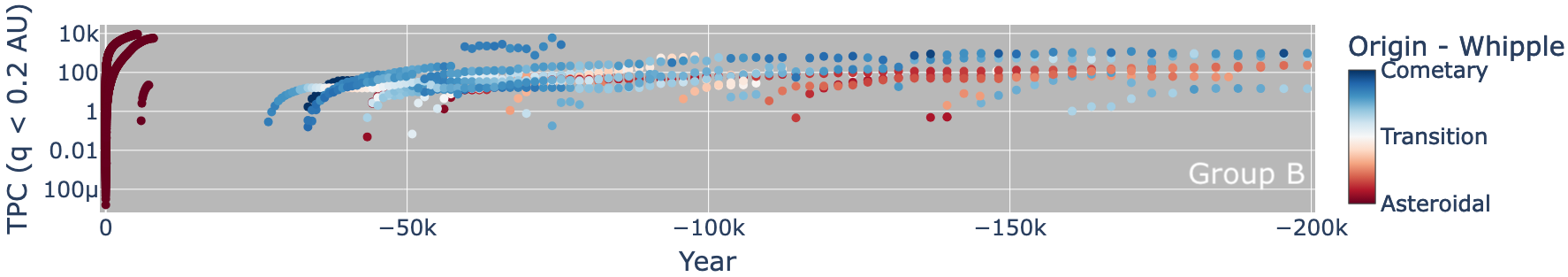}}
    \addtocounter{subfigure}{-2}
    \subfigure[]{\includegraphics[width=\textwidth]{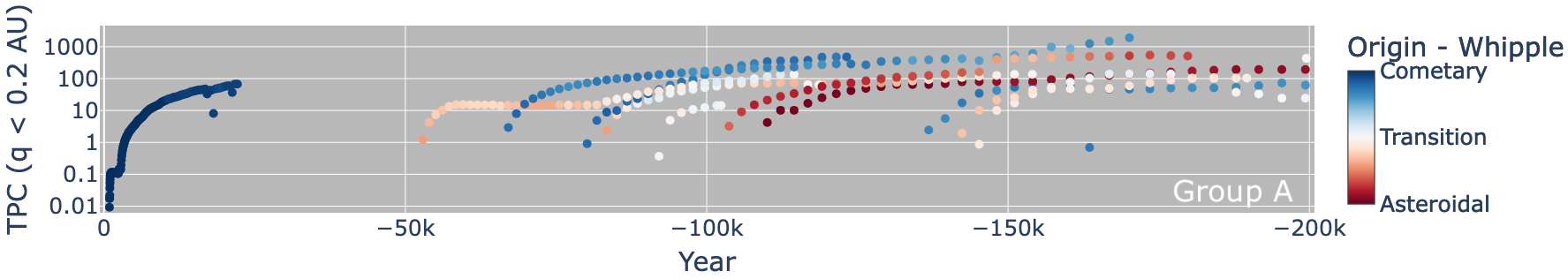}}
    \caption{The thermal processing coefficient (TPC) with a perihelion threshold of 0.2 AU, plotted over time for the 4 $k_c$ groups (see \autoref{subsubsec:popdemo}) that had non-zero TPC: a) Carbonaceous chondritic material from either asteroids or comets (37 meteoroids), b) Dense cometary material (76 meteoroids), c) Regular cometary material (211 meteoroids), and d) Soft cometary material (25 meteoroids). The colour of the markers indicates the dynamic origin, as determined by the $K$ criterion at each time step of the integration. Meteoroids with a TPC of zero over the 200 kyr window are not shown on these plots.}
    \label{fig:tpc2}
\end{figure*}

How does the thermal history of a meteoroid affect its apparent $k_c$ for meteoroids with different dynamical origins? To address this question, the evolution of the TPC over time for our dataset, separated into $k_c$ groupings, is shown in \autoref{fig:tpc2}.

Within our dataset of observed meteors, just under half (163/351) were found to have a non-zero TPC for the $q_{thresh} = 0.2$ AU threshold. This means at least one clone for each event spent time with perihelion lower than $q_{thresh}$. For simplicity, the following analysis takes the average over the lifetime of the surviving clones for each meteoroid (or the full 200 kyr we examine, whichever is longer) whether the majority of clones were deemed asteroidal or cometary for each criteria. Among these thermally processed meteoroids, 107 had a cometary orbital designation for the majority of their collisional lifetime as determined using the $Pe$-criterion and 134 as determined by the $K$-criterion.

This tendency for meteoroids on orbits dynamically similar to comets to experience low perihelion distances is consistent with prior work that finds comets evolve frequently to sungrazing states \citep{Bailey1996, Levison1994}, particularly for high inclinations \citep{Bailey1992}. Of the 188 meteoroids that never have any clones with $q < q_{thresh}$, only 1 had a cometary designation for the majority of their collisional lifetime as determined using \citet{1967kresak}'s criteria and 20 as determined by \citet{1954whipple}'s criteria, confirming our findings that the thermally processed population is mainly cometary. 

\begin{figure}[h!]
    \centering
    \subfigure[]{\includegraphics[width=0.49\textwidth]{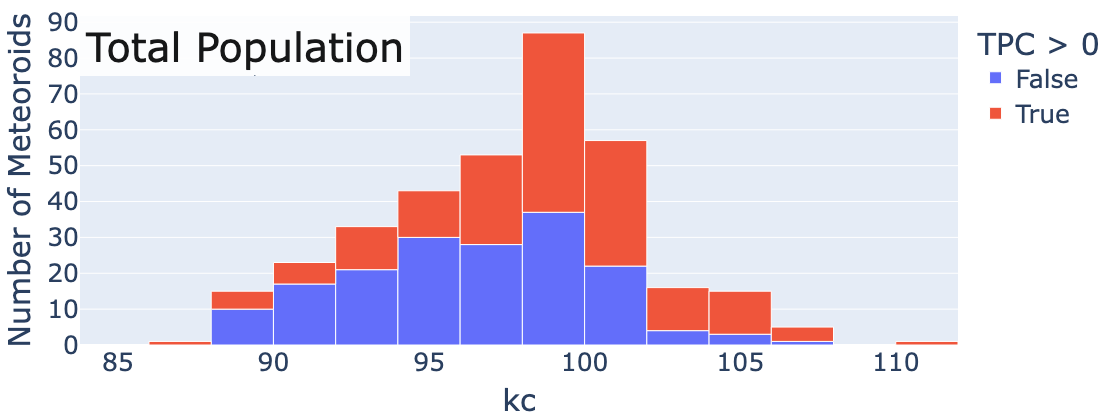}}
    \subfigure[]{\includegraphics[width=0.49\textwidth]{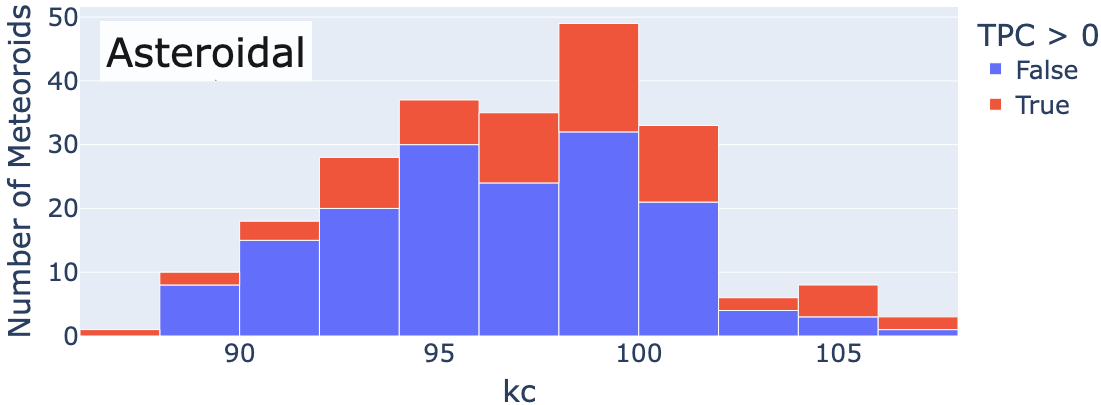}}
    \subfigure[]{\includegraphics[width=0.49\textwidth]{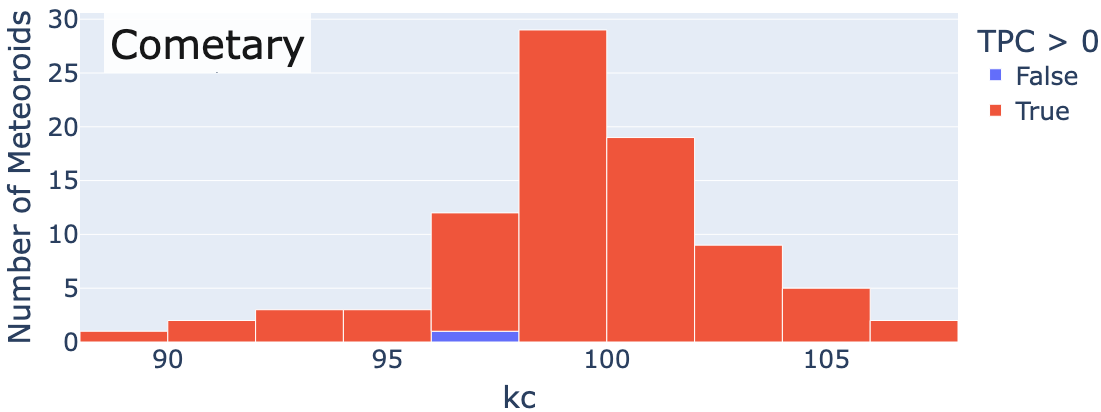}}
    \subfigure[]{\includegraphics[width=0.49\textwidth]{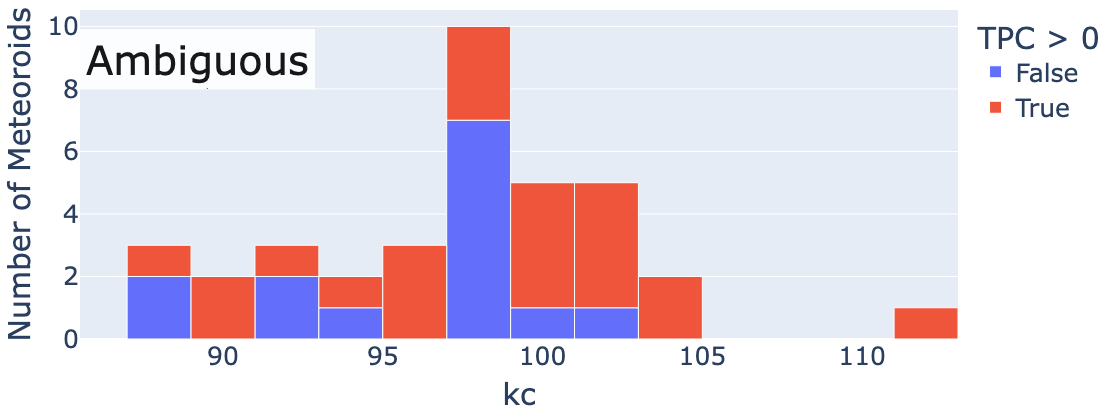}}
    \caption{Meteor populations showing non-zero thermal processing coefficient (TPC) (red) with a perihelion threshold of 0.2 AU compared to those not experiencing thermal heating, binned by $k_c$ for a) the full population, b) the dynamically asteroidal population, c) the dynamically cometary population, and d) the population classified as asteroidal by the $Pe$-criterion but cometary by the $K$-criterion. Note that for b) and c), we only include the meteoroids with classifications agreed upon by both $K$ and $Pe$ criteria.}
    \label{fig:tpc-kc}
\end{figure}

\autoref{fig:tpc-kc} shows the distribution of $k_c$ for all backward integrated meteors in our dataset grouped by agreement between $Pe$ and $K$ criteria. The ``ambiguous" group shown in subplot d of \autoref{fig:tpc-kc} are those that had a majority of clones classified as asteroidal by the $Pe$-criterion and cometary by the $K$-criterion over the lifetime/200 kyr time period. These were categorized in colour as either having the potential for thermal processing based on a threshold of $q<0.2$ AU or never having a perihelion below that threshold. 

\begin{figure}[h!]
    \centering
    \includegraphics[width=\linewidth]{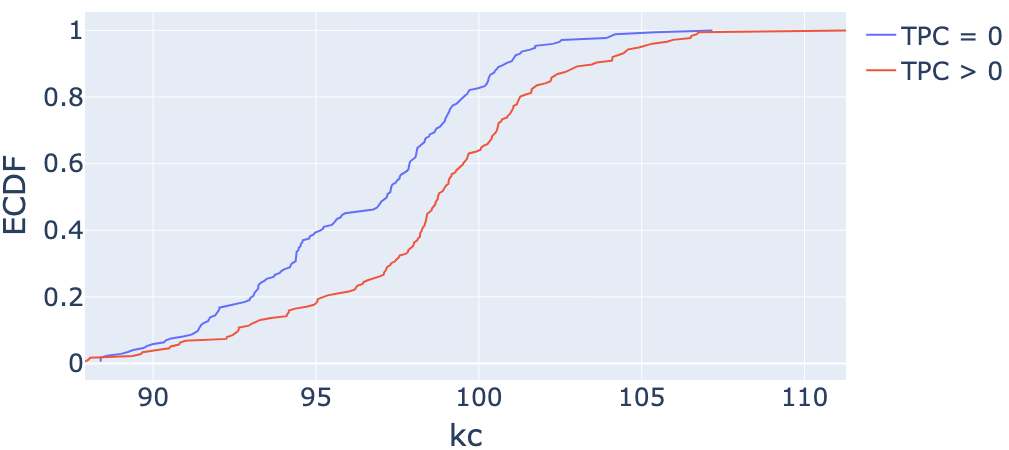}
    \caption{The Empirical Cumulative Distribution Function (ECDF) for the full dataset of meteoroids that have potentially experienced thermal processing in the past (red) compared to those that have not (blue).}
    \label{fig:ecdf-tcp}
\end{figure}

We apply a two-sample Kolmogorov-Smirnov (KS) test \citep{Kolmogorov1933, Smirnov1948} to determine if the distributions of $k_c$ between the thermally processed and unprocessed populations shown in \autoref{fig:tpc-kc} (subplot A) are statistically different. The KS statistic showed that the maximum vertical separation between the two cumulative distributions was $D = 0.28$, indicating that at some $k_c$ value, the fraction of meteoroids that experienced thermal processing below that $k_c$ value differs by ${\sim} 28$ percentage points compared with those that had not experienced thermal processing. A $p$-value of 0.05 or lower is required to reject the null hypothesis that both $k_c$ distributions are the same. The low $p$-value of $1.82\times10^{-6}$, indicates that the two samples (TPC = 0 and TPC > 0) are highly unlikely to be drawn from the same parent distribution. We see from the Empirical Cumulative Distribution Function (ECDF) plot shown in \autoref{fig:ecdf-tcp} that thermally processed meteoroids exhibit systematically higher $k_c$ values. 

We consider that our TPC values may not be a good metric for thermal processing as there is the potential for a few clones to experience small perihelia when the majority of clones for that meteoroid never pass the threshold and we see small TPC values over long time periods that are misleading. \autoref{fig:tpc2} has a logarithmic scale on the TPC axis, showing that there are a few meteoroids in the higher $k_c$ groups that have a TPC of less than 1 full passage (average time clones spent with $q < q_{thresh}$ divided by average orbital period) over long time periods.

We further tested whether the degree of thermal processing is correlated with the $k_c$ distribution by comparing populations with increasing thresholds in TPC (TPC $> 0$, $> 5$, and $> 10$). Across the full dataset, we find that while meteoroids that have experienced any thermal processing are statistically distinct from those that have not, there is no significant difference between populations with increasing levels of thermal exposure. This suggests that thermal processing does not act as a continuous modifier of fragmentation behaviour, but instead produces a threshold effect. 

When restricting the analysis to dynamically asteroidal meteoroids, this behaviour persists, but with a gradual increase in statistical significance when comparing unprocessed meteoroids to increasingly processed subsets (e.g., $p = 0.060$ for TPC $> 0$, $p = 0.045$ for TPC $> 5$, and $p = 0.032$ for TPC $> 10$). However, comparisons between processed populations themselves yield consistently high $p$-values, indicating no measurable difference in $k_c$ once thermal processing has occurred. This is consistent with a scenario in which a small number of perihelion passages is sufficient to induce structural modification, such as devolatilization, micro-fracturing, or the formation of a porous surface layer. However, after these few passages, ongoing thermal cycling produces little further change in structure prior to complete disruption. 

\section{Discussion}
\label{sec:discussion}
\subsection{Interpretation of TPC and $k_c$ Correlations}

The connection between thermal processing and the observed $k_c$ distributions is not straightforward, as thermal processing is strongly coupled to orbital dynamics. In particular, meteoroids on dynamically cometary orbits are far more likely to experience sustained periods at low perihelion distances ($q<0.2$ au) and therefore undergo significant thermal processing, while dynamically asteroidal meteoroids rarely do. At the same time, it has been suggested that cometary material is intrinsically more fragile and porous at the time of ejection. Thus young or unaltered cometary meteoroids should produce systematically higher $k_c$ values during atmospheric entry \citep{Levasseur-Regourd_etal_2018}. As a result, thermal processing is not an independent variable but is preferentially applied to meteoroids that already exhibit the suggested cometary-like fragility. This coupling naturally produces a population-level correlation in which thermally processed meteoroids display higher $k_c$, even if thermal processing itself is not the cause of the fragmentation behaviour.

To isolate the physical effect of thermal processing, we performed an additional analysis restricted to meteoroids classified as dynamically asteroidal, as the populations of both thermally processed and unprocessed meteoroids were large enough for meaningful comparison (unlike the subset of dynamically cometary meteoroids). Within this subset, we compared the distributions of $k_c$ for meteoroids that experienced significant thermal processing (TPC $> 10$) and those that did not. Our KS test yields $D = 0.22$ and $p = 0.032$, indicating that the two samples are unlikely to be drawn from the same parent distribution at the 95\% confidence level. This indicates that thermal processing has a measurable effect on meteoroid fragmentation behaviour even when dynamical origin is held fixed. However, the magnitude of this shift is modest compared to the separation between asteroidal and cometary populations, suggesting that thermal processing acts as a secondary modifier of meteoroid structure rather than the primary determinant of $k_c$.

One limitation in this analysis to be noted is that as we cease evaluating TPC once the temporal sampling of the integration outputs becomes larger than the Lidov-Kozai oscillation period, our TPC values should be interpreted as lower limits for cumulative thermal processing. Meteoroids classified here as unprocessed or weakly processed may have undergone additional thermal processing outside the interval over which TPC can be reliably measured. This introduces uncertainty in the separation between processed and unprocessed populations, although within the resolvable timeframe, we find that thermally processed meteoroids exhibit systematically higher $k_c$ values even when dynamical origin is held fixed. If this trend extends to unresolved earlier epochs, then any misclassifications would tend to reduce the observed contrast between populations, implying that the measured difference may be conservative.

An interpretation consistent with this result is that thermal processing leads to structural weakening rather than simple densification. While solar heating can drive devolatilization and compaction in the interior, repeated thermal cycling also induces micro-fracturing and the formation of a porous, low-conductivity surface layer. \citet{capek2012_streesesII} show that even a small number of perihelion passages can produce such an outer rind, which can reduce internal thermal gradients and delay catastrophic disruption. Meteoroids that survive this process may therefore develop a mechanically weak outer structure that promotes early fragmentation during atmospheric entry, leading to systematically higher $k_c$ values. Once this structural modification has occurred, additional thermal processing may have limited incremental effect until the meteoroid is ultimately destroyed. In this framework, thermal processing acts as a structural re-sculpting process, rather than one which purely changes bulk density. This reinforces the interpretation of $k_c$ as a composite diagnostic of fragmentation behaviour as opposed to a direct proxy for bulk density.

\begin{figure}[h!]
    \centering
    \includegraphics[width=\linewidth]{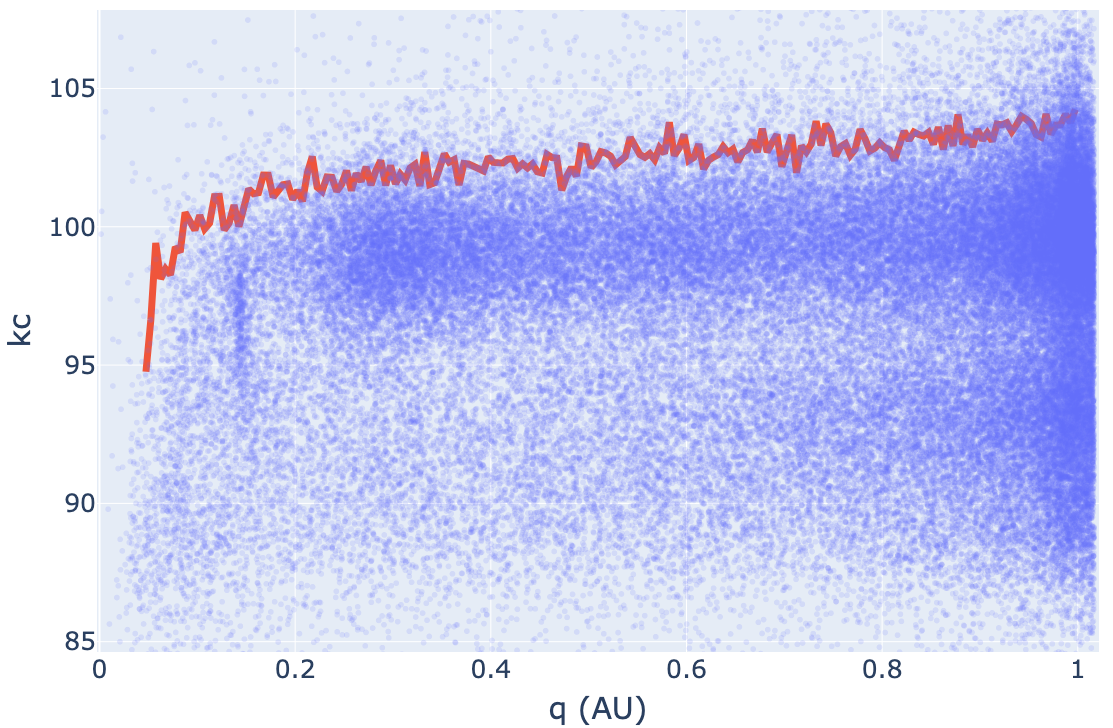}
    \caption{The $k_c$ parameter versus current perihelion distance for all 113928 EMCCD meteors. Overlaid as a red line, we show the 95th percentile line; that is 95\% of the population resides below the line.}
    \label{fig:all-kc}
\end{figure}

\begin{figure}[h!]
    \centering
    \subfigure[]{\includegraphics[width=0.5\textwidth]{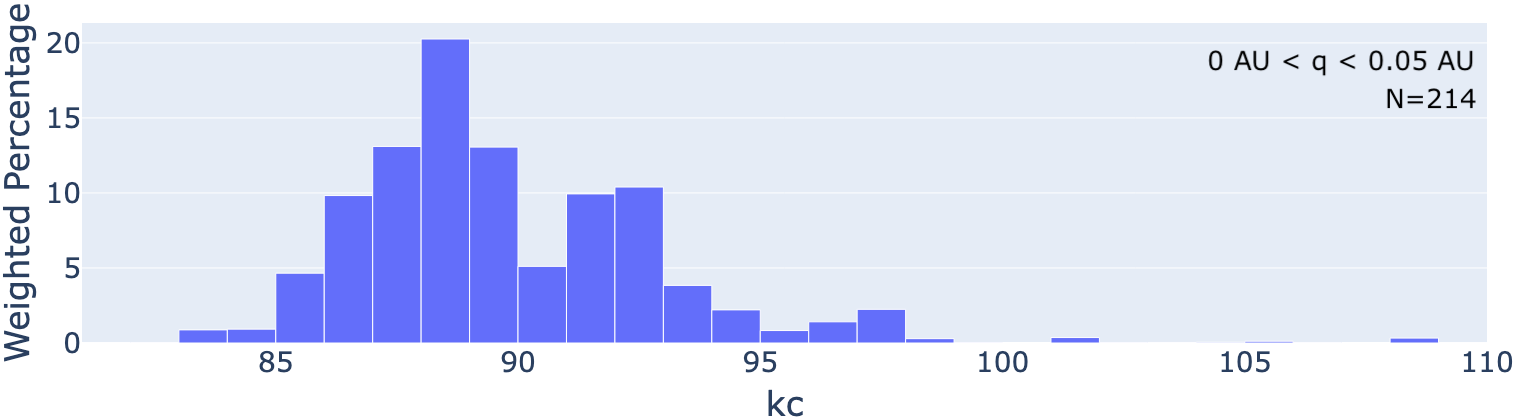}}
    \subfigure[]{\includegraphics[width=0.5\textwidth]{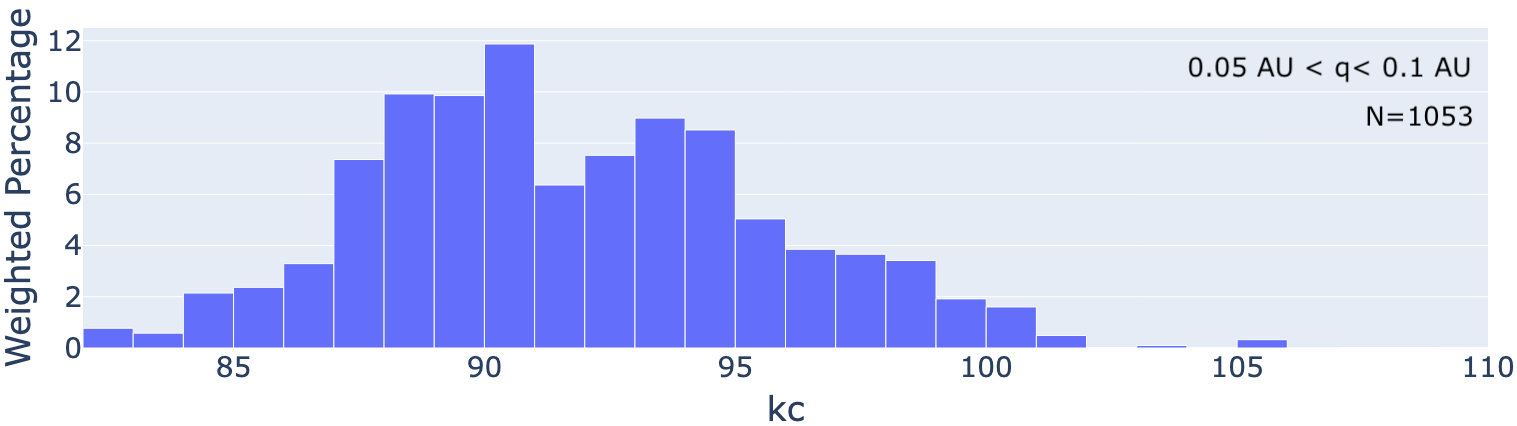}}
    \subfigure[]{\includegraphics[width=0.5\textwidth]{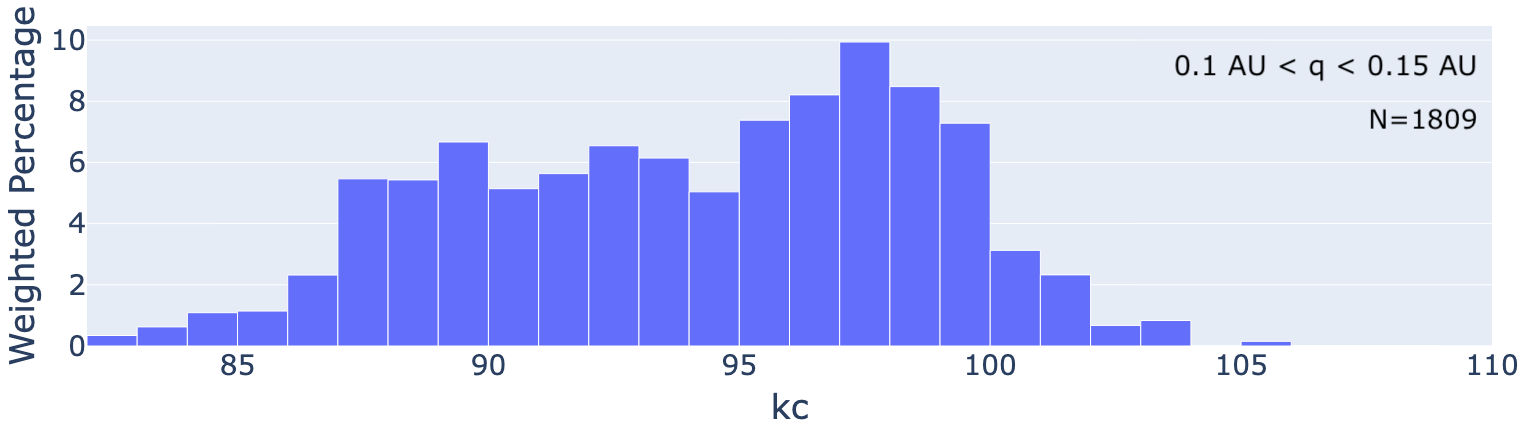}}
    \subfigure[]{\includegraphics[width=0.5\textwidth]{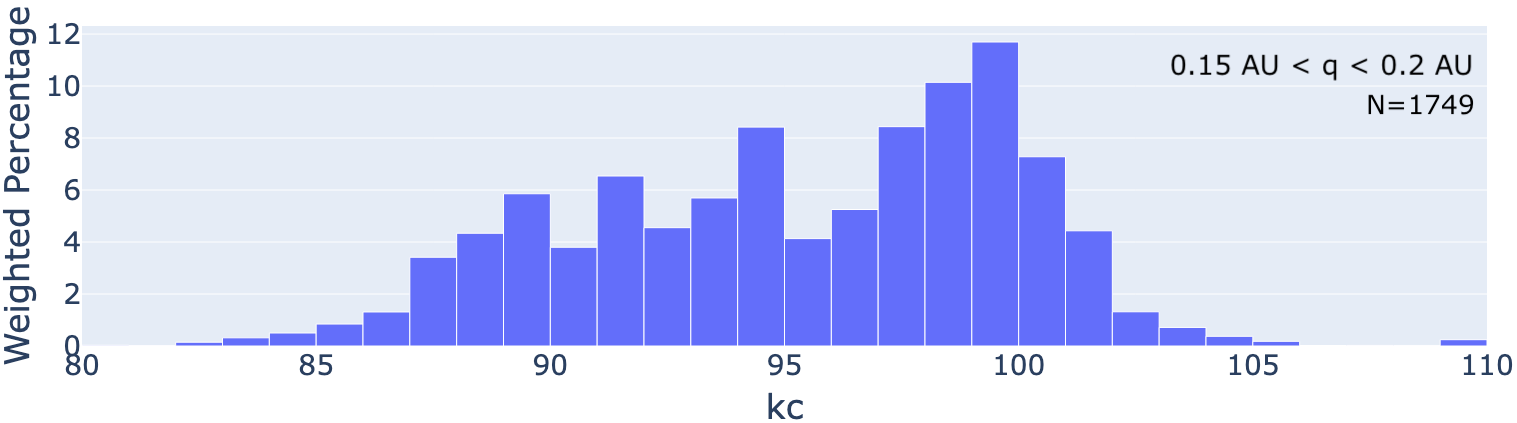}}
    \subfigure[]{\includegraphics[width=0.5\textwidth]{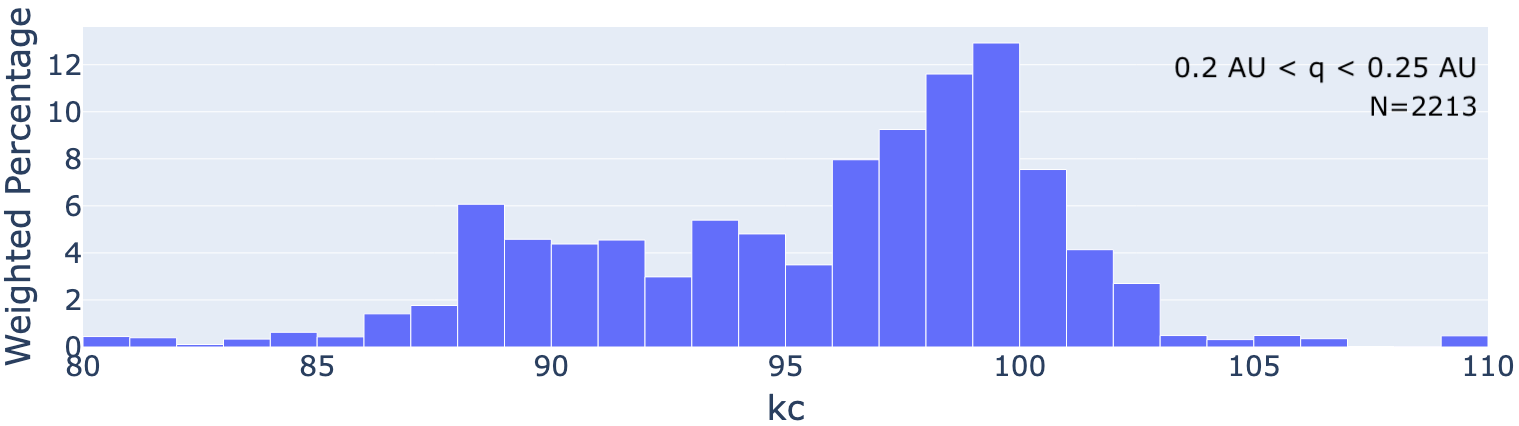}}
    \subfigure[]{\includegraphics[width=0.5\textwidth]{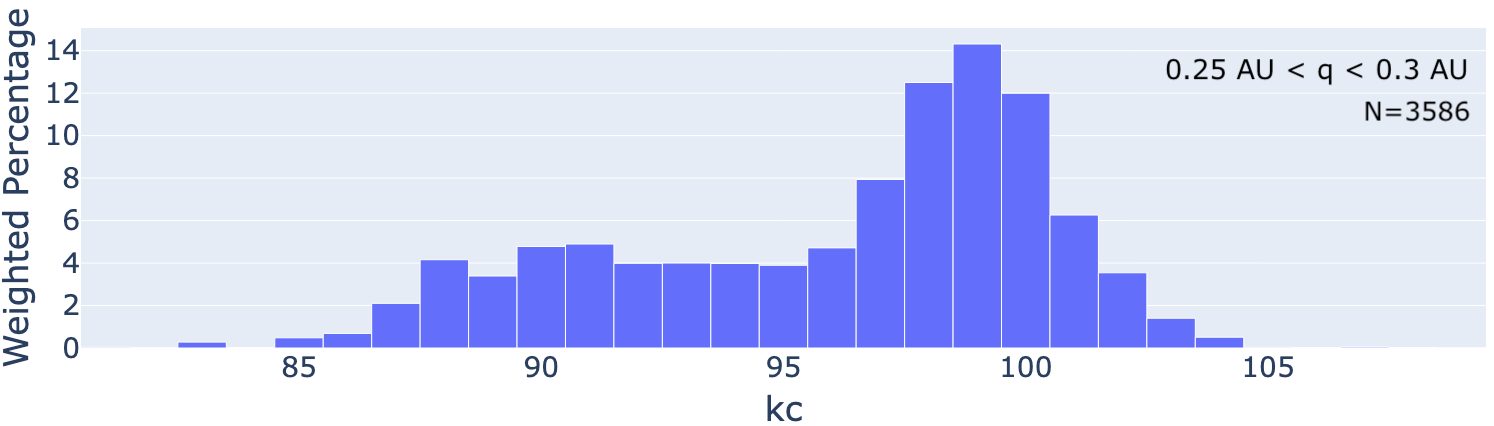}}
    \caption{The lower $q$ subpopulation of the full EMCCD population (113928 meteors) binned in 0.05 AU widths in $q$, and re-weighting the number of events in each $k_c$ bin based on the inverse EIP \citep{2013pokorny} for each meteoroid using its orbit at the time of Earth impact.}
    \label{fig:kc-q-hists}
\end{figure}

To examine this further, in \autoref{fig:all-kc} we plot the full EMCCD dataset of 113928 meteors in $k_c-q$ space. A clear trend toward lower $k_c$ emerges at small perihelion distances, beginning near $q \approx 0.3$ AU and becoming more pronounced below $q \approx 0.1$ AU. We overlay the 95th percentile envelope (red line) to highlight this decline, noting that the small number of meteors at very low $q$ introduces additional uncertainty in this regime. Using binned windows in $q$ and re-weighting the number of events in each $k_c$ bin by the inverse EIP \citep{2013pokorny}, we further demonstrate a systematic shift in the distribution toward lower $k_c$ with decreasing $q$ (\autoref{fig:kc-q-hists}). 

Our interpretation of these trends, together with the results of \citet{2010capek}, suggest that meteoroids subjected to extreme thermal processing are progressively removed from the observable millimetre-centimetre population at Earth through catastrophic disruption. Because the orbits that enable such extreme thermal processing are predominantly dynamically cometary, this removal mechanism preferentially depletes the most fragile cometary material. This provides a natural explanation for why, in \autoref{fig:total-results-whipple} and \autoref{fig:total-results-kresak}, some higher $k_c$ bins are unexpectedly dominated by dynamically asteroidal meteoroids at short inferred dynamical ages: the most weakly bound cometary meteoroids have already been removed from the observable population.

An alternative but not mutually exclusive, interpretation is that many meteoroids presently on dynamically asteroidal orbits may have evolved from cometary progenitors over long timescales. As shown in \autoref{fig:total-results-whipple} and \autoref{fig:total-results-kresak}, the fraction of cometary clones increases when integrating backward in time, leading to a clearer separation in which higher $k_c$ bins are dominated by cometary material at earlier epochs. In this scenario, the present-day mixing of dynamical classes reflects orbital evolution rather than intrinsic compositional similarity. Together, these results suggest that both dynamical evolution and thermal processing contribute to shaping the observed $k_c$ distribution. In this interpretation dynamical origin sets the baseline structural properties of meteoroids, while thermal processing and long-term evolution act to modify and selectively remove material, producing the complex correlations observed in the present-day mm-sized meteoroid population impacting Earth.

\subsection{Interpretation of Origin Classification}

A central motivation for this study is the long-standing uncertainty surrounding the relative contributions of asteroidal and cometary material to the meteoroid complex, specifically within the low-velocity dynamically unassociated sporadic population. Previous studies have reached differing conclusions depending on whether classifications were based primarily on orbital dynamics, material properties, or mass-balance arguments. By validating and applying two prominent dynamical classification schemes to a large observational dataset of millimetre-sized meteoroids, with well characterized uncertainties, our results help clarify which observable parameters most robustly encode origin as either asteroidal or cometary. 

Across both the $K$ and $Pe$ criteria, we see a strong correlation of asteroidal-dominated material with initial velocities 17 km/s and below. However, this result presumes mm-sized cometary meteoroids are less than ${\sim}150-200$ kyr in age. JFC-related meteoroids at our sizes which are older would have evolved into orbits that are now indistinguishable from those initially of asteroidal origin. Hence our results suggest either young asteroidal meteoroids or a mix of older JFC and asteroidal origin meteoroids comprising the bulk of the millimetre-sized particle population reaching the Earth at low velocities. 

In contrast, we have a more secure result from our backward integrations suggesting that at higher velocities ($\geq 27$ km/s) we increasingly sample cometary reservoirs. We do not see a sharp compositional divide based on the $k_c$ distributions of origins, indicating that within the asteroidal and cometary domains, there is both mixing and likely meteoroid evolution affecting the $k_c$.

From both the behaviour of observed meteoroids in our backwards integrations and in the virtual particles forward integrated from source regions in our validation integration discussed in \autoref{sec:petrsims}, we see that meteoroids classified as asteroidal at times close to Earth impact (younger ages) can transition to cometary classification at older ages. These transitions occur over long enough timescales that the corresponding meteoroids will lose their dynamical coherence with their parent body due to planetary perturbations and radiation pressure progressively reshaping their orbits. This suggests that without a known age for an individual meteoroid, orbital resemblance alone becomes insufficient to identify origin and to be confident in the dynamical classification, age limits based on other physical factors are needed.

The $k_c$ parameter, often tied to meteoroid bulk density and by extension, origin, shows substantial overlap between asteroidal and cometary populations by both orbital classification criteria. Although the nominal boundary separating asteroidal and cometary material shifts between $k_c \approx 93$ and $k_c \approx 100$ depending on the adopted criterion, neither threshold produces a clean separation based on orbital history. Further, the $k_c$–velocity phase space reveals diffuse boundaries in material properties but sharp divisions in velocity. This finding directly supports recent conclusions that material properties alone are insufficient for determining meteoroid parentage and should instead be interpreted as indicators of both evolutionary processing and parent body origin.

\subsection{Orbital Divergence and Classification Robustness}

Although the collisional lifetime of a meteoroid sets an upper limit on its physical age, we are primarily concerned with how long its orbital evolution can be followed backward in time in a way that preserves meaningful information about its dynamical origin. Rather than focusing on the precise reconstruction of an individual meteoroid's past orbit, our analysis is concerned with whether its classification as asteroidal or cometary remains robust over time.

Orbital divergence due to chaotic dynamics is often quantified using the Lyapunov Characteristic Exponent (LCE), which describes the exponential rate at which nearby trajectories diverge. The inverse of the LCE defines the Lyapunov time, a commonly used estimate of the timescale over which an orbit becomes unpredictable. In principle, this provides a limit on the reliability of backward integrations. However, for Earth-crossing meteoroids, Lyapunov times are typically short due to frequent planetary perturbations, and their calculation incorporates divergence in all orbital elements.

For the purposes of this study, such a definition of orbital divergence is not directly relevant. Our classification relies only on specific combinations of orbital elements through the $Pe$ and $K$ criteria, and therefore the divergence that matters is not the total phase-space separation between trajectories, but whether that divergence leads to a change in dynamical classification. In other words, even if clones of a meteoroid diverge significantly in orbital element space, this does not affect our analysis so long as they remain within the same orbital regime to be clearly classified as comet or asteroidal.

This distinction leads to two scenarios of interest when considering the evolution of clone populations. First, a meteoroid may have a population of clones that straddles the classification boundary for the majority of its integration time, with no clear trend toward either asteroidal or cometary classification. Such cases represent intrinsically ambiguous dynamical origins, where classification is uncertain, regardless of orbital divergence. Second, a meteoroid may exhibit a transition in classification over time, with its clone population evolving from one regime to another. These transitions are physically meaningful and reflect the dynamical evolution of meteoroid orbits, for example through the effects of radiation forces driving cometary orbits toward asteroidal-like configurations.

To assess the impact of orbital divergence on our results, we examined the evolution of $Pe$ and $K$ values for each meteoroid and its associated clone population throughout the integration. We identified ``straddler'' meteoroids as those whose clones remain distributed across the classification boundary for most of their lifetime without a clear transition. This identification was performed independently for each classification criterion, as straddling one boundary does not necessarily imply straddling the other.

We then repeated our analysis with these straddler meteoroids removed from the sample. We find that their exclusion does not produce a significant change in the overall trends of dynamical origin as a function of velocity. This indicates that ambiguous cases do not bias our results, and that the statistical conclusions of our study are robust to the effects of orbital divergence.

In this framework, the effective backward time horizon of our analysis is not set by a single dynamical timescale such as the Lyapunov time, but instead by the persistence of a consistent classification across the clone population. As long as the majority of clones occupy a well-defined dynamical regime, the meteoroid's origin can be inferred with confidence, even in the presence of substantial orbital divergence.

\section{Conclusions}

In this work, we conducted a quantitative validation of several orbit-based classification schemes designed to assign parent body origins to small meteoroids.  Our focus was primarily on the $K$-criterion proposed originally by \citet{1954whipple} and the $Pe$-criterion proposed by \citet{1967kresak}. We also examined the $Q$-cutoff classification scheme proposed by \citep{Borovicka2022physical} and the \citet{1890Tisserand} invariant. 

Using backward integrations that include both gravitational and radiation pressure forces, we find that the $K$ and $Pe$ criteria are the most robust for recovering broad parent-body origin classes over timescales relevant to millimetre-sized meteoroids. These validated criteria were then applied to a large, well-characterized dataset of meteoroids observed by the CAMO and EMCCD systems to assess the relative asteroidal and cometary contributions to the sporadic meteoroid population as a function of entry velocity.

Across both the $K$ and $Pe$ criteria, entry velocity consistently emerged as the strongest discriminator of origin, with a transition near 17 km/s, separating predominantly asteroidal meteoroids from those increasingly dominated by cometary sources, assuming meteoroid ages were less than 200 kyr. In contrast, the ablative fragmentation parameter $k_c$ exhibits substantial overlap between asteroidal and cometary populations and does not provide a clean separation under either criterion. While $k_c$ remains a valuable descriptor of atmospheric fragmentation behaviour, our results reinforce that material properties alone are insufficient for uniquely identifying parent-body origin and must be interpreted in conjunction with dynamical information.

Our long-term integrations further demonstrate that dynamical origin classifications are inherently time-dependent. Meteoroids that appear dynamically asteroidal near the time of Earth impact may have evolved from cometary-like orbital regimes at earlier epochs as a result of planetary perturbations and radiation pressure. 

In particular, we find that JFC meteoroids older than ${\sim}150-200$ kyr can occupy orbital phase space indistinguishable from that of asteroidal material at the time of impact with Earth. As a consequence, millimetre-sized meteoroids on asteroidal-like orbits today may represent either genuinely asteroidal material or older cometary debris, highlighting a fundamental degeneracy between age and origin that cannot be resolved through orbital information alone in the absence of a clear meteoroid age.

We may divide our main findings by speed and assumed ejection age as judged by each of \citet{1954whipple}'s $K$-criterion and \citet{1967kresak}'s $Pe$-criterion.

\begin{enumerate}
    \item For low velocities of 17 km/s or less, most meteoroids appear to have an asteroidal origin provided they are less than $\sim150-200$ kyr in age. Older meteoroids appear to be dynamically cometary in origin but transition into dynamically asteroidal orbits before impacting Earth. However, our simulations suggest this is also roughly the cutoff age where comet vs. asteroidal origins become most unreliable.
    \item For velocities 18 $\leq v <$ 27 km/s, origins are a mix of potentially asteroidal or cometary with increasingly more dominant cometary origins for older material. 
    \item For velocities 27 km/s and above, origins are strongly cometary, although the $Pe$-criterion shows potentially asteroidal origins if the meteoroids are only a few tens of thousands of years old (unlikely for sporadic meteoroids).
    \end{enumerate}

Thermal processing may play a particularly important role in shaping meteoroid structure but not in the manner initially expected. Contrary to our expectations that repeated solar heating should lead to compaction and lower $k_c$ values as bulk density increases, we find that meteoroids which have experienced any degree of thermal processing exhibit statistically higher $k_c$ values than those that have not. We suggest this difference is primarily driven by the strong coupling between thermal exposure and dynamical origin, as cometary meteoroids are both more likely to undergo thermal processing and compositionally more prone to fragmentation. 

However, when controlling for dynamical origin, we find that thermal processing still produces a statistically significant shift towards higher $k_c$, indicating that it acts as a secondary modifier of meteoroid structure. This effect does not scale with the degree of processing, suggesting that a limited number of perihelion passages is sufficient to transition meteoroids into a structurally altered state, after which additional thermal cycling produces little further change prior to complete disruption.

Taken together, these results indicate that thermal processing operates as a threshold process that modifies meteoroid structure and fragmentation behaviour, while dynamical origin sets the primary baseline for $k_c$. As such, $k_c$ should be interpreted as a composite diagnostic, reflecting both intrinsic material properties and evolutionary history, rather than a direct proxy for bulk density or parent body type.

In conclusion, these results support a probabilistic, multi-parameter framework for meteoroid origin classification in which entry velocity provides the most reliable primary indicator of parentage (provided the meteoroids are sufficiently young), while material properties and thermal history provide critical context for understanding structural evolution. Our study finds that without independent constraints on meteoroid age, neither orbital dynamics nor material properties alone can uniquely determine origin. Future work incorporating improved thermal-mechanical modelling, compositional measurements, and larger observational samples will be essential for breaking this degeneracy and for refining our understanding of how asteroidal and cometary material is delivered to Earth.

\section{Acknowledgements}
We thank Dr. Patrick Shober and an anonymous referee for very insightful comments which helped to improve this manuscript. This work was supported in part by NASA co-operative agreement 80NSSC24M0060. The authors would like to thank Zbyszek Krzemenski, Emma Harmos and Maximilian Vovk for their work manually reducing the events from CAMO and some from the EMCCD observations, Auriane Egal for providing her code that was used to calculate the MOID between meteoroid and planet orbits, as well as Michael Mazur for his contribution of Figure 10 showing the differences in collecting efficiency between the EMCCD camera pairs. Dr. Pokorn\'y acknowledges support provided by NASA’s Planetary Science Division Research Program, through ISFM work packages EIMM and Planetary Geodesy at NASA Goddard Space Flight Center, NASA award numbers 80GSFC24M0006.
\printcredits

\bibliographystyle{cas-model2-names}


\appendix

\clearpage
\section*{Appendix}

\section{Data Availability for Observed Meteors} 
\label{appendixa}

The raw data in CSV form can be found on Zenodo under the doi:\href{10.5281/zenodo.19615720}{10.5281/zenodo.19615720} or the direct link: \href{https://zenodo.org/records/19615720}{https://zenodo.org/records/19615720}. This data was used to run the backwards/forwards integrations in REBOUND. The format for the CSV file is shown in \autoref{tab:dataset-descr}.

\begin{table*}[h!]
    \centering
    \begin{tabular}{m{25mm}|m{30mm}|m{95mm}}
        Column Name & Units & Sample Value \\
        \hline
        solution\_id & & 20200427\_065016\_mc\_report.txt \\
        \hline
        velocities & cm/s & 2445907.0 \\
        \hline
        state\_vec & Epoch of Date ECI: [$x$ (m), $y$ (m), $z$ (m), $v_x$ (m/s), $v_y$ (m/s), $v_z$ (m/s)] & [-2552011.57, -3905769.15, 4471125.63, -2092.57, -9899.01, 22268.29] \\
        \hline
        cov\_matrix & Cov($x, y, z, v_x, v_y, v_z$) & [[0.7979437, 0.4260085, -3.595427, 2.178033, 0.3822699, -8.338652],\newline [0.4260085, 1.202905, -3.234345, 0.4124117, 0.07243447, -1.641519],\newline [-3.595427, -3.234345, 23.7437, -8.553311, -2.545831, 5.315135],\newline [2.178033, 0.4124117, -8.553311, 8.159881, 1.261265, -3.044799], \newline [0.3822699, 0.07243447, -2.545831, 1.261265, 0.4155618, -8.737847], \newline [-8.338652, -1.641519, 53.15135, -30.44799, -8.737847, 1.968425]] \\
        \hline
        date & UTC & 2020-04-27 06:50:16.320875\\
        \hline
        begin\_height & km & 94.26405\\
        \hline
        mass & kg & 2.5e-05 \\
        \hline
        a & AU & 2.511536 \\
        \hline
        i & degrees & 35.399508\\
        \hline
        q & AU & 1.006324 \\
        \hline
        e & & 0.599319\\
        \hline
        ascending\_node & degrees & 37.2095 \\
        \hline
        arg\_of\_peri & degrees & 182.467619 \\
        \hline
        julian\_date & & 2458966.785 \\
        \hline
        kc & km & 95.6222219 \\
        \hline
        rho & kg/m$^3$ & 1000.0 \\
        \hline
        diameter & m & 0.003627832 \\
        \hline
        collisional\_lifetime & years & 347355.0 \\
        \hline
        tisserand & $T_J$ & 2.978599\\
        
    \end{tabular}
    \caption{An overview of the raw data contained in the CSV file in Zenodo.}
    \label{tab:dataset-descr}
\end{table*}

\section{EMCCD Dataset Quality Testing}
\label{appendixb}

While the wide-field CAMO dataset employing narrow-field mirror tracking is captured with extremely high precision and each event is manually reduced, the EMCCD dataset is of higher sensitivity but lower precision and most events have automatic trajectory solutions, without manual picks. As such, beyond our initial data quality elimination criteria (a minimum of 6 frames for at least one camera capturing the event, a convergence angle between stations of $>5\degree$, that the meteor begins and ends entirely within the field of view of at least one camera and that the meteor is not captured at the edge of the frame), we compared the 41 events that have both manual reductions and automatic solutions.

\begin{figure*}[h!]
    \centering
    \subfigure[]{\includegraphics[width=0.37\textwidth]{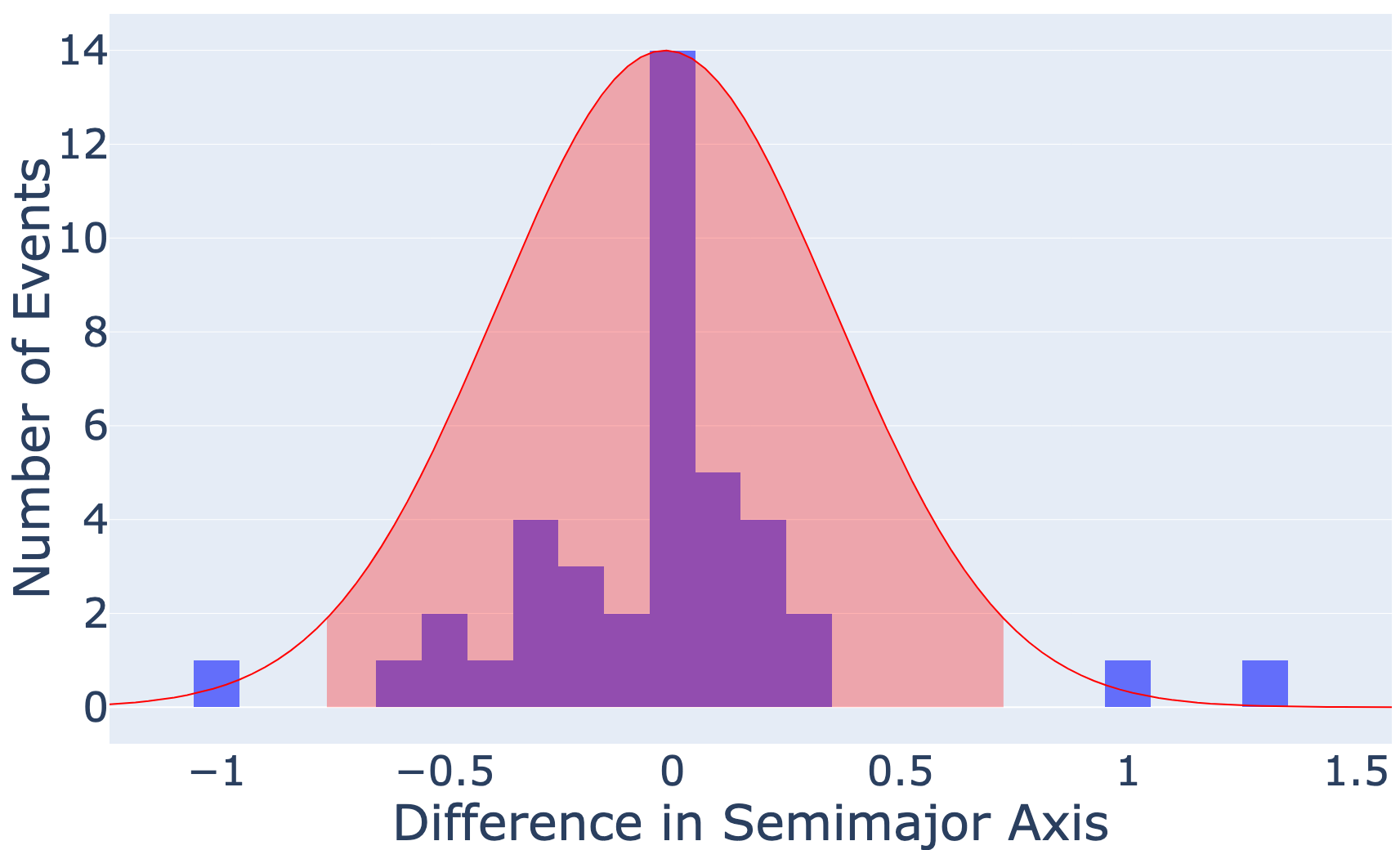}}
    \subfigure[]{\includegraphics[width=0.41\textwidth]{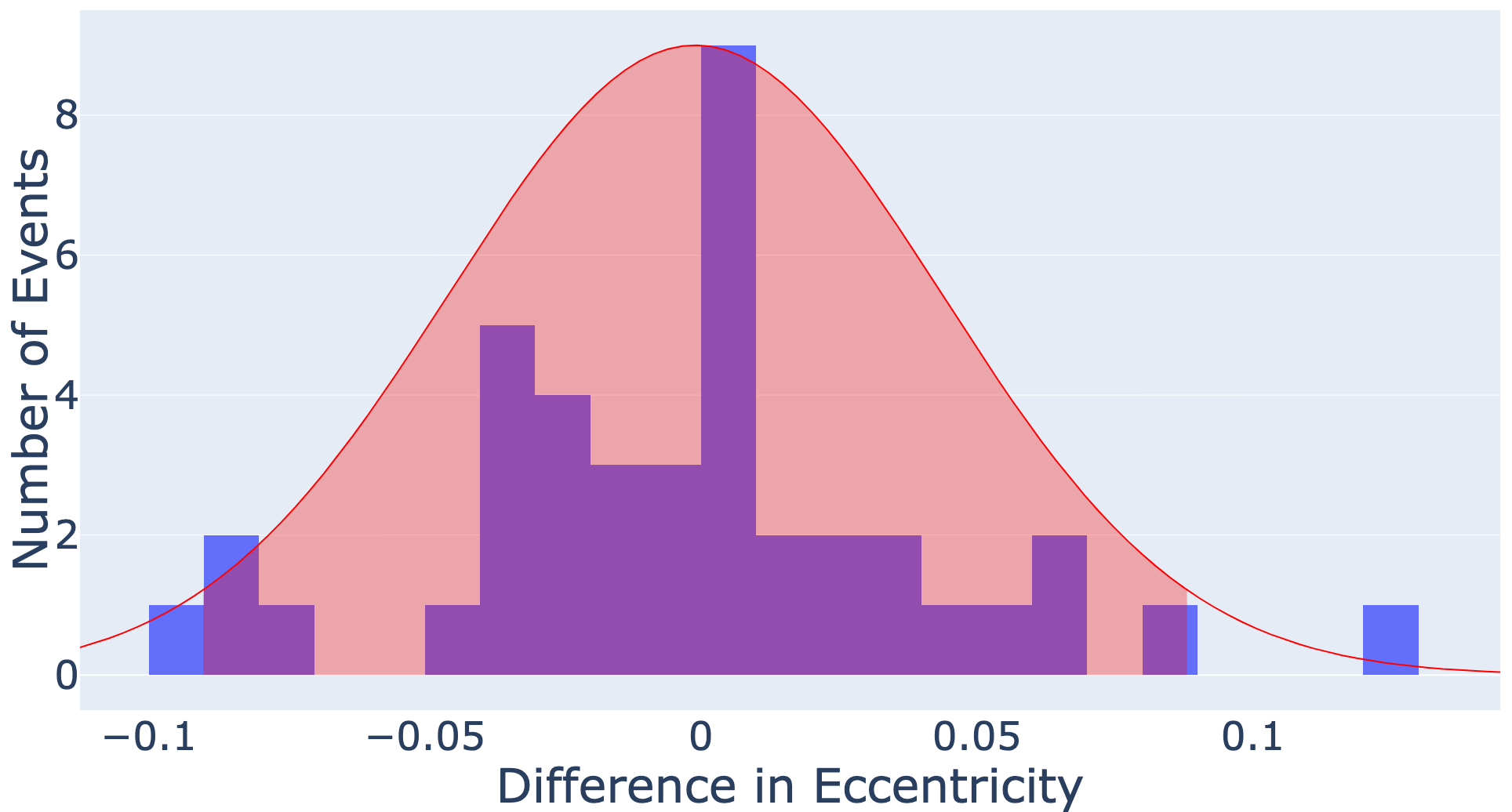}}
    \subfigure[]{\includegraphics[width=0.41\textwidth]{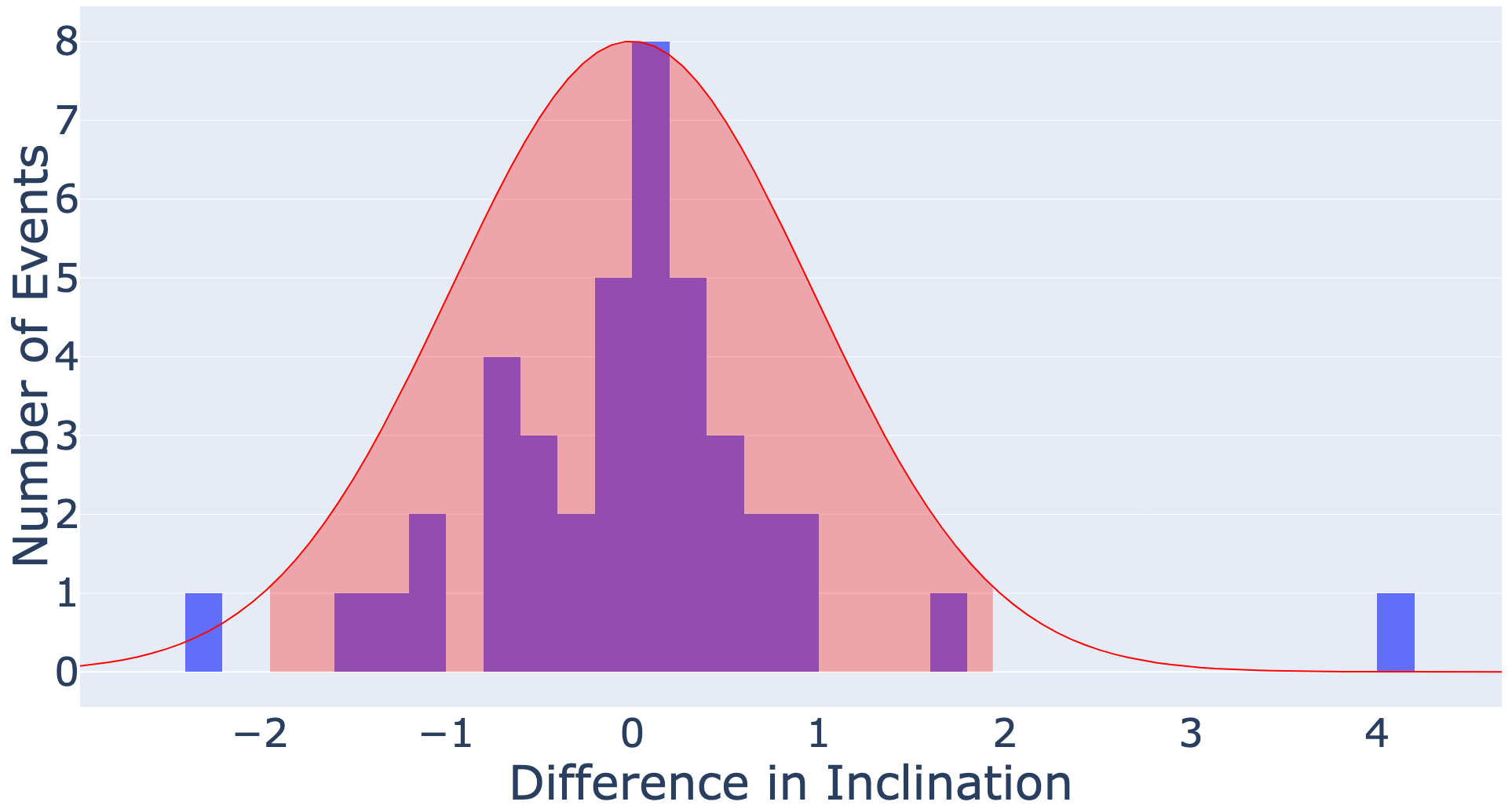}}
    \subfigure[]{\includegraphics[width=0.38\textwidth]{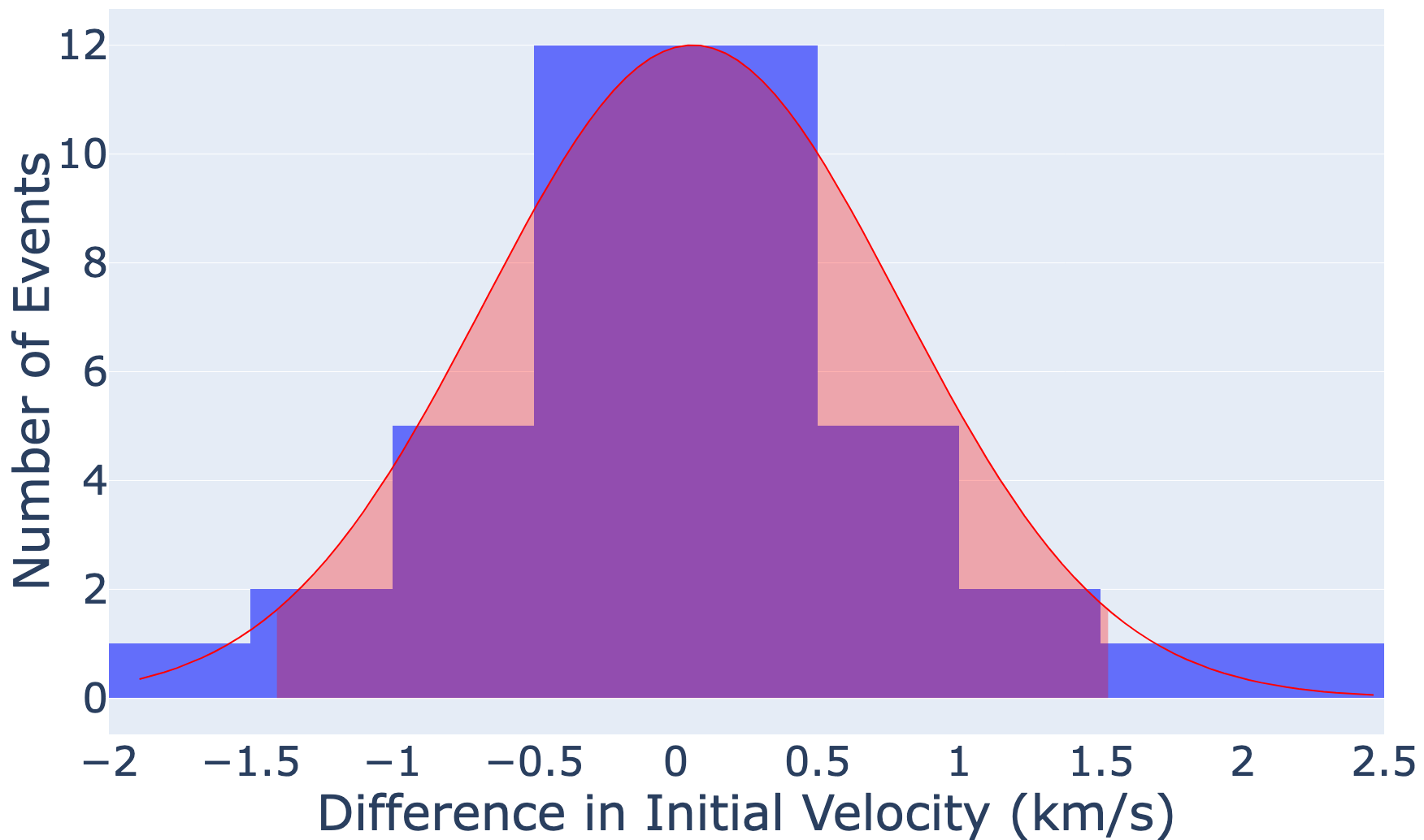}}
    \caption{Histograms of the absolute differences in orbital elements and starting velocity between manually reduced EMCCD events and the automatically generated solution for the same events. A red Gaussian curve is overlaid with the 2-$\sigma$ region shaded in red.}
    \label{fig:diff_emccd}
\end{figure*}

\begin{figure*}[h!]
    \centering
    \subfigure[]{\includegraphics[width=0.38\textwidth]{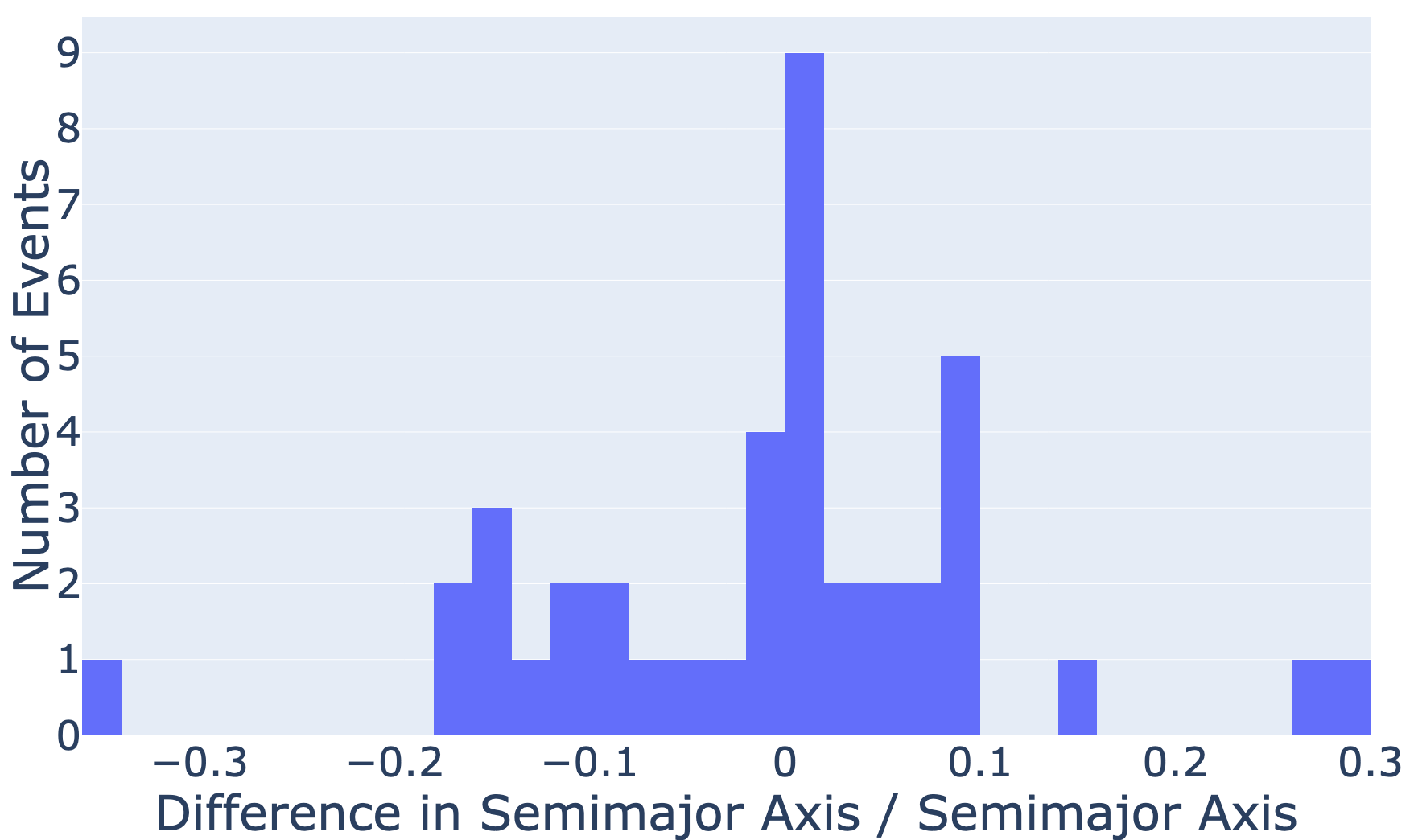}}
    \subfigure[]{\includegraphics[width=0.38\textwidth]{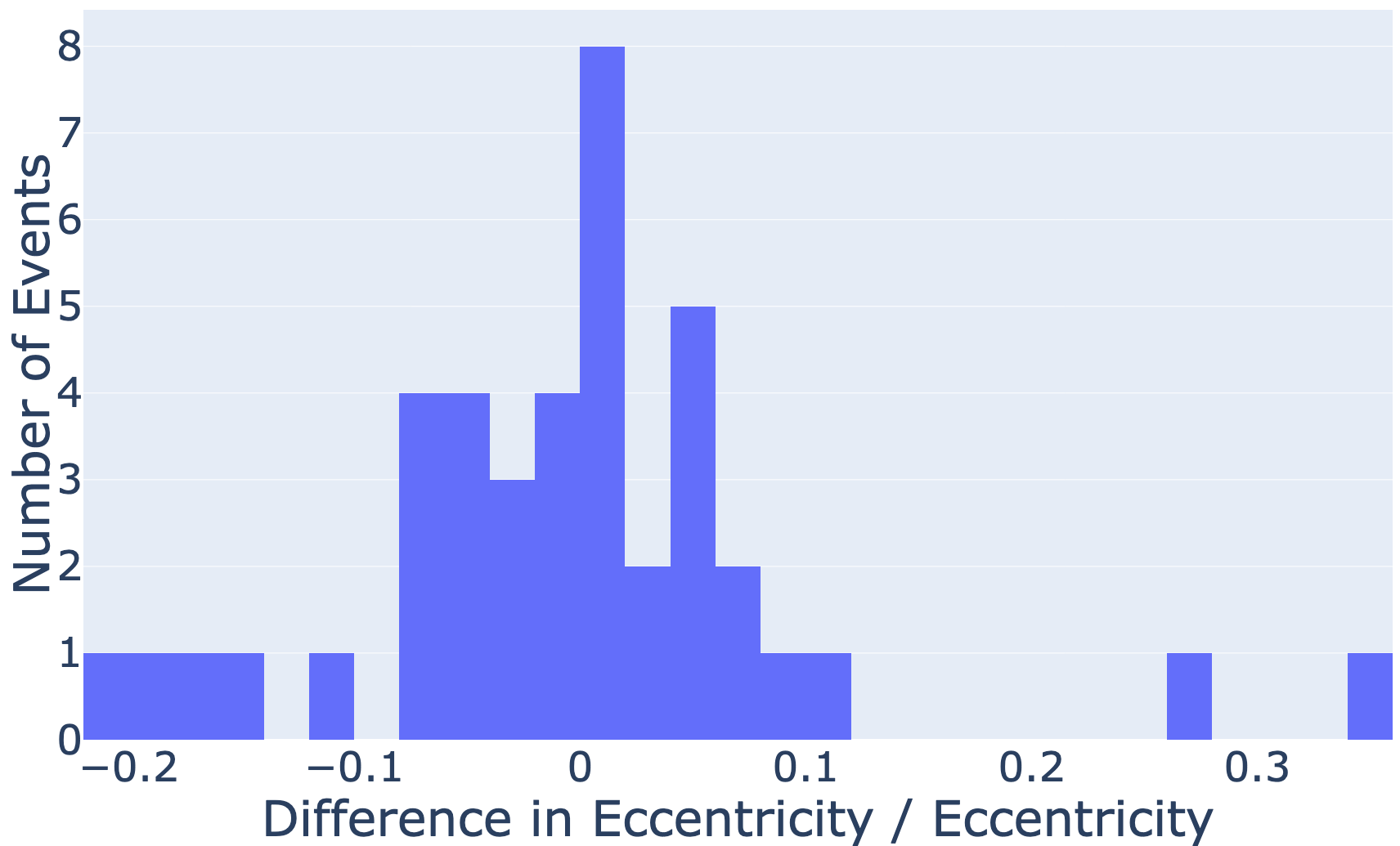}}
    \subfigure[]{\includegraphics[width=0.38\textwidth]{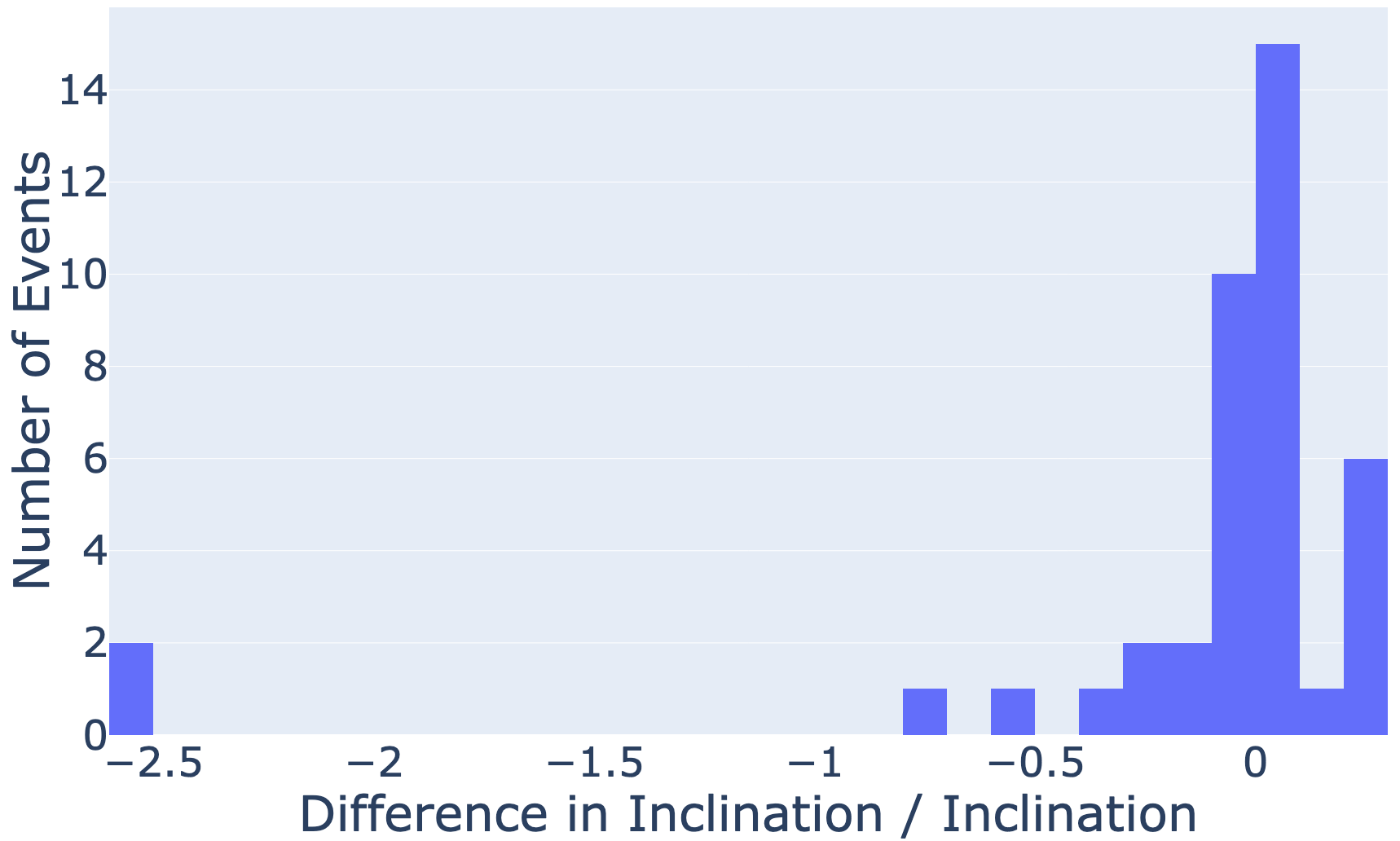}}
    \subfigure[]{\includegraphics[width=0.38\textwidth]{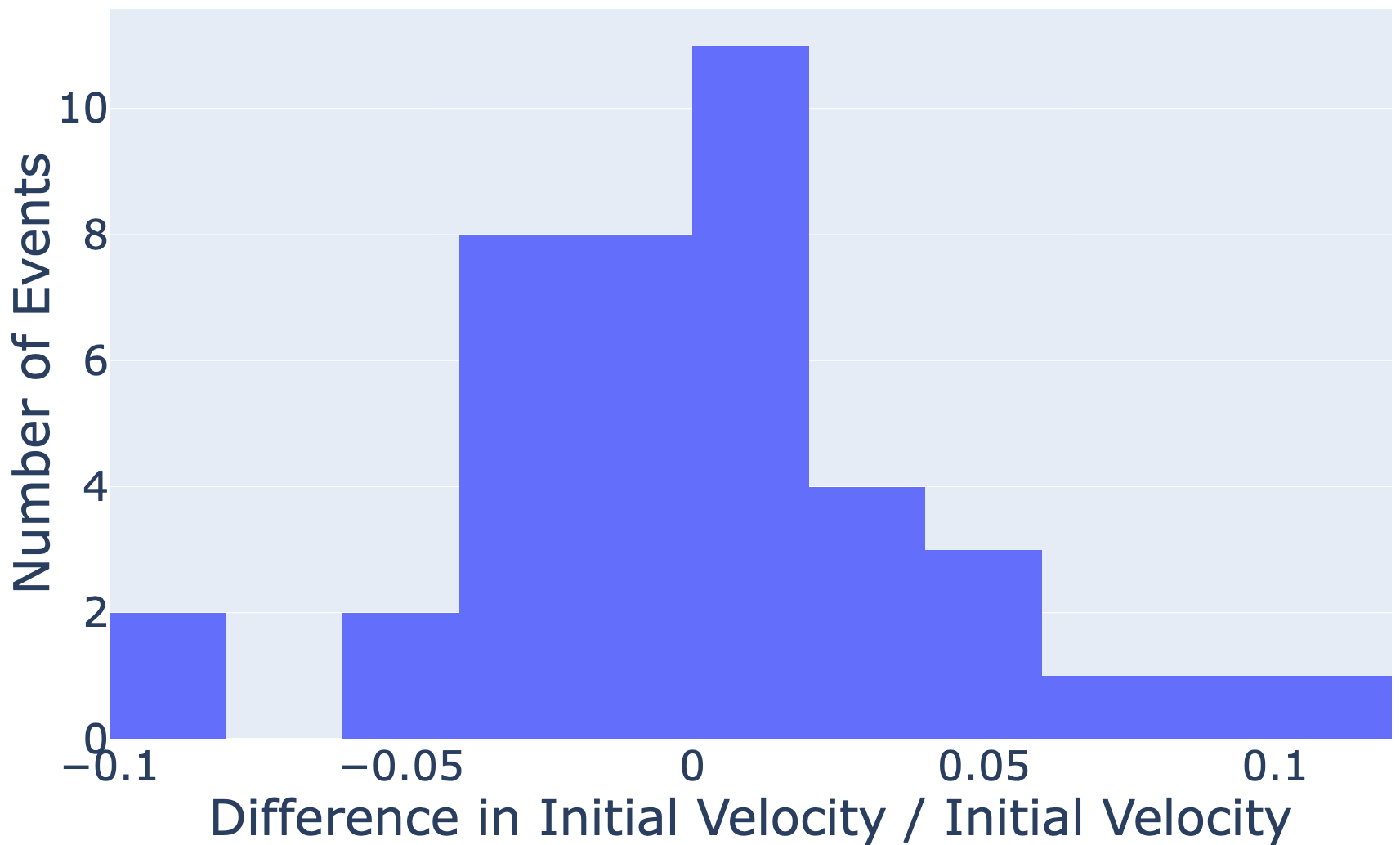}}
    \caption{Histograms of the relative differences ($\Delta a/a$, $\Delta i/i$, etc.) in orbital elements and starting velocity between manually reduced EMCCD events and the automatically generated solution for the same events. Events with relative differences within 10\% are acceptable.}
    \label{fig:diff_rel_emccd}
\end{figure*}

Automatic solutions are generated using frame picks found from the DetApp algorithm \citep{2022gural} which is then used in the meteor trajectory solving software called \verb|pylig| \citep{pylig}. \footnote{Publicly available at https://github.com/wmpg/WesternMeteorPyLib/}. The absolute and relative differences between manual and automatic solutions for the same events are shown in \autoref{fig:diff_emccd} and \autoref{fig:diff_rel_emccd}. 

We then run simulations for both the manually and automatically reduced events using their respective starting conditions. Despite the slight changes in starting conditions, the determination of parent body remains largely the same for the set of 41 meteoroids, with the exception of two events which switch from cometary to asteroidal between the manual and automatic reductions.

We use these 41 comparison cases to validate our ability to accurately judge the quality of an automatically reduced EMCCD solution. Individually and blinded to the results of the manual/automatic comparison results, authors Brown and Do went through the lag and velocity plots of the meteor events generated by the automatic reduction, deciding which events seem acceptable to include and which are not accurately represented. With significantly more experience analyzing meteor event data, Brown's picks are used as a reference standard for Do's ability as she was responsible for going through the automatic events that have no manual reduction counterpart to determine which events will be included in the final dataset. Looking at three criteria for acceptance by the manual/automatic reduction comparison, we consider acceptance of events with a) initial velocities within 0.45 km/s of each other, b) initial velocities within 5\% of each other, and c) initial semimajor axis within 0.5 AU of each other. Out of the 41 events:

\begin{itemize}
    \item There were 14 cases where all criteria were passed and both Brown and Do approved the automatic solutions.
    \item There were 9 cases where Brown and Do disagreed, of which:
    \begin{itemize}
        \item 3 cases pass all criteria and Do approves but Brown rejects
        \item 1 case that passes all criteria and Brown approves but Do rejects
        \item 3 cases that pass criteria b) and c), Do approves but Brown rejects and criteria a) is not passed
        \item 1 case where b) is passed, Do approves but a) and c) are failed and Brown rejects
        \item 1 case where a) and b) are failed, Do rejects, but c) is passed and Brown accepts
    \end{itemize}
    \item 4 cases where both Brown and Do reject but all three criteria are passed
    \item 1 case where both Brown and Do accept but all three criteria are failed
    \item 7 cases that pass criteria b) and c), fail criteria a), both Brown and Do reject
    \item 1 case that passes criteria a) and b) but fails c), Brown and Do both reject
    \item 1 case that passes criteria c), fails criteria a) and b), Brown and Do both approve.
    \item 1 case that passed b) and c) but failed a), Brown and Do both approve
    \item 3 cases that failed a) and b), passed c) and Brown and Do both reject
\end{itemize}

The important takeaway from this validation step is that only 2 cases that failed 2/3 criteria and 1 case that failed all 3 criteria were approved through Do's quality check. From this, we extrapolate that our method of determining which events have a high enough quality solution via automatic \verb|pylig| reduction is accurate for the majority of cases, with a 7.3\% false positive rate of accepting bad data.

\begin{figure}[h!]
    \centering
    \subfigure[]{\includegraphics[width=0.8\linewidth]{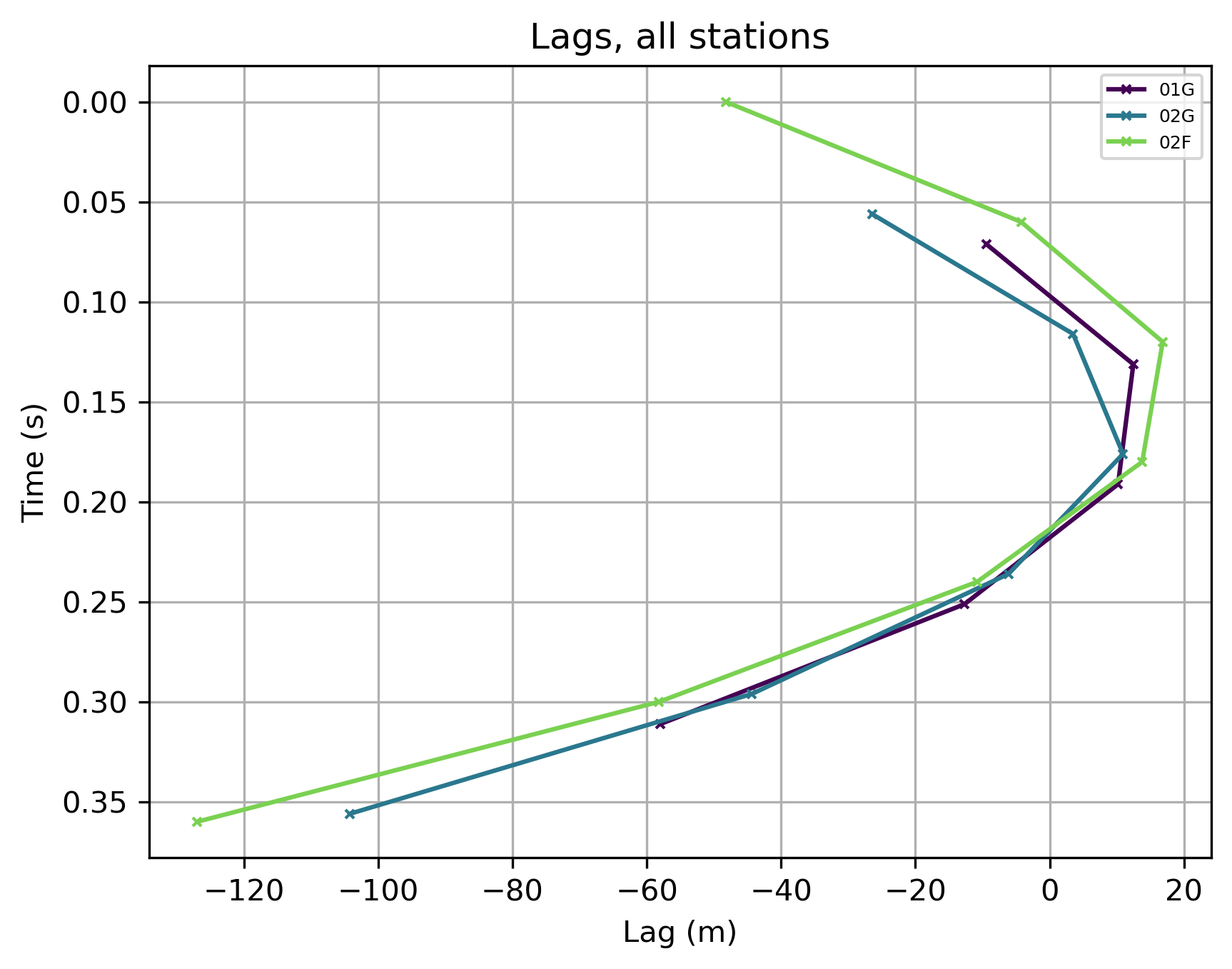}}
    \subfigure[]{\includegraphics[width=0.8\linewidth]{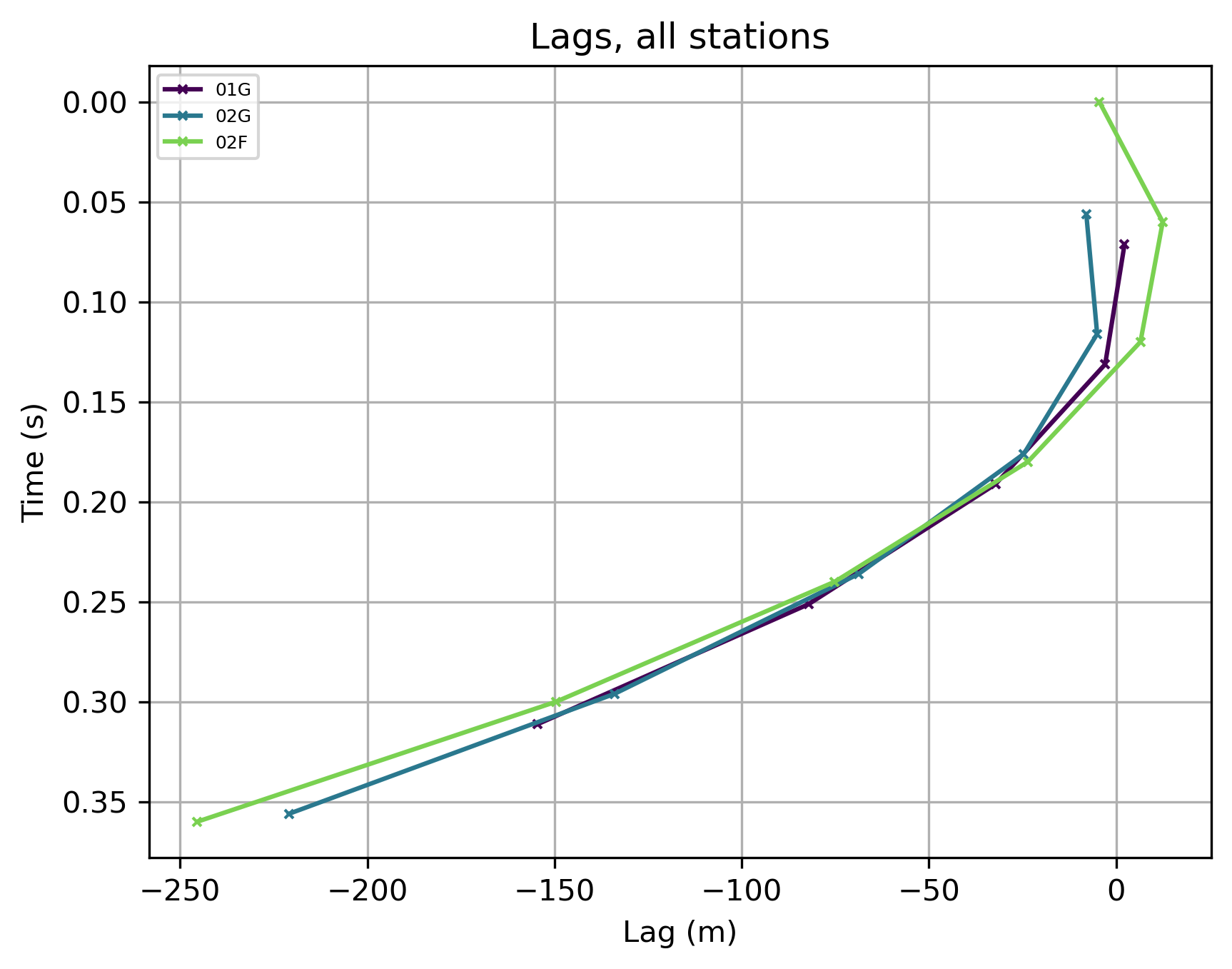}}
    \caption{An example showing lags for automatic vs manual initial velocity determination. In a), we see the lags of the automatic initial velocity determination which is curved such that the initial velocity is underestimated. In b), we've determined that the "knee" of the lag curve in a) is at around 0.15 s, corresponding to a height of 91.9 km, so averaging only the points above that height to determine the initial velocity produces a lag curve that it more vertical at the start of the event that then decelerates around 0.15 s.}
    \label{fig:initvel-lag}
\end{figure}

The next quality check we implemented was to compare the automatically determined initial velocity that \verb|pylig| calculates based on a sliding window with a manually chosen minimum height selected based on the shape of the lag curve. In \autoref{fig:initvel-lag}, we show the difference between the lags of the observation points for a single event when \verb|pylig| selects the window to average over for the initial velocity vs when we choose the minimum height above which the initial velocities are averaged. When applied to the CAMO dataset, differences between the automatic and manual initial velocities were minimal (on the order of tens of m/s). However, when applied to the EMCCD events, differences were more significant. In the 41 events that were manually reduced, we saw an average of 240 m/s difference in initial velocity, with two events over 1 km/s of difference with the \verb|pylig| automatic velocity determination.

To determine the acceptable threshold of difference between the manually and automatically determined initial velocities, we ran simulations for events that had varying differences and analyzed whether they returned the same determined parental origins using our methodology. We found that the most sensitive indicator was for the $T_J$ value, specifically for objects sitting near the 3.05 boundary line between cometary and asteroidal orbits. As such, all meteor events with Tisserand between 2.5 and 3.5 were redone with manually chosen initial velocity windows. A total of 35 EMCCD events were redone out of 269 events that had been initially selected for analysis. 

\end{document}